\documentclass[
   aps,prd,         
   twocolumn,       
   10pt,            
   nofootinbib,     
   superscriptaddress,  
   amsmath,amssymb, 
   floatfix,         
   longbibliography
]{revtex4-2}

\usepackage{amsfonts,amsmath,amssymb,amsthm, mathtools}
\usepackage{bm}               
\usepackage{physics}         
\usepackage{slashed}         
\usepackage{booktabs}       
\usepackage{mathrsfs}        
\usepackage{graphicx}        
\usepackage[english]{babel}
\usepackage{microtype}
\usepackage{comment, xcolor}
\colorlet{RED}{black} 
\usepackage{hyperref}
\hypersetup{colorlinks=true, linkcolor=black, filecolor=black, urlcolor=black, allcolors=black}
\usepackage{silence}
\begin{document}

\title{Holographic heat engines for Schwarzschild black holes}

\author{Ripunjay Dwivedi}
\affiliation{Department of Physics, Indian Institute of Technology Bombay, Mumbai, Maharashtra 400076, India}

\author{Manus R. Visser}
\affiliation{Institute for Mathematics, Astrophysics and Particle Physics, and Radboud Center for Natural Philosophy, Radboud University, 6525 AJ Nijmegen, The Netherlands}

\begin{abstract}

We construct reversible black hole heat engines in asymptotically flat
spacetime using quasi-local gravitational thermodynamics. We enclose a Schwarzschild black hole in a finite spherical cavity and
identify the working substance with the holographically dual thermal system on the cavity boundary. The
surface pressure and the boundary area define a thermodynamic pressure-volume
pair, while all coupling constants of the gravitational theory remain fixed.
We derive exact efficiencies for the Carnot, Otto, Diesel, Brayton, and Stirling
engines and compare them numerically. The non-Carnot efficiencies probe
the quasi-local equations of state of the Schwarzschild black hole.  
{\color{black}Both the regenerative and non-regenerative Stirling efficiencies on the large
black hole branch approach the Carnot value in the high-temperature limit,
with regeneration reducing the leading deviation from the Carnot bound.
}
\end{abstract}

\maketitle

\section{Introduction}
\label{sec:intro}

Black holes are thermodynamic objects. They possess a temperature and entropy, and obey laws directly analogous to the four laws of thermodynamics \cite{Bardeen1973,Bekenstein1973,Hawking1975}. This suggests a simple but far-reaching question: can a black hole be used as the working substance of a heat engine?

Early thought experiments considered cyclic engine processes involving black holes, but typically with the black hole acting as a heat sink rather than as the working substance itself \cite{Geroch1971,Bekenstein1973,SCIAMA1976385,Unruh:1982ic}. Related perspectives on black hole thermodynamic cycles and engines were developed in later work \cite{Kaburaki1991,Landsberg1992,Deng2009,Richterek,Curiel:2014zua,Bravetti:2015xsp,Prunkl:2019wdw}. The problem of using the black hole itself as the working substance of a heat engine was put on a systematic footing in extended black hole thermodynamics, where the cosmological constant is promoted to a thermodynamic pressure and its conjugate is interpreted as a thermodynamic volume \cite{Kastor2009,Dolan2011,Cvetic2011,Kubiznak2012}. This framework led to holographic black hole heat engines in asymptotically anti-de Sitter (AdS) spacetime \cite{Johnson2014}, with subsequent work computing efficiencies for a variety of black holes and cycles \cite{Johnson2016,Chakraborty:2016ssb,Hennigar2015,Wei2014,Johnson:2019olt}.

However, the pressure in extended black hole thermodynamics carries a conceptual subtlety. Since $P\propto-\Lambda$ \cite{Kastor2009,Dolan2011}, varying the pressure during a cycle means varying the cosmological constant. In holography this changes the central charge, or equivalently the number of field theory degrees of freedom, of the dual conformal field theory (CFT) \cite{Karch2015,Visser2022}. Such a cycle therefore does not describe a path through the thermodynamic state space of one fixed theory, but rather a trajectory through a family of theories \cite{Johnson2016,Mancilla:2024spp}.

A recent construction avoids this issue by taking the working substance to be a thermal CFT state dual to an AdS black hole \cite{LilaniVisser2025}. In that setting, pressure and volume are defined directly in the boundary theory, so that the cosmological constant and number of degrees of freedom remain fixed. However, this construction is intrinsically tied to the AdS/CFT correspondence and is therefore restricted to asymptotically AdS spacetimes.

In this work, we develop a construction of heat engines for black holes in asymptotically flat spacetime. The key step is to place the black hole inside a finite timelike boundary and use the quasi-local gravitational formalism pioneered by York \cite{York1990}. In this framework, thermodynamic quantities are assigned at the boundary of the finite cavity, with the Brown-York stress-energy tensor providing the quasi-local energy and surface pressure~\cite{Brown1993}. \textcolor{black}{We identify the working substance with the thermal system on this timelike boundary, which we interpret holographically as being encoded by the black hole inside the cavity.} The surface pressure and boundary area then provide a thermodynamic pressure-volume pair for the boundary system \cite{Borsboom:2026ash}. Unlike in extended black hole thermodynamics, this pressure is not obtained by varying the cosmological constant. The heat engine cycle therefore takes place within the thermodynamic state space of a single gravitational theory, with all coupling constants held fixed.

We illustrate the construction with a four-dimensional Schwarzschild black hole enclosed in a spherical cavity. This is the simplest asymptotically flat example in which thermal equilibrium is well-defined, and the thermodynamic state space can be parametrized explicitly. The same logic is not restricted to asymptotically flat spacetimes, but can be applied whenever a suitable timelike boundary and quasi-local thermodynamic description are available, including black holes with other asymptotics.

Within this framework, we construct the Schwarzschild-cavity analogues of the standard reversible heat engines. We derive exact efficiencies for the Carnot, Otto, Diesel, Brayton, and Stirling cycles (\textcolor{black}{regenerative and non-regenerative}),  compare them numerically, and display their paths in the $P$-$V$ and $T$-$S$ planes. Unlike the   CFT engines of~\cite{LilaniVisser2025}, whose efficiencies are often fixed by the universal scale-invariant equation of state, the non-Carnot efficiencies found here probe the specific quasi-local equations of state of the Schwarzschild black hole in a cavity. In this sense, these heat engines become a diagnostic of quasi-local gravitational thermodynamics: their non-Carnot efficiencies encode how the black hole equation of state differs from that of ordinary matter.

\section{Holographic thermodynamics of a Schwarzschild black hole in a  cavity}
\label{sec:thermo}

We consider a four-dimensional, asymptotically flat Schwarzschild black hole enclosed in a finite spherical cavity. The spacetime metric is given by
\begin{align}
    ds^2 &= - f(r)\, dt^2 + f^{-1}(r)\,dr^2 + r^2 d\Omega_{2}^2\,,\\
    f(r) &= 1 - \frac{r_h}{r}\,,
\end{align}
where $r_h$ is the horizon radius. The spherical cavity is located at the radius   $r_B>r_h$. 
{\color{black}
We assume that the quasi-local thermodynamic system defined on the timelike
boundary at $r=r_B$ admits a holographic interpretation: the boundary thermal
state is encoded by the Schwarzschild geometry inside the cavity. In this
sense, the timelike boundary acts as a holographic screen \cite{Bousso1999}.
In the membrane paradigm \cite{Damour1978,Thorne1986}, in which the boundary is
taken toward the near-horizon limit $r_B\to r_h$, this screen approaches the
stretched horizon.

We restrict attention to thermal equilibrium configurations rather than
evaporating black holes. The Schwarzschild black hole is maintained in
equilibrium by an incoming thermal flux that balances the outgoing Hawking
flux, as in the Hartle-Hawking construction \cite{Hartle:1976tp}. The
thermodynamic state functions used below are evaluated at leading
semiclassical order on the Schwarzschild black hole saddle of the Euclidean
gravitational path integral.
}

The equilibrium thermodynamics of the boundary system can be described in terms of the thermodynamic state variables $(E,T,S,P,V)$. These comprise the quasi-local internal energy~$E$, the boundary temperature~$T$, the entropy~$S$, the holographic pressure~$P$, and the thermodynamic volume~$V$ associated with the cavity wall. These quantities are not independent: the Schwarzschild cavity equilibrium family is two-dimensional, as is already apparent from the two geometric parameters $r_h$ and $r_B$. A choice of two independent thermodynamic variables specifies a thermodynamic representation~\cite{callen1985thermodynamics}. In each representation, the corresponding thermodynamic potential expressed as a function of the chosen variables is the fundamental relation; the equations of state are then obtained from its first derivatives \cite{Borsboom:2026sex}.

Taking the entropy~$S$ and volume~$V$ as independent variables, the thermodynamic potential is the internal energy~$E(S, V)$, known as the \emph{energy representation}. Its differential  yields
\begin{equation}\label{eq:firstlaw}
    d E = T d S - P d V\,.
\end{equation}
In the present spherically symmetric setup, the entropy and volume are determined solely by the horizon radius $r_h$ and the cavity radius $r_B$, respectively. The entropy is given by the Bekenstein-Hawking formula and the thermodynamic volume is identified with the   area of the spherical cavity 
\begin{equation}\label{eq:SV}
    S = \frac{\pi r_h^{2}}{G}\,, \qquad V = 4\pi r_B^{2}\,.
\end{equation}
{\color{black}
Although $V$ is geometrically the area of the cavity wall from the bulk
perspective, it is the spatial volume of the boundary thermodynamic system,
which has two spatial dimensions. We therefore refer to $V$ as the
 boundary volume. Its conjugate $P$ is a surface pressure, so that
$PdV$ has the dimensions and  interpretation of mechanical work.
}

 The internal energy $E$ is equal to the quasi-local Brown-York energy \cite{York1990,Brown1993}, which can be written in terms of the state variables $(S,V)$ as
\begin{equation}\label{eq:U}
    E(S,V) = \frac{1}{2G}\sqrt{\frac{V}{\pi}}\left[\,1 - \sqrt{1 - \sqrt{\tfrac{4GS}{V}}}\,\right]. 
\end{equation}
Similarly, the local temperature at the boundary is   the   redshifted Hawking
temperature seen by an observer at $r=r_B$, also known as the Tolman temperature, which as a function of $(S,V)$ is
\begin{equation}\label{eq:T}
    T(S,V) = \frac{1}{4\sqrt{\pi G S}\,\sqrt{1 - \sqrt{4GS/V}}}\,.
\end{equation}
According to \cite{Banihashemi:2024yye,Borsboom:2026ash}, the holographic pressure $P$ is the Brown-York  surface pressure  \cite{York1990,Brown1993} on the cavity wall 
\begin{equation}\label{eq:P}
    P(S,V) = \frac{1}{4G\sqrt{\pi V}}\left[ \frac{1 - \tfrac{1}{2}\sqrt{4GS/V}}{\sqrt{1 - \sqrt{4GS/V}}} - 1 \right]\,.
\end{equation}
A flat-space background subtraction is applied to both $E$ and $P$, so that they vanish identically in   Minkowski spacetime, for which  $r_h = 0$. In the infinite-volume limit at fixed entropy ($r_B \to \infty$, $r_h$ fixed), the  energy becomes   the ADM mass $E \to M$, the boundary temperature reduces to the Hawking temperature $T \to T_{\rm H}$, and the pressure vanishes, $P \to 0$. Consequently, the work term $P dV$ in~\eqref{eq:firstlaw} vanishes, and the differential fundamental relation reduces to $d M = T_{\rm H} d S$.

Two further thermodynamic representations are relevant for the cycle calculations below. In the \emph{Helmholtz (canonical) representation}, the independent variables are the temperature $T$ and the volume $V$, and the thermodynamic potential is the Helmholtz free energy 
\begin{equation}
F = E - TS
\end{equation}
and its variation yields 
\begin{equation}
d F = -S d T - P d V .
\end{equation}
To express thermodynamic quantities in terms of $(T,V)$, we must invert the Tolman temperature equation \eqref{eq:T} to obtain the entropy equation of state $S(T,V)$. This inversion is also required to parametrize the isothermal processes that appear in the Carnot and Stirling cycles.

In four spacetime dimensions, the Tolman temperature equation \eqref{eq:T} can be rearranged into a cubic equation for the horizon radius,
\begin{equation}\label{eq:cubic}
    r_h^3 - r_B\, r_h^2 + \frac{r_B}{(4\pi T)^2} = 0\,.
\end{equation}
York \cite{York1985a,York1990} showed that this cubic equation admits two distinct positive, real roots in the physical range $0<r_h<r_B$ when $r_B T>\sqrt{27}/(8\pi)$. Introducing $x\equiv r_h/r_B=\sqrt{4GS/V}$, these roots correspond to the small black hole branch $0<x<2/3$ and the large black hole branch $2/3<x<1$. {\color{black}
Formally, thermodynamic cycles can be drawn on either branch. For the
reservoir driven reversible cycles considered here, however, we
restrict to the large black hole branch. The fixed volume heat capacity is
positive for $2/3<x<1$, so this branch is locally stable under quasi-static heat
exchange at fixed boundary volume. By contrast, the small black hole branch has
$C_V<0$: a temperature fluctuation induced by heat exchange is amplified rather
than damped, making this branch unsuitable as a  passively  stable working
substance.

}

In closed form, the horizon radius for the large black hole is given by  
\begin{equation}\label{eq:rh_large}
    r_h(T, V) = \frac{r_B}{3} \left[ 1 + 2\cos\!\left(\frac{\alpha(T,V)}{3}\right) \right], 
\end{equation}
where the auxiliary angle $\alpha$ is defined by
\begin{equation}\label{eq:alpha}
    \alpha(T, V) = \arccos\!\left[1 - \frac{27}{8\pi V T^2}\right], \qquad 0 \leq \alpha \leq \pi\,.
\end{equation}
The horizon radius for the  small black hole
branch takes the same form, except that the   cosine argument in \eqref{eq:rh_large} is shifted by
$4\pi/3$, i.e. $\cos [(\alpha + 4\pi)/3]$.

In the \emph{enthalpy representation}, the two independent variables are the entropy $S$ and the pressure $P$, and the thermodynamic potential is the enthalpy
\begin{equation}\label{eq:H_r}
    H = E + PV  \, .
\end{equation}
Its differential satisfies 
\begin{equation}\label{eq:dH}
    d H = T d S + V d P\,.
\end{equation}
The enthalpy is particularly useful for the Brayton and Diesel cycles because, along a reversible isobaric stroke, the  heat exchanged is equal to the change in enthalpy between the endpoints.

To express the thermodynamic quantities in terms of $(S,P)$, the pressure equation \eqref{eq:P} must be inverted for the volume.   Following \cite{Borsboom:2026sex}, we introduce the dimensionless redshift parameter
\begin{equation}\label{eq:y}
    y \equiv \sqrt{1 - \frac{r_h}{r_B}}\,, \qquad 0 < y < 1\,,
\end{equation}
which measures the gravitational redshift between the horizon and the cavity wall. In terms of $y$,  the pressure equation can be rewritten as 
\begin{equation}\label{eq:yconstraint}
    (1-y)^3(1+y) = 16\,G P \sqrt{\pi G S}\; y\,.
\end{equation}
For $S>0$ and $P>0$ this admits a unique physical root $y(S,P)\in(0,1)$, which can be obtained in closed form via Ferrari's method  as \cite{Borsboom:2026sex}
\begin{equation}\label{eq:ySP}
    y(S, P) = \frac{1}{2} - \frac{1}{2}w + \frac{1}{2}\sqrt{3 - w^2 - 2w + 2\sqrt{w^4 - 2w^2 + 5}} \,,
\end{equation}
where $w=w(S,P)$. In terms of the dimensionless combination $Z \equiv GP\sqrt{\pi GS}$, the function $w(S,P)$ is  
\begin{equation}\label{eq:wSP}
\begin{split}
    &w(S, P) \\
    =& \Bigg[ 1 + \sqrt[3]{32Z\left( 1 + 4Z + \sqrt{1 + \tfrac{184}{27}Z + 16Z^2} \right)} \\
    &+ \sqrt[3]{32Z\left( 1 + 4Z - \sqrt{1 + \tfrac{184}{27}Z + 16Z^2} \right)} \Bigg]^{1/2} .
\end{split}
\end{equation}
 The equations of state then take the compact form
\begin{equation}\label{eq:TVofSP}
    T(S,P) = \frac{1}{4\sqrt{\pi G S}\, y(S,P)}\,, \,\,\, V(S,P) = \frac{4GS}{\left[1 - y(S,P)^2\right]^2}\,,
\end{equation}
and the enthalpy is
\begin{equation}\label{eq:HSP}
    H(S,P) = \sqrt{\frac{S}{\pi G}}\;\frac{1 + 3\,y(S,P)}{4\,y(S,P)\left[1 + y(S,P)\right]}\,.
\end{equation}
These closed-form expressions for the enthalpy and equations of state in this representation are all that is required to evaluate the isobaric strokes of the Brayton and Diesel cycles below.

\section{Reversible Heat Engines for Schwarzschild Black Holes}
\label{sec:engines}

A heat engine is a device that operates in a closed thermodynamic cycle, absorbs heat from a   source, converts part of this heat into mechanical work, and releases the remaining heat to a   sink. We denote by $Q_{\mathrm{in}}$ the total heat absorbed from the source, by $Q_{\mathrm{out}}$ the total heat released to the sink, and by $W$ the net work performed by the working substance (system) during one complete cycle. The efficiency is defined as
\begin{equation}\label{eq:effdef}
\eta
=
\frac{W}{Q_{\mathrm{in}}}
=
1-\frac{Q_{\mathrm{out}}}{Q_{\mathrm{in}}}\,,
\end{equation}
where $Q_{\mathrm{in}}$ and $Q_{\mathrm{out}}$ are taken to be positive quantities.

We consider idealized reversible heat engines: the working substance evolves quasi-statically through equilibrium states, and no entropy is produced. The heat source and heat sink are modelled as one or more thermal reservoirs, large enough that their temperatures remain unchanged during heat exchange. Isothermal strokes involve heat exchange with reservoirs at a fixed temperature. More generally, when heat is exchanged along a reversible non-isothermal stroke, it should be understood as quasi-static exchange with a continuous family of auxiliary reservoirs whose temperatures match the instantaneous temperature of the working substance.

In this paper, the working substance is not an ideal gas in a cylinder, but the thermal system on the spherical cavity wall, described holographically by a   Schwarzschild black hole inside the cavity. The thermodynamic volume $V$ is the area of the cavity wall, and the pressure $P$ is the corresponding surface pressure. Along a reversible stroke, heat exchange is given by $\delta Q = T dS$, while the mechanical work performed by the working substance is $\delta W = P dV$. The quasi-local energy $E$ accounts for the balance between these heat and work contributions.

This gives a direct geometric interpretation of the cycle. Heat exchange changes the black hole entropy and hence the horizon size: when heat is absorbed, $S$ increases and the black hole grows; when heat is rejected, $S$ decreases and the black hole shrinks. Physically, heat rejection can be viewed as a controlled quasi-static extraction of energy through the Hawking radiation reaching the cavity wall, which is absorbed by the cold reservoir, rather than as uncontrolled black hole evaporation. Mechanical work, on the other hand, is associated with rescaling the cavity wall. An expansion increases $V$ and corresponds to work performed by the working substance, whereas a compression corresponds to work performed on it. The relation between heat flow and cavity motion depends on the type of stroke.

The reversible cycles considered below are built from four elementary strokes: isothermal, adiabatic, isochoric, and isobaric processes. In an \emph{isochoric} stroke, $dV=0$, the cavity wall is held fixed. No mechanical work is performed, and heat exchange changes only the entropy. Isochoric heat input  grows the black hole, while isochoric heat rejection shrinks it. The associated temperature change is branch-dependent. At fixed $V$, one finds
\begin{equation}
\left(\frac{\partial T}{\partial S}\right)_V
=
\frac{T(3x-2)}{4S(1-x)}\,.
\end{equation}
 Thus, heat input at fixed volume raises the boundary temperature on the large black hole branch ($x>2/3$), but lowers it on the small black hole branch  ($x<2/3$). Conversely, heat rejection lowers the   temperature on the large branch but raises it on the small branch.  The sign change in $\left(\partial T/\partial S\right)_V$ is the same sign change as that of the fixed-volume heat capacity $C_V$ at $x=2/3$ \cite{York1990}.

In an \emph{adiabatic} stroke, no heat is exchanged. Since the process is reversible, $dS=0$, so the horizon size remains fixed. The only geometric change is the rescaling of the cavity wall, and the corresponding change in quasi-local energy is entirely due to mechanical work. This is independent of the large/small branch distinction.

In an \emph{isobaric} stroke, $dP=0$, the holographic pressure is held fixed. The horizon and cavity wall must then vary in a correlated way: as heat exchange changes~$S$, the volume~$V$ adjusts so that the state remains on the isobaric curve $P(S,V)=P_0$. At positive pressure this response does not change sign between the small and large black hole branches: heat input increases the entropy   and is accompanied by an increase in $V$. {\color{black}
However, the boundary temperature decreases along such a stroke, reflecting the
negative fixed-pressure heat capacity $C_P$ \cite{Comer:1992pc,Borsboom:2026sex}.
Consequently, the isobaric path should not be interpreted as a passively stable
equilibrium under unconstrained pressure and temperature fluctuations. In the
Diesel and Brayton cycles below, an isobar is instead treated as an actively
controlled quasi-static path: the cavity wall and the heat exchange are adjusted
continuously so that the Brown-York pressure remains fixed while the system
passes through equilibrium states.
}

In an \emph{isothermal} stroke, $dT=0$, the boundary temperature at the cavity wall is held fixed. The cavity wall expands or contracts while the working substance remains in thermal contact with a reservoir at the same boundary temperature, so both $S$ and $V$ vary along the isothermal curve $T(S,V)=T_0$. Here the relation between heat flow and cavity motion is branch-dependent. Along an isotherm one finds
\begin{equation}
\left(\frac{\partial V}{\partial S}\right)_T
=\frac{4G(3x-2)}{x^3}\,.
\end{equation}
Thus $\left(\partial V/\partial S\right)_T>0$ on the large black hole branch, while $\left(\partial V/\partial S\right)_T<0$ on the small black hole branch. Isothermal heat input increases the entropy and grows the horizon on both branches. On the large branch, this is accompanied by expansion of the cavity, whereas on the small branch, it is accompanied by compression. Conversely, isothermal heat rejection decreases the entropy and shrinks the horizon on both branches; the cavity is compressed on the large branch but expanded on the small branch.


{\color{black}
At fixed boundary temperature, the large black hole branch has negative
isothermal bulk modulus and is therefore mechanically unstable under
unconstrained fluctuations of the cavity volume
\cite{Borsboom:2026sex}. The isothermal strokes in the Carnot and Stirling
cycles are instead treated as externally constrained quasi-static paths: the
cavity radius is varied slowly according to a prescribed protocol, while heat
is exchanged with a reservoir to keep the boundary temperature fixed.
}

Combining these four strokes gives the Schwarzschild-cavity analogues of the standard reversible engines.

\section{Efficiencies of standard holographic heat engines}
\label{sec:efficiencies}

\begin{figure*}[!ht]
    \centering
    \begin{minipage}[t]{0.48\textwidth}
        \vspace{0pt}\centering
        \includegraphics[height=0.22\textheight]{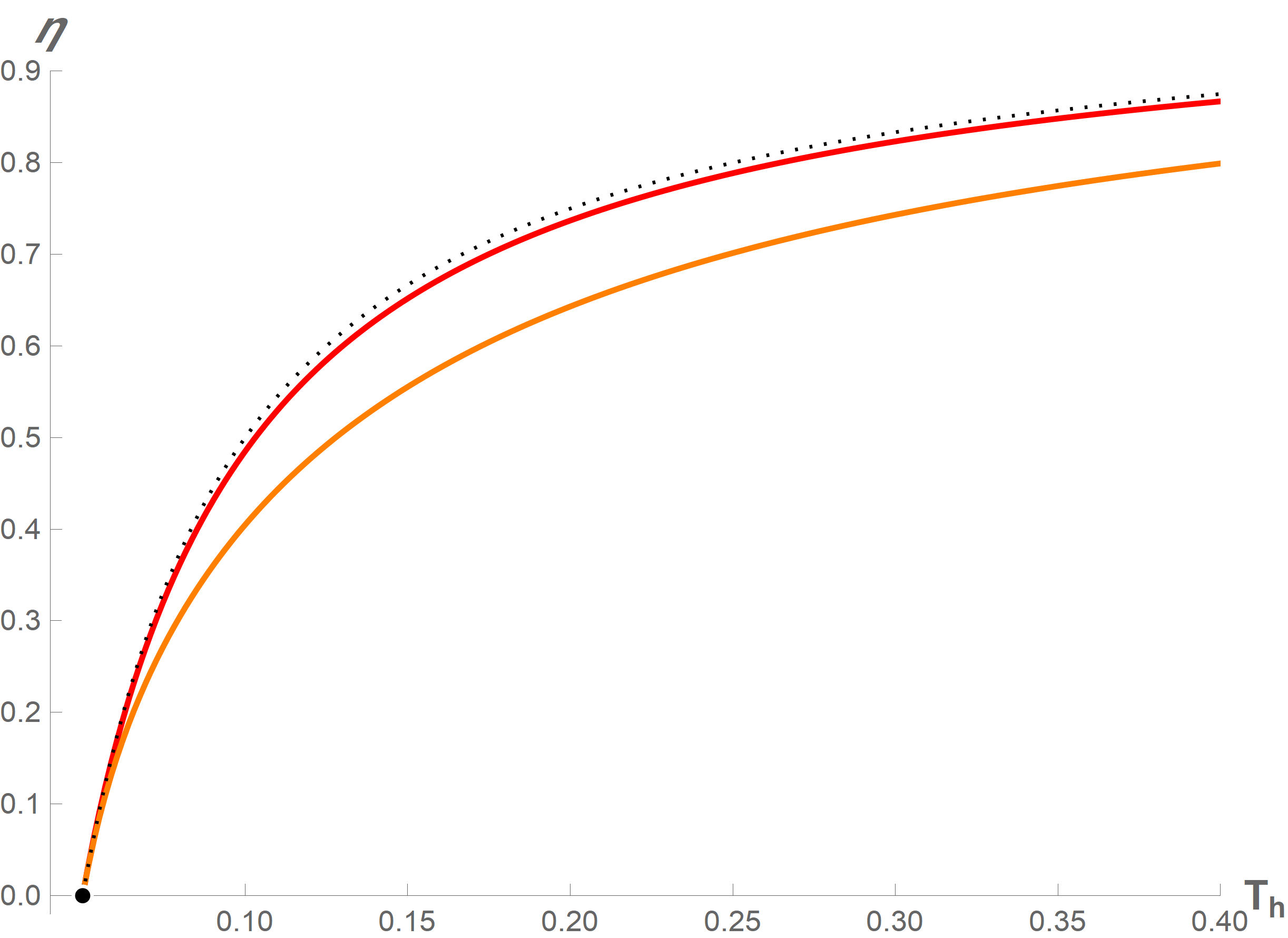}
        \caption{\textit{Efficiency vs.\ hot reservoir temperature for the large black hole branch.}
       \textcolor{black}{
The regenerative (red) and non-regenerative (orange) Stirling
efficiencies are shown as functions of $T_{\rm h}$ at fixed
$T_{\rm c}=0.05$, $V_1=V_{\min}=260$, and
$V_2=V_{\max}=440$.
The black dotted curve is the Carnot bound.
Both Stirling efficiencies approach zero as
$T_{\rm h}\to T_{\rm c}^{+}$; at $T_{\rm c}=T_{\rm h}$, marked by a dot, the cycle is
degenerate.
As $T_{\rm h}$ increases, both efficiencies approach the Carnot
bound from below, with the regenerative efficiency remaining closer
to it.
There is no finite upper branch endpoint: at fixed volume, the
large branch solution approaches $x=1$ only asymptotically as
$T_{\rm h}\to\infty$.
}
        }
        \label{fig:eff_th}
    \end{minipage}\hfill
    \begin{minipage}[t]{0.48\textwidth}
        \vspace{0pt}\centering
        \includegraphics[height=0.22\textheight]{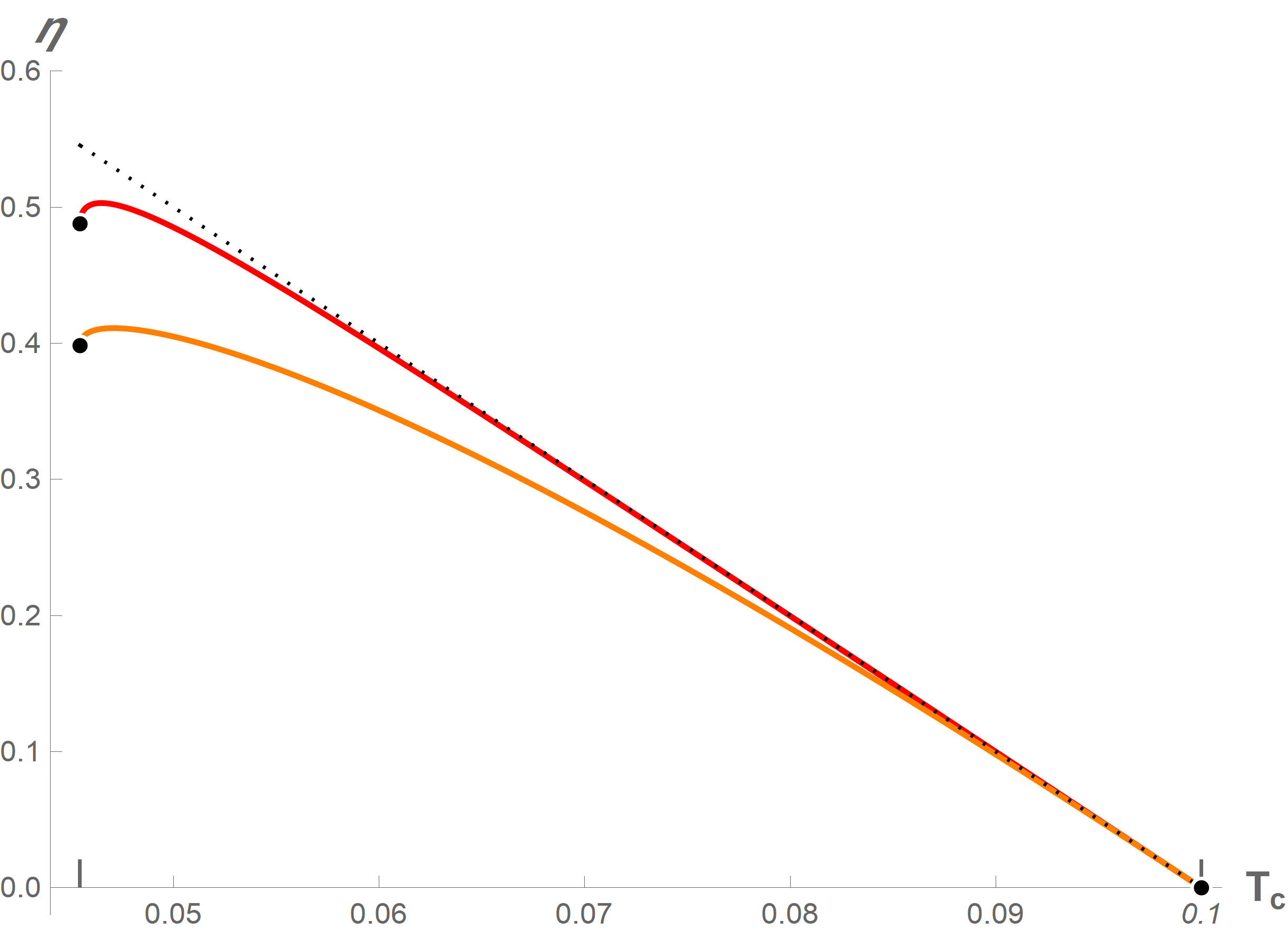}
        \caption{\textit{Efficiency vs.\ cold reservoir temperature for the large black hole branch.}
    \textcolor{black}{
The regenerative (red) and non-regenerative (orange) Stirling
efficiencies are shown as functions of $T_{\rm c}$ at fixed
$T_{\rm h}=0.10$, with the same volumes as
Figure~\ref{fig:eff_th}.
The black dotted curve is the Carnot bound.
Both Stirling efficiencies approach zero as
$T_{\rm c}\to T_{\rm h}^{-}$; at $T_{\rm h}=T_{\rm c}$, marked by a dot, the cycle is
degenerate.
The large branch domain is bounded on the left by
$T_{\rm c}=\sqrt{27/(16\pi V_{\min})}$, where the state
$(T_{\rm c},V_{\min})$ approaches the branch merger point $x=2/3$;
the black dots mark the limiting efficiencies at this boundary.
The regenerative efficiency is non-monotonic and reaches a maximum
of approximately $0.503$ near $T_{\rm c}=0.047$.
}}
        \label{fig:eff_tc}
    \end{minipage}
\end{figure*}

\begin{figure*}[!ht]
    \centering
    \begin{minipage}{0.33\textwidth}
        \centering
        \includegraphics[width=\linewidth]{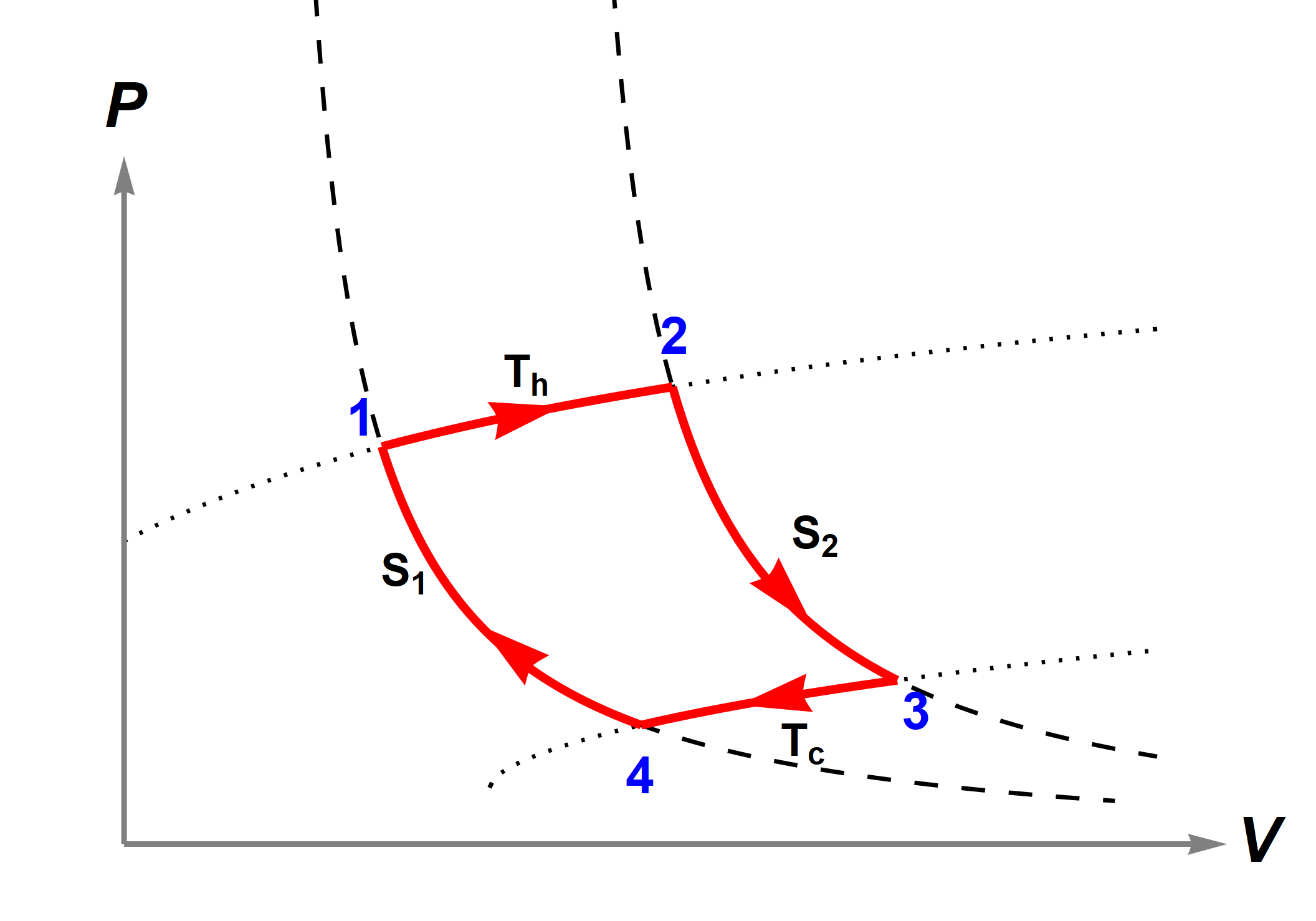}
        \par\vspace{4pt}
        (a) Carnot cycle
    \end{minipage}\hfill
    \begin{minipage}{0.33\textwidth}
        \centering
        \includegraphics[width=\linewidth]{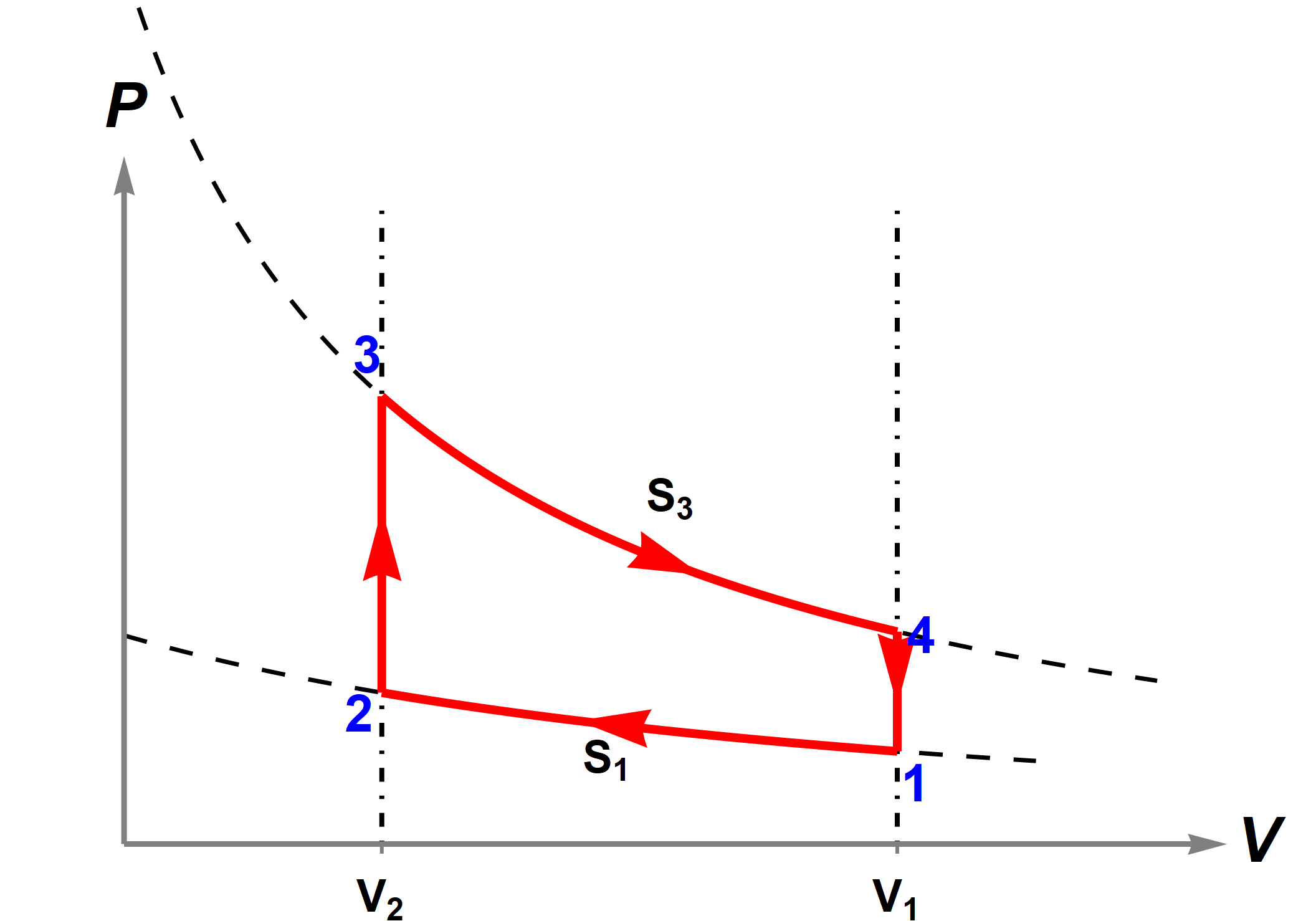}
        \par\vspace{4pt}
        (b) Otto cycle
    \end{minipage}\hfill
    \begin{minipage}{0.33\textwidth}
        \centering
        \includegraphics[width=\linewidth]{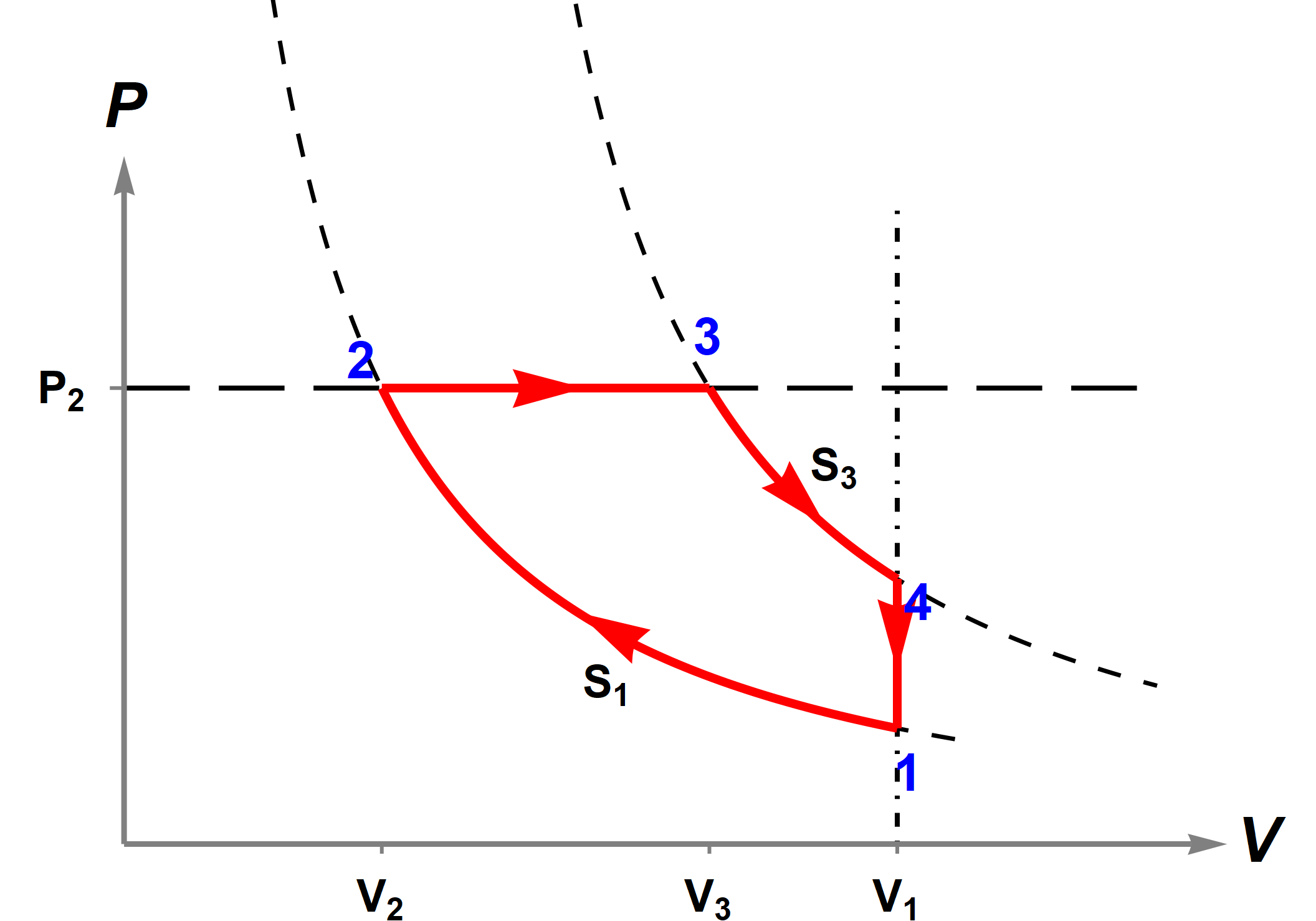}
        \par\vspace{4pt}
        (c) Diesel cycle
    \end{minipage}

    \vspace{6mm} 
    \begin{minipage}{0.33\textwidth}
        \centering
        \includegraphics[width=\linewidth]{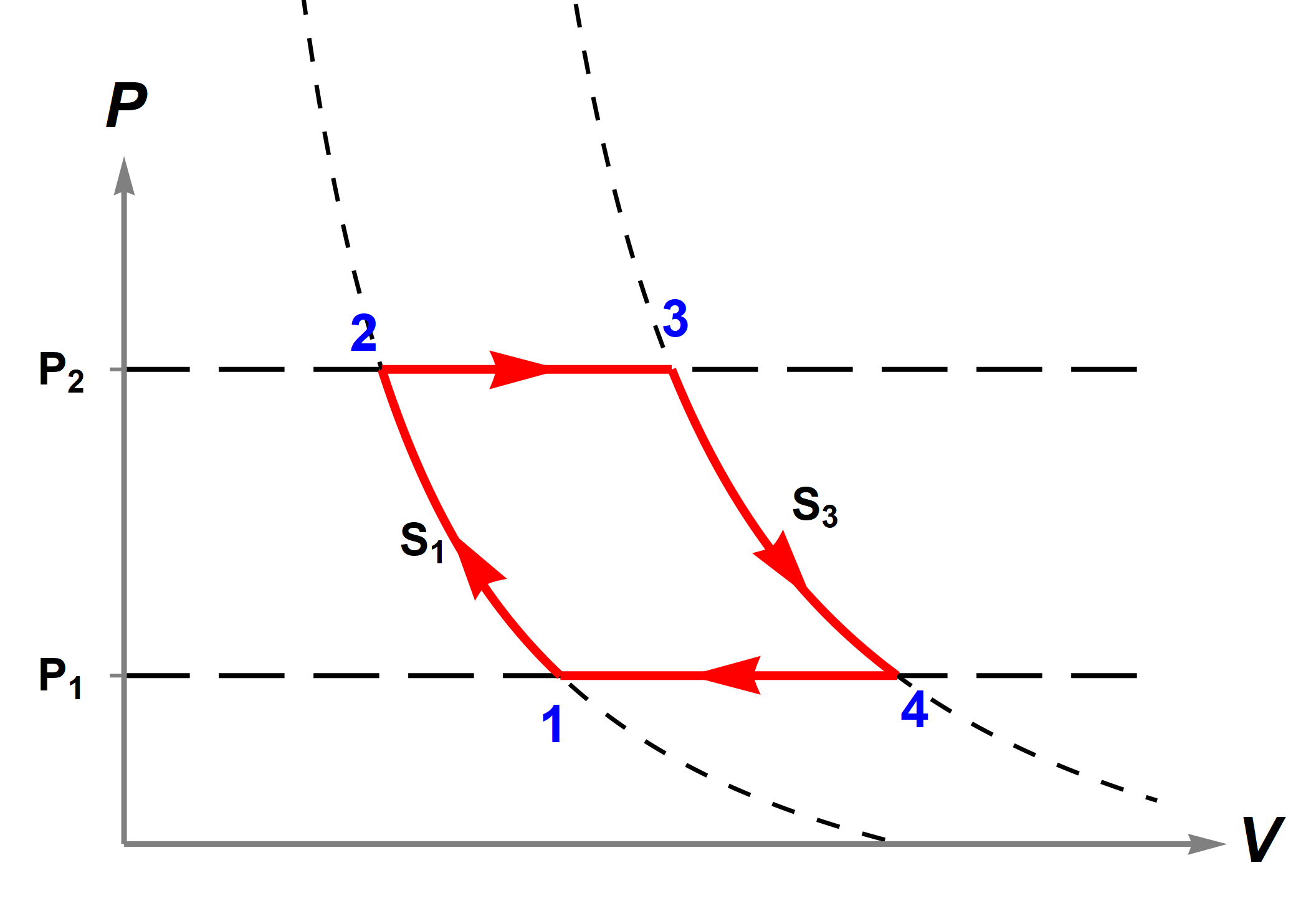}
        \par\vspace{4pt}
        (d) Brayton cycle
    \end{minipage}\hspace{0.034\textwidth}%
    \begin{minipage}{0.33\textwidth}
        \centering
        \includegraphics[width=\linewidth]{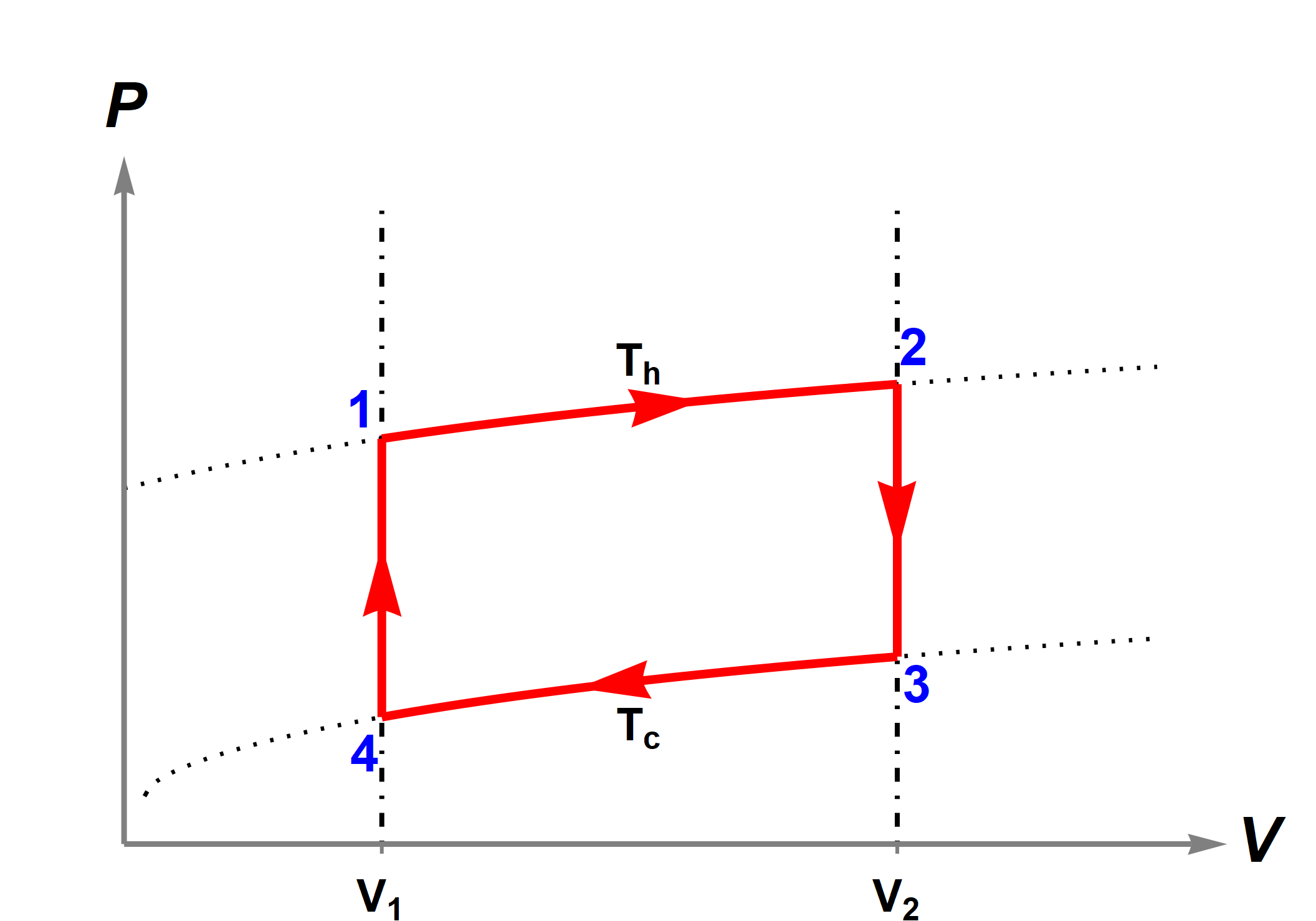}
        \par\vspace{4pt}
        (e) Stirling cycle
    \end{minipage}
   \caption{\textit{Pressure-volume diagrams for the large black hole branch.}
Exact $P$-$V$ cycles for a thermal system on a spherical cavity that is dual to a four-dimensional Schwarzschild black hole on the large branch. The panels show, in reading order, the Carnot,  Otto, Diesel, Brayton, and Stirling cycles. In each panel, the solid red curve traces the working substance through the vertices $1\to2\to3\to4\to1$ clockwise. The auxiliary constant-property curves are shown with different line styles: adiabats are short dashed, isotherms dotted, isochores dot-dashed, and isobars long dashed. The curved strokes of the cycles follow the corresponding adiabatic or isothermal curves, while the isochoric and isobaric strokes appear as straight segments of constant $V$ and constant $P$, respectively. Unlike the ideal gas case, the adiabats and isotherms are not simple power-law curves, such as $PV^\gamma=\mathrm{constant}$ or $PV=\mathrm{constant}$. The adiabats are determined by $P(S,V)$ at fixed $S$, while the isotherms are determined by the branch-dependent relation $P(T,V)$ at fixed $T$.}
    \label{fig:pv_cycles}
\end{figure*}

{\color{black}
We now state the efficiencies of the following idealized cycles: the Carnot, Otto, Diesel, Brayton, and Stirling cycles. Detailed derivations are given in Appendix~\ref{app:efficiency_derivations}. We take each cycle to run through the states $1\to2\to3\to4\to1$. The efficiencies are plotted as functions of the relevant thermodynamic control parameters in Figures~\ref{fig:eff_vmax} to \ref{fig:eff_smin}. The corresponding cycle paths in the $P$-$V$ and $T$-$S$ planes are shown in Figures~\ref{fig:pv_cycles} and~\ref{fig:ts_cycles}. The constant-property curves used to construct these diagrams are given explicitly in Appendix~\ref{app:cycle_paths}.

All efficiencies and cycle diagrams in the main text are restricted to the   large black hole branch, $x=r_h/r_B>2/3$. The branch condition is tested at every state entering an efficiency, and a curve is drawn only over the range for which all of its states remain on that branch. Since $x$ varies monotonically along each of the four stroke types used here, it suffices to test the four corner states of each cycle. On the large branch, every state obeys $4GS<V<9GS$, so an Otto cycle confined to it must satisfy
\begin{equation}
    V_{\min}>4G S_{\max}\ ,\qquad V_{\max}<9G S_{\min}\ ,
    \label{eq:otto-large-branch}
\end{equation}
where the first condition keeps the state $(S_{\max},V_{\min})$ below $x=1$, while the second keeps $(S_{\min},V_{\max})$ above $x=2/3$. For completeness, corresponding efficiency plots and cycle diagrams for the   small black hole branch, $x<2/3$, are presented and discussed in the Supplemental Material. In all numerical plots we work in units in which $G=1$.
}\\
 

\noindent \textbf{Carnot engine.}
The Carnot cycle consists of two isothermal strokes at temperatures $T_{\rm h}$ and $T_{\rm c}<T_{\rm h}$, connected by two adiabatic strokes. 

Along the hot isotherm $1\to2$, the working substance absorbs heat from the hot reservoir at fixed boundary temperature $T_{\rm h}$: \emph{isothermal heat input}. The entropy increases and the horizon grows. On the large black hole branch this stroke is an expansion of the cavity, so $dV>0$ and the working substance performs positive work.

The stroke $2\to3$ is an \emph{adiabatic expansion}. The cavity is thermally isolated, so no heat is exchanged and the horizon size remains fixed. The cavity wall moves outward, so the working substance performs positive work. The    temperature of the system  decreases from $T_{\rm h}$ to $T_{\rm c}$.

Along the cold isotherm $3\to4$, the working substance rejects heat to the cold reservoir at fixed boundary temperature $T_{\rm c}$: \emph{isothermal heat output}. The entropy decreases and the horizon shrinks. On the large black hole branch this stroke is a compression of the cavity, so $dV<0$ and work is done on the working substance.

Finally, $4\to1$ is an \emph{adiabatic compression}. The entropy and horizon size remain fixed, the cavity wall moves inward, work is done on the system, and the   temperature of the working substance rises back to $T_{\rm h}$.

For a reversible engine operating between two reservoirs at temperatures
$T_{\rm h}$ and $T_{\rm c}$, Carnot's theorem fixes the efficiency independently
of the working substance. Hence the Schwarzschild-cavity Carnot cycle has
\begin{equation}
\eta_{\rm Carnot}
=
1-\frac{T_{\rm c}}{T_{\rm h}}\,.
\end{equation}
The nontrivial information is therefore not in the value of the efficiency,
but in the shape of the cycle in the $P$-$V$ plane, which differs from that
of an ideal gas.\\

\noindent \textbf{Otto engine.}
The Otto cycle consists of two adiabatic strokes and two isochoric strokes. It is specified by two volumes $V_1>V_2$ and two entropy levels fixed by the adiabats, with $S_1=S_2$, $S_3=S_4$, and $S_3>S_1$. 

The stroke $1\to2$ is an \emph{adiabatic compression}. No heat is exchanged, so the horizon size remains fixed, while the cavity wall moves inward from $V_1$ to $V_2$. Since $dV<0$, work is done on the working substance, and the   temperature of the working substance increases.

Along $2\to3$, the cavity wall is held fixed at $V_2$ and the working substance absorbs heat: \emph{isochoric heat input}. Since $dV=0$, no mechanical work is performed. As heat is absorbed, the entropy increases from $S_1$ to $S_3$, and the black hole grows. On the large black hole branch, this is accompanied by an increase in the boundary temperature.

The stroke $3\to4$ is an \emph{adiabatic expansion}. The horizon size remains fixed, while the cavity wall moves outward from $V_2$ back to $V_1$. Hence $dV>0$, the working substance performs positive work, and the temperature decreases.

Finally, along $4\to1$, the cavity wall is held fixed at $V_1$, and the working substance rejects heat: \emph{isochoric heat rejection}.  No mechanical work is performed. As heat is rejected, the entropy decreases from $S_3$ to $S_1$ and the black hole shrinks, returning the system to its initial state. On the large black hole branch this is accompanied by a decrease in the boundary temperature.

The heat exchange takes place only along the two isochoric strokes. Since $dV=0$ there, the heat exchanged is equal to the change in the internal energy. The exact efficiency is therefore
\begin{equation}
\eta_{\rm Otto}
=
1
-
\frac{E(S_3,V_1)-E(S_1,V_1)}
{E(S_3,V_2)-E(S_1,V_2)}\,,
\end{equation}
where the function $E(S,V)$ is given in~\eqref{eq:U}. For an ideal gas the corresponding expression reduces to a function of the compression ratio $V_1/V_2$ alone. For the Schwarzschild-cavity system this cancellation does not occur, so the efficiency depends on the four endpoint data $(S_1,S_3,V_1,V_2)$.\\

\noindent \textbf{Diesel engine.}
The Diesel cycle differs from the Otto cycle by replacing the isochoric heat input stroke with an isobaric one. It is specified by $S_1=S_2$, $S_3=S_4$, $P_2=P_3$, and $V_4=V_1$.

The stroke $1\to2$ is an \emph{adiabatic compression}. The horizon size remains fixed, while the cavity wall moves inward from $V_1$ to $V_2$. Work is done on the working substance, and the boundary temperature increases.

Along $2\to3$, the  pressure is held fixed, $P_2=P_3$, and the working substance absorbs heat: \emph{isobaric heat input}. The entropy increases from $S_1$ to $S_3$, so the black hole grows. The \textcolor{black}{boundary volume} adjusts along the isobar, so this stroke is accompanied by an expansion. Unlike the isochoric heat input stroke of the Otto cycle, the boundary temperature decreases along this isobaric stroke, which is related to the negative fixed-pressure heat capacity.

The stroke $3\to4$ is an \emph{adiabatic expansion}. The horizon size remains fixed, while the cavity wall moves outward until the volume reaches $V_1$. The working substance performs positive work, and the boundary temperature decreases.

Finally, along $4\to1$, the cavity wall is held fixed at $V_1$, and the working substance rejects heat: \emph{isochoric heat rejection}.   As heat is rejected, the entropy decreases from $S_3$ to $S_1$, and the black hole shrinks, returning the system to its initial state. On the large black hole branch, this is accompanied by a decrease in the temperature.

The relevant heat exchanges follow directly from the thermodynamic potentials: along the isobar $dH=T\,dS$, while along the isochore $dE=T\,dS$. Hence the heat absorbed is $H(S_3,P_2)-H(S_1,P_2)$, while the heat rejected is $E(S_3,V_1)-E(S_1,V_1)$. The exact efficiency is therefore
\begin{equation}
\eta_{\rm Diesel}
=
1
-
\frac{E(S_3,V_1)-E(S_1,V_1)}
{H(S_3,P_2)-H(S_1,P_2)}\,.
\end{equation}
Here $E(S,V)$ and $H(S,P)$ are the Schwarzschild-cavity energy and enthalpy given in~\eqref{eq:U} and \eqref{eq:HSP}, respectively. The intermediate volumes are fixed by $V_2=V(S_1,P_2)$ and $V_3=V(S_3,P_2)$, with $V(S,P)$ determined by \eqref{eq:TVofSP}. The endpoint data must be chosen so that $V_1>V_3>V_2$, \textcolor{black}{ensuring that the adiabatic stroke $3\to4$ is an expansion to $V_1$.}\\

\noindent \textbf{Brayton engine.}
The Brayton cycle consists of two adiabatic strokes and two isobaric strokes. It is specified by $S_1=S_2$, $S_3=S_4$, $P_2=P_3$, and $P_4=P_1$, with $S_3>S_1$ and $P_2>P_1$. We consider the Brayton cycle without a regenerator: any internal recovery of heat between the two isobaric strokes is not included in the heat balance below.

The stroke $1\to2$ is an \emph{adiabatic compression}. The horizon size remains fixed, while the cavity wall moves inward and the pressure rises from $P_1$ to $P_2$. Work is done on the system, and the temperature increases.

Along $2\to3$, the pressure is held fixed, $P_2=P_3$, and the working substance absorbs heat: \emph{isobaric heat input}. The entropy increases from $S_1$ to $S_3$, so the black hole grows. The cavity wall moves outward so as to keep the Brown-York pressure fixed. As for the Diesel cycle, the   temperature decreases along this isobaric heat input stroke, reflecting the negative $C_P$.

The stroke $3\to4$ is an \emph{adiabatic expansion}. The horizon size remains fixed, while the cavity wall moves outward and the pressure decreases from $P_2$ back to $P_1$. The system does work, and the temperature decreases.

Finally, along $4\to1$, the pressure is held fixed, $P_4=P_1$, and the working substance rejects heat: \emph{isobaric heat output}. The entropy decreases from $S_3$ to $S_1$, and the black hole shrinks. The \textcolor{black}{boundary volume} decreases along the isobar, returning the system to its initial state, and the temperature increases.

The heat exchange takes place only along the two isobars. Since $dP=0$, the heat exchanged is equal to the change in enthalpy. Hence the heat absorbed is $H(S_3,P_2)-H(S_1,P_2)$, while the heat rejected is $H(S_3,P_1)-H(S_1,P_1)$. The exact efficiency is therefore
\begin{equation}
\eta_{\rm Brayton}
=
1
-
\frac{H(S_3,P_1)-H(S_1,P_1)}
{H(S_3,P_2)-H(S_1,P_2)}\,.
\end{equation}
The enthalpy $H(S,P)$ is given in~\eqref{eq:HSP}. The volumes at the four vertices are determined by $V_1=V(S_1,P_1)$, $V_2=V(S_1,P_2)$, $V_3=V(S_3,P_2)$, and $V_4=V(S_3,P_1)$, with $V(S,P)$ given in \eqref{eq:TVofSP}.\\

\noindent \textbf{\textcolor{black}{Non-regenerative Stirling engine.}}
The Stirling cycle consists of two isothermal strokes and two isochoric strokes. It is specified by two temperatures $T_{\rm h}>T_{\rm c}$ and two volumes $V_1<V_2$, with $T_1=T_2=T_{\rm h}$, $T_3=T_4=T_{\rm c}$, $V_1=V_4$, and $V_2=V_3$.
This volume ordering makes the hot isotherm $1\to2$ a heat input stroke on the large black hole branch.

Along $1\to2$, the working substance absorbs heat from the hot reservoir at fixed boundary temperature~$T_{\rm h}$: \emph{isothermal heat input}. On the large black hole branch, this is an isothermal expansion: the cavity wall moves outward from $V_1$ to $V_2$, the entropy increases, and the black hole grows.

Along $2\to3$, the cavity wall is held fixed at $V_2$, and the working substance rejects heat: \emph{isochoric heat output}. Since $dV=0$, no mechanical work is performed. On the large black hole branch, this heat rejection is accompanied by a decrease in temperature from $T_{\rm h}$ to $T_{\rm c}$. The entropy decreases, and the black hole shrinks.

Along $3\to4$, the working substance rejects heat to the cold reservoir at fixed boundary temperature $T_{\rm c}$: \emph{isothermal heat output}. On the large black hole branch, this is an isothermal compression: the cavity wall moves inward from $V_2$ to $V_1$, the entropy decreases, and the black hole shrinks further.

Finally, along $4\to1$, the cavity wall is held fixed at $V_1$, and the working substance absorbs heat: \emph{isochoric heat input}.   On the large black hole branch, this heat input is accompanied by an increase in temperature from $T_{\rm c}$ to~$T_{\rm h}$. The entropy increases, and the black hole grows back to its initial state.

The isothermal heat exchanges are proportional to $T\Delta S$, while the isochoric heat exchanges are energy differences at fixed volume. Without regeneration, both isochoric contributions enter the external heat balance. Defining
\begin{equation} \label{notationdeltas}
\Delta S|_T\equiv S(T,V_2)-S(T,V_1)\,,
\end{equation}
the exact efficiency is
\begin{equation}
{\color{black}\eta_{\mathrm{Stirling}}^{\mathrm{nonreg}}}
=
1
-
\frac{
T_{\rm c}\,\Delta S|_{T_{\rm c}}
+
E(T_{\rm h},V_2)-E(T_{\rm c},V_2)
}{
T_{\rm h}\,\Delta S|_{T_{\rm h}}
+
E(T_{\rm h},V_1)-E(T_{\rm c},V_1)
}\,.
\end{equation}
Here $S(T,V)$ is the entropy on the large black hole branch, obtained from \eqref{eq:rh_large}-\eqref{eq:alpha}, and $E(T,V)\equiv E(S(T,V),V)$ is the energy \eqref{eq:U} evaluated on the same branch.

The small black hole branch can be treated formally, but its heat flow assignments differ. Since $(\partial S/\partial V)_T<0$ and $C_V<0$ on the small branch, a clockwise cycle with heat input along the hot isotherm requires the opposite volume ordering, $V_2<V_1$. Then $1\to2$ and $2\to3$ are heat input strokes, while $3\to4$ and $4\to1$ are heat output strokes. We derive the efficiency in this case in Appendix \ref{app:efficiency_derivations}.\\

\noindent \textbf{\textcolor{black}{Regenerative Stirling engine.}}
{\color{black}
\textcolor{black}{We now include an ideal regenerator with unit effectiveness. The regenerator is
an internal heat exchanger that stores heat rejected during one isochoric stroke
and returns it to the working substance during the other. Unit effectiveness
means that all heat made available to the regenerator is recovered; it does not
imply that the heat rejected and required along the two isochores are
intrinsically equal.}

{\color{black}
On the large black hole branch, heat is rejected during the isochoric stroke
$2\to3$ at volume $V_2$ and is required during the isochoric stroke $4\to1$ at
volume $V_1$. Following \cite{LilaniVisser2026}, we define the signed local
isochoric heat mismatch as the heat released during isochoric cooling minus the
heat required during isochoric heating,
\begin{equation}\label{eq:local-isochoric-mismatch}
    dQ_{\mathrm{loc}}(T)
    \equiv
    \left[
        C_V(T,V_2)-C_V(T,V_1)
    \right]dT \,,
\end{equation}
for a common positive temperature increment $dT>0.$
Thus, $dQ_{\mathrm{loc}}(T)>0$ denotes a local heat surplus, whereas
$dQ_{\mathrm{loc}}(T)<0$ denotes a local heat deficit. The non-negative
integrated mismatch is
\begin{align}\label{eq:regenerator-mismatch}
    Q_{\mathrm{mis}}
    &\equiv
    \left|
        Q_{\mathrm{out}}^{2\to3}
        -
        Q_{\mathrm{in}}^{4\to1}
    \right|
    =
    \left|
        \int_{T_{\rm c}}^{T_{\rm h}}
        dQ_{\mathrm{loc}}(T)
    \right| \,.
\end{align}
As shown in Appendix~\ref{app:efficiency_derivations}, $C_V(T,V)$ decreases with
$V$ at fixed $T$ on the large black hole branch. Since $V_2>V_1$, one has
\begin{equation}\label{eq:positive-local-deficit}
    C_V(T,V_1)>C_V(T,V_2),
    \qquad
    T_{\rm c}\leq T\leq T_{\rm h}\,,
\end{equation}
and therefore
\begin{equation}\label{eq:positive-mismatch}
    dQ_{\mathrm{loc}}(T)<0,
    \qquad
    Q_{\mathrm{mis}}>0\,.
\end{equation}
The local mismatch consequently has a fixed deficit sign throughout the
temperature interval. In this case the absolute value in
Eq.~\eqref{eq:regenerator-mismatch} may be evaluated explicitly, giving
\begin{align}
    Q_{\mathrm{mis}}
    &=
    \int_{T_{\rm c}}^{T_{\rm h}}
    \left[
        C_V(T,V_1)-C_V(T,V_2)
    \right]dT
   \\
    &=
    \left[E(T_{\rm h},V_1)-E(T_{\rm c},V_1)\right]
    -
    \left[E(T_{\rm h},V_2)-E(T_{\rm c},V_2)\right]\,.  \nonumber
\end{align}
Because $dQ_{\mathrm{loc}}(T)<0$ throughout the cycle,
$
Q_{\mathrm{mis}}
$
is the total heat deficit that cannot be supplied by the regenerator,
rather than merely a net integrated difference. A reversible implementation
therefore requires an external heat supply
$-dQ_{\mathrm{loc}}(T)>0$ at the corresponding instantaneous temperatures.
Thus, the regenerator reduces, but does not completely remove, the external
heat input along the isochoric heating stroke.
}
}

{\color{black}
The small black hole branch can be treated formally, but the roles of the
isochores are reversed. With the branch-dependent ordering $V_2<V_1$, heat is
rejected during $4\to1$ and required during $2\to3$. Using the same signed
convention, $dQ_{\mathrm{loc},\mathrm{s}}(T)<0$ throughout the temperature
interval, so the cycle is again in the deficit regime. This case is treated in
Appendix~\ref{app:efficiency_derivations}.
}

{\color{black}
The net work is unchanged by the regenerator, because regeneration redistributes
heat internally without changing the path in the $P$-$V$ plane. For the
Schwarzschild-cavity engine, the local mismatch has the deficit sign, so the
regenerator leaves a heat shortfall $Q_{\mathrm{mis}}$ that must be supplied
externally. The external heat input is therefore
\begin{equation}\label{eq:regenerative-heat-input}
    Q_{\mathrm{in}}^{\mathrm{reg}}
    =
    Q_{\mathrm{in}}^{1\to2}
    +
    Q_{\mathrm{mis}}{\color{black}\,,}
\end{equation}
whereas its external heat output is the heat rejected along the cold isotherm,
\begin{equation}\label{eq:regenerative-heat-output}
    Q_{\mathrm{out}}^{\mathrm{reg}}
    =
    Q_{\mathrm{out}}^{3\to4}\,.
\end{equation}
Using the notation in equation~\eqref{notationdeltas}, the regenerative Stirling
efficiency is thus
\begin{align}\label{eq:eta_stirling_reg}
    \eta_{\mathrm{Stirling}}^{\mathrm{reg}}
    &=
    1-
    \frac{Q_{\mathrm{out}}^{3\to4}}
         {Q_{\mathrm{in}}^{1\to2}+Q_{\mathrm{mis}}}
     =
    1-
    \frac{T_{\rm c}\Delta S|_{T_{\rm c}}}
         {T_{\rm h}\Delta S|_{T_{\rm h}}+Q_{\mathrm{mis}}}\,.
\end{align}
This is the fixed-sign deficit formula for an ideal regenerative Stirling engine
\cite{LilaniVisser2026}.   The specific Schwarzschild cavity thermodynamics enters
through the state functions $S(T,V)$ and $E(T,V)$, which determine
$\Delta S|_T$ and $Q_{\mathrm{mis}}$.
}

{\color{black}
A sufficient condition for an ideal regenerative Stirling engine to attain the
Carnot efficiency is that the fixed-volume heat capacity be independent of
volume \cite{LilaniVisser2026},
\begin{equation}\label{eq:volume-independent-cv}
    C_V(T,V)=C_V(T)\,.
\end{equation}
This condition ensures pointwise matching of the isochoric heat exchanges: the
heat released at each temperature during $2\to3$ is exactly equal to the heat
required at the same temperature during $4\to1$. Consequently, the regenerator
returns to its initial energy and entropy, and no external heat exchange is
required along either isochore.

{\color{black}
Indeed, under condition~\eqref{eq:volume-independent-cv}, $ Q_{\mathrm{mis}}=0.$
}
Moreover, since
\begin{equation}
    C_V
    =
    T\left(\frac{\partial S}{\partial T}\right)_V\,,
\end{equation}
one obtains
\begin{equation}
    \frac{\partial}{\partial T}\Delta S|_T
    =
    \frac{1}{T}
    \left[
        C_V(T,V_2)-C_V(T,V_1)
    \right]
    =
    0\,.
\end{equation}
Hence $\Delta S|_{T_{\rm h}}=\Delta S|_{T_{\rm c}}$, and
equation~\eqref{eq:eta_stirling_reg} reduces to
\begin{equation}
    \eta_{\mathrm{Stirling}}^{\mathrm{reg}}
    =
    1-\frac{T_{\rm c}}{T_{\rm h}}
    =
    \eta_{\rm Carnot}\,.
\end{equation}
The volume independence of $C_V$ is a sufficient, but not necessary, condition
for pointwise matching on a particular cycle. Conversely, the weaker condition
$Q_{\mathrm{mis}}=0$ by itself guarantees only equality of the integrated
isochoric energies. It does not ensure that heat is returned at the same
temperatures at which it was stored, and therefore does not by itself guarantee
reversible regeneration.
}

{\color{black}
For a classical ideal gas, $C_V$ is independent of volume, so the isochoric heat
exchanges match pointwise and an ideal regenerative Stirling engine attains the
Carnot efficiency exactly. For the Schwarzschild cavity working substance,
however, $C_V$ depends on the boundary volume. More specifically, on the large
black hole branch $C_V(T,V)$ decreases monotonically with $V$ at fixed $T$.
{\color{black}
Consequently, for $V_2>V_1$ the local mismatch has the fixed deficit sign
$dQ_{\mathrm{loc}}(T)<0$, as shown in Eq.~\eqref{eq:positive-mismatch}.
The regenerative Schwarzschild Stirling cycle therefore remains strictly
sub-Carnot for every large branch cycle with $T_{\rm h}>T_{\rm c}$ and
$V_2>V_1$,}
\begin{equation} \label{inequalitysubcarnot}
    \eta_{\mathrm{Stirling}}^{\mathrm{reg}}
    <
    \eta_{\rm Carnot}\,.
\end{equation}
\textcolor{black}{As shown in Appendix~\ref{app:stirling_limit}, at fixed
$T_{\rm c}$, $V_1$, and $V_2$, the gap between the regenerative efficiency and
the Carnot bound is suppressed at large $T_{\rm h}$,}
\begin{equation}
    \eta_{\rm Carnot}
    -
    \eta_{\mathrm{Stirling}}^{\mathrm{reg}}
    =
    \mathcal{O}\!\left(\frac{1}{T_{\rm h}}\right)\,,
    \qquad
    T_{\rm h}\to\infty\,.
\end{equation}
The large branch regenerative efficiency therefore approaches the Carnot bound
from below. The   small branch cycle is likewise sub-Carnot. However, unlike on the
large branch, the horizon shrinks as $T_{\rm h}$ increases rather than
approaching the cavity wall, so the efficiency remains a finite distance below
the Carnot bound as $T_{\rm h}\to\infty$. This limiting separation is derived
explicitly in Appendix~\ref{app:stirling_limit}.
}


{\color{black}
The comparison with the non-regenerative Stirling cycle follows directly from
the external heat balance. Both cycles follow the same path in the $P$-$V$
plane and therefore perform the same net work. Without regeneration, the full
isochoric heat input
$
Q_{\mathrm{in}}^{4\to1}
=
Q_{\mathrm{out}}^{2\to3}
+
Q_{\mathrm{mis}}
$
must be supplied externally. With regeneration,
$Q_{\mathrm{out}}^{2\to3}$ is recycled internally, so only the remaining
deficit $Q_{\mathrm{mis}}$ must be supplied externally. Hence
$Q_{\mathrm{in}}^{\mathrm{reg}}
=
Q_{\mathrm{in}}^{1\to2}+Q_{\mathrm{mis}}$ is smaller than
$Q_{\mathrm{in}}^{\mathrm{nonreg}}
=
Q_{\mathrm{in}}^{1\to2}+Q_{\mathrm{in}}^{4\to1}$, since
$Q_{\mathrm{out}}^{2\to3}>0$. Because the net work is the same, regeneration
therefore increases the efficiency. Combining this with the sub-Carnot result \eqref{inequalitysubcarnot}
gives
\begin{equation}\label{eq:stirling-hierarchy}
    \eta_{\mathrm{Stirling}}^{\mathrm{nonreg}}
    <
    \eta_{\mathrm{Stirling}}^{\mathrm{reg}}
    <
    \eta_{\rm Carnot}
\end{equation}
for every large branch cycle with $T_{\rm h}>T_{\rm c}$ and $V_2>V_1$. At fixed $T_{\rm c}$, $V_1$, and $V_2$, both Stirling efficiencies
approach the Carnot efficiency from below as $T_{\rm h}\to\infty$,
\begin{equation}
    \lim_{T_{\rm h}\to\infty}
    \eta_{\rm Stirling}^{\rm nonreg}
    =
    \lim_{T_{\rm h}\to\infty}
    \eta_{\rm Stirling}^{\rm reg}
    =
    \lim_{T_{\rm h}\to\infty}
    \eta_{\rm Carnot}
    =1 \,.
\end{equation}
The inequalities in \eqref{eq:stirling-hierarchy} remain strict at every finite
$T_{\rm h}$.
}

The hierarchy \eqref{eq:stirling-hierarchy} is visible in Figures~\ref{fig:eff_th} and~\ref{fig:eff_tc}. The
regenerative curve lies above the non-regenerative one and remains close to the
Carnot bound throughout the displayed parameter range. In Figure~\ref{fig:eff_th}, increasing $T_{\rm h}$ at fixed $T_{\rm c}$
causes both Stirling efficiencies to approach the Carnot bound from below,
without the cycle collapsing or its work output  vanishing. In Figure~\ref{fig:eff_tc}, all three plotted curves approach zero as
$T_{\rm c}\to T_{\rm h}^{-}$. At $T_{\rm c}=T_{\rm h}$ the two
reservoir temperatures coincide and the Stirling cycles are degenerate,
with both the net work and heat input vanishing.\\

\section{Conclusion}

We have constructed reversible heat engines whose working substance is the thermal system on a finite spherical screen, described holographically by a four-dimensional Schwarzschild black hole in a cavity. Using quasi-local gravitational thermodynamics, the Brown-York surface pressure and the cavity area provide a natural pressure-volume pair. This gives a fixed-theory realization of black hole heat engines in asymptotically flat spacetime.

{\color{black}
\textcolor{black}{For the Otto, Diesel, Brayton, and Stirling cycles, we derived exact
efficiency formulae in terms of the Schwarzschild-cavity equations of state.
These non-Carnot efficiencies are therefore not universal numbers, but probes
of the specific quasi-local thermodynamics of the black hole. The Carnot
efficiency, by contrast, retains its universal form $1-T_{\rm c}/T_{\rm h}$.}
}

{\color{black}
For the Stirling engine, the two branches (large and small black hole)   have qualitatively different
high-temperature behavior. On the   large branch, both the
regenerative and non-regenerative efficiencies approach the Carnot value, with
regeneration reducing the leading deviation. On the   small branch, the
hot-isothermal heat input instead vanishes at high temperature, and the
regenerative efficiency remains a finite distance below the Carnot bound.
}

\begin{figure*}[!ht]
    \centering
    \begin{minipage}[t]{0.48\textwidth}
        \vspace{0pt}
        \centering
        \includegraphics[height=0.22\textheight]{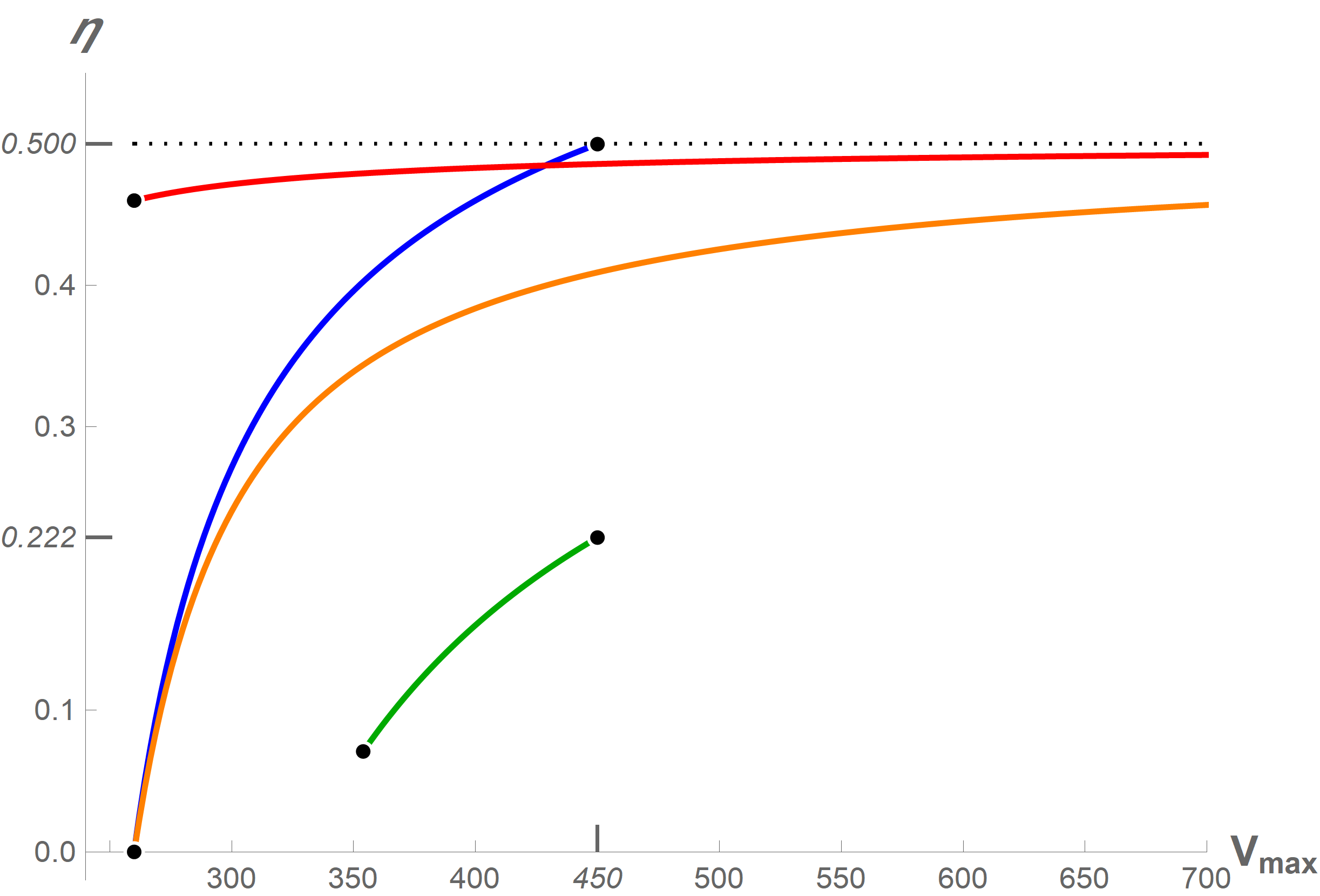}
        \caption{\textit{Efficiency vs.\ maximum boundary volume for the large black hole branch.}
        \textcolor{black}{
        The Otto (blue), Diesel (green), and regenerative (red) and
        non-regenerative (orange) Stirling efficiencies are shown as functions of
        $V_{\max}$ at fixed $V_{\min}=260$, with $S_{\min}=50$,
        $S_{\max}=60$, $P_2=3.0\times10^{-3}$, $T_{\rm c}=0.05$, and
        $T_{\rm h}=0.10$.
        The black dotted line is the Carnot efficiency
        $1-T_{\rm c}/T_{\rm h}=1/2$ associated with the Stirling reservoirs.
        The Otto and Diesel cycles have different temperature extrema and are
        therefore not bounded by this particular reference value.
        Curves are shown only where all cycle vertices lie on the large black hole
        branch.
        The Otto and Diesel curves terminate as
        $V_{\max}\to(9GS_{\min})^{-}$, where the state
        $(S_{\min},V_{\max})$ approaches the branch merger point $x=2/3$.
       The lower boundary of the Diesel domain is
$V_{\max}=V(S_{\max},P_2)$, where vertices 3 and 4 coincide and the
adiabatic expansion $3\to4$ collapses to a point.
        The Stirling curves have no upper branch endpoint because, at fixed
        temperature, increasing the  volume drives $x$ toward $1$ rather
        than toward $2/3$.
        }}
        \label{fig:eff_vmax}
    \end{minipage}
    \hfill
    \begin{minipage}[t]{0.48\textwidth}
        \vspace{0pt}
        \centering
        \includegraphics[height=0.22\textheight]{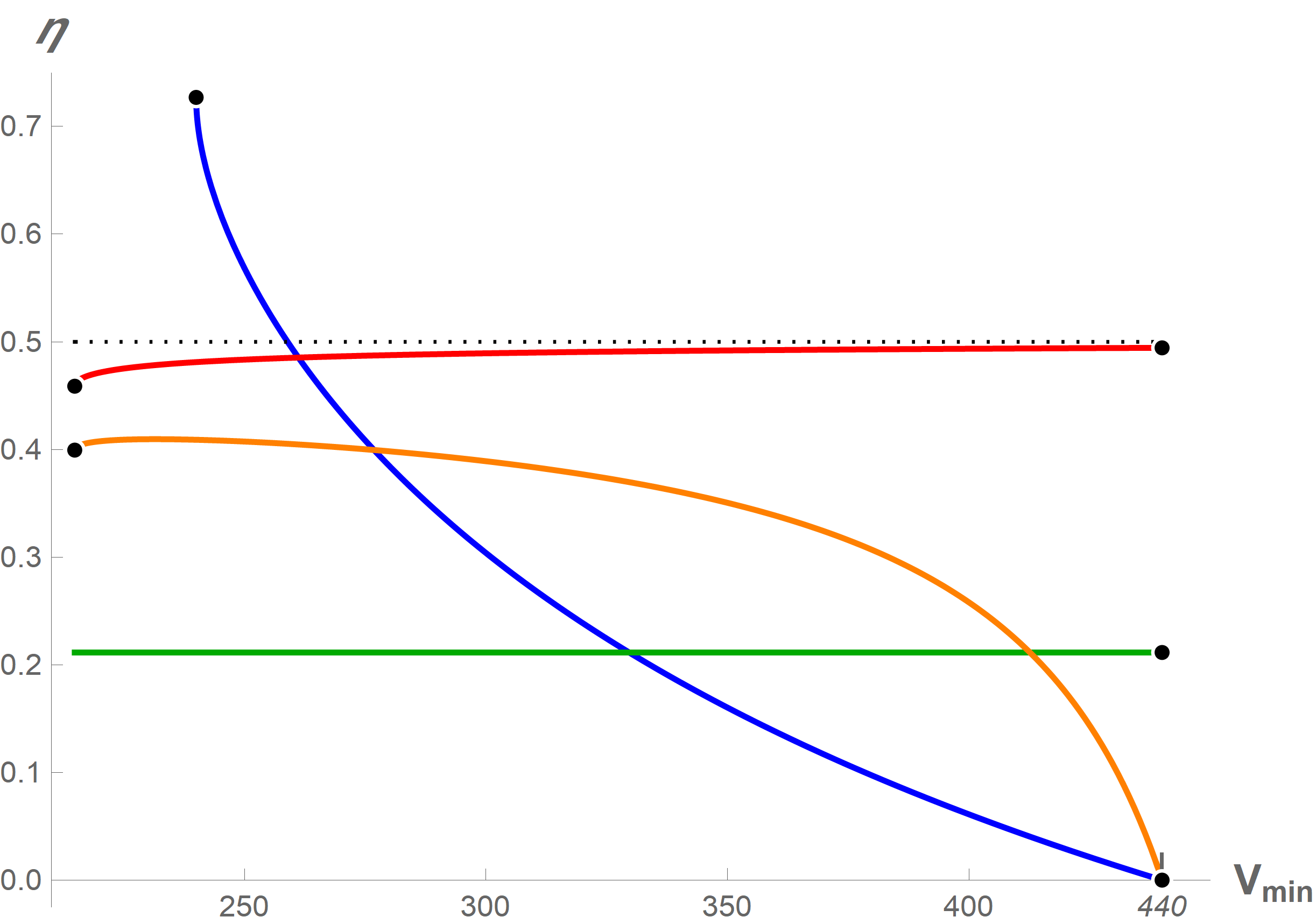}
        \caption{\textit{Efficiency vs.\ minimum boundary volume for the large black hole branch.}
     \textcolor{black}{
The Otto (blue) and regenerative (red) and non-regenerative
(orange) Stirling efficiencies are shown as functions of $V_{\min}$ at
fixed $V_{\max}=440$, with the remaining parameters as in
Figure~\ref{fig:eff_vmax}.
The Diesel efficiency (green) is independent of $V_{\min}$ and is included
as a horizontal comparison curve.
The black dotted line is the Carnot efficiency
  associated with the Stirling reservoirs.
The Stirling curves begin as
$V_{\min}\to[27/(16\pi T_{\rm c}^{2})]^{+}$, where the
cold-isothermal vertex $(T_{\rm c},V_{\min})$ approaches the
branch merger point $x=2/3$.
The Otto curve begins as
$V_{\min}\to(4GS_{\max})^{+}$, where the vertex
$(S_{\max},V_{\min})$ approaches the near-wall limit $x=1$.
As $V_{\min}\to V_{\max}^{-}$, the Otto and non-regenerative Stirling
efficiencies vanish as their cycles become degenerate.
The regenerative Stirling efficiency instead approaches a finite limiting
ratio; at $V_{\min}=V_{\max}$, however, both its work and external heat
input vanish, so this value is not the efficiency of a non-degenerate
cycle. }}
        \label{fig:eff_vmin}
    \end{minipage}
\end{figure*}

\begin{figure*}[!ht]
    \centering
    \begin{minipage}[t]{0.48\textwidth}
        \vspace{0pt}\centering
        \includegraphics[height=0.22\textheight]{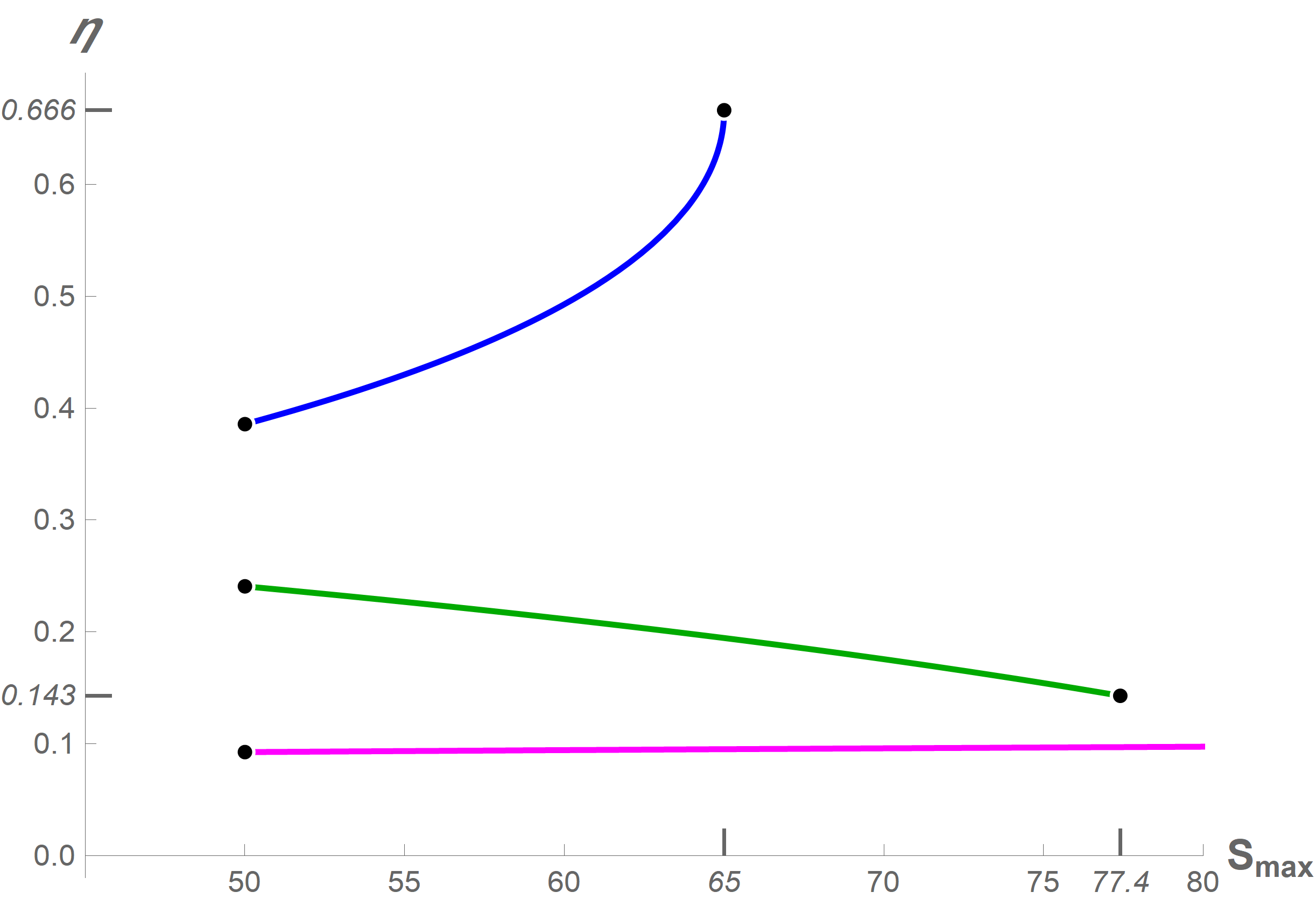}
        \caption{\textit{Efficiency vs.\ maximum entropy for the large black hole branch.}
       \textcolor{black}{
The Otto (blue), Diesel (green), and Brayton (magenta) efficiencies
are shown as functions of $S_{\max}$ at fixed $V_{\min}=260$,
$V_{\max}=440$, $S_{\min}=50$, $P_1=2.2\times10^{-3}$, and
$P_2=3.0\times10^{-3}$.
As $S_{\max}\to S_{\min}^{+}$, the work and heat input vanish
together, while the efficiency ratios approach finite limiting values.
The Otto curve terminates as
$S_{\max}\to[V_{\min}/(4G)]^{-}$, where the state
$(S_{\max},V_{\min})$ approaches the near-wall limit $x=1$ and
the Tolman temperature diverges; its efficiency approaches $2/3$.
The Diesel curve terminates when
$V(S_{\max},P_2)=V_{\max}$, where the vertices 3 and 4 coincide
and the adiabatic expansion $3\to4$ collapses to a point.
For the Brayton cycle, all four vertex volumes are determined by
$V(S,P)$, so the curve has no volume-induced endpoint and approaches
$\eta=1-P_1/P_2=0.267$ as $S_{\max}\to\infty$.
        }}
        \label{fig:eff_smax}
    \end{minipage}\hfill
    \begin{minipage}[t]{0.48\textwidth}
        \vspace{0pt}\centering
        \includegraphics[height=0.22\textheight]{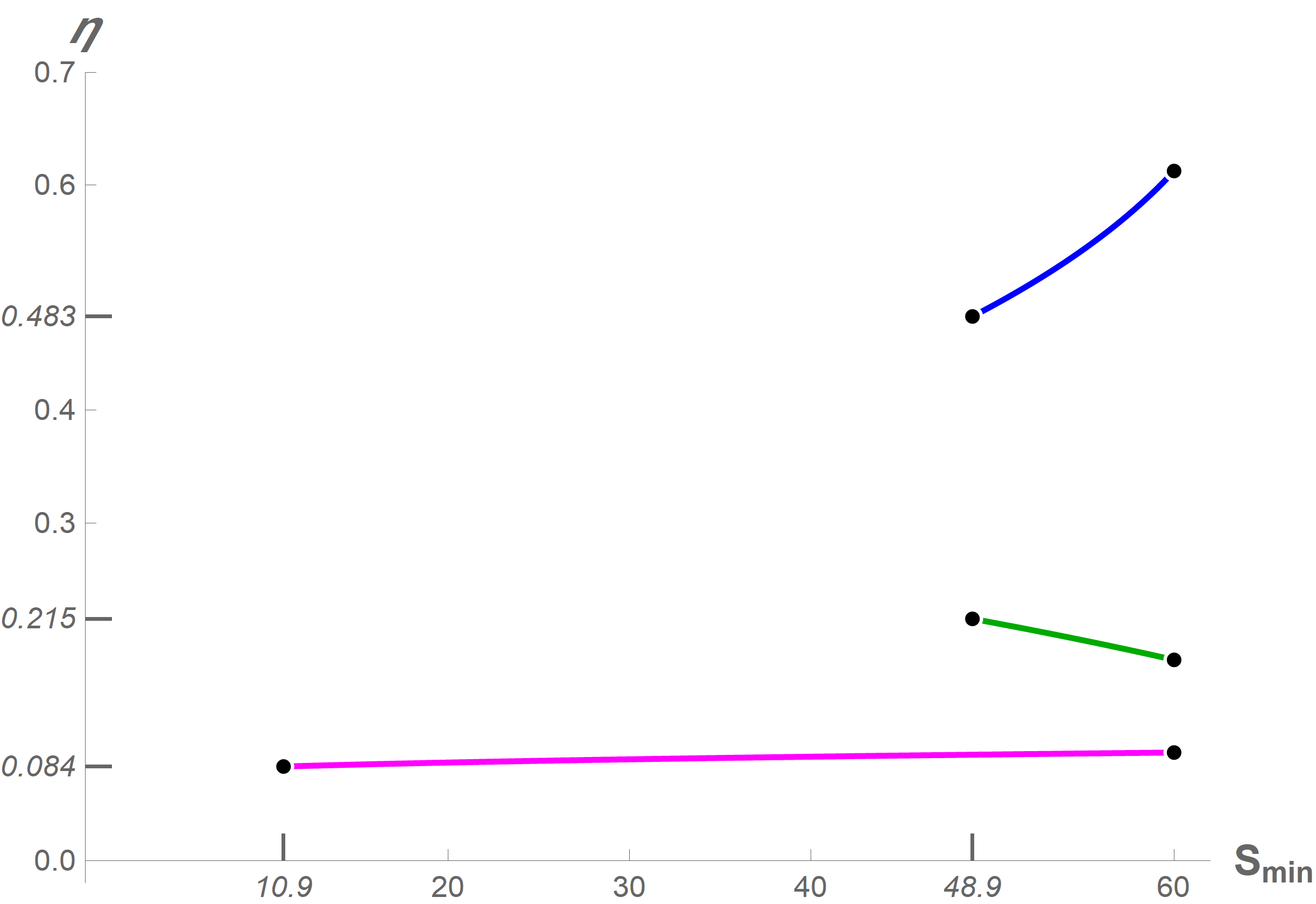}
        \caption{\textit{Efficiency vs.\ minimum entropy for the large black hole branch.}
   \textcolor{black}{
The Otto (blue), Diesel (green), and Brayton (magenta) efficiencies
are shown as functions of $S_{\min}$ at fixed $S_{\max}=60$, with
the volumes and pressures as in Figures~\ref{fig:eff_smax}.
The Otto efficiency increases with $S_{\min}$, whereas the Diesel
efficiency decreases. Throughout their common domain, the Otto
efficiency is the highest, the Diesel efficiency is intermediate,
and the Brayton efficiency is the lowest.
The Otto and Diesel curves begin as
$S_{\min}\to[V_{\max}/(9G)]^{+}$, where their shared state
$(S_{\min},V_{\max})$ approaches the branch merger point $x=2/3$.
No Brayton vertex is fixed at $V_{\max}$; its four vertex volumes are
determined by $V(S,P)$. Its curve therefore begins at the smaller
value of $S_{\min}$ for which the state $(S_{\min},P_1)$ approaches
  $x=2/3$.
As $S_{\min}\to S_{\max}^{-}$, the work and heat input vanish for all
three cycles, while the efficiency ratios approach finite limiting values.  }}
        \label{fig:eff_smin}
    \end{minipage}
\end{figure*}

\textcolor{black}{The volume-dependent comparisons in
Figures~\ref{fig:eff_vmax} and~\ref{fig:eff_vmin} do not exhibit a fixed
hierarchy among the displayed efficiencies over their full parameter ranges.
By contrast, for the common entropy endpoints and fixed control parameters used
in Figures~\ref{fig:eff_smax} and~\ref{fig:eff_smin}, the efficiencies satisfy
$\eta_{\rm Otto}>\eta_{\rm Diesel}>\eta_{\rm Brayton}$
throughout their common domain. Since the cycles use different volume and
pressure constraints, these orderings are features of the chosen cycle
families rather than universal rankings.} 

{\color{black}
Several extensions are natural. The same quasi-local construction can be
applied to charged or rotating black holes, black holes in de Sitter or
anti-de Sitter backgrounds with a finite timelike boundary,
higher-dimensional solutions, and black holes in higher-curvature theories.
It would be particularly interesting to determine how local and global
stability constrain the admissible cycle domain in these examples, and whether
the local regenerator mismatch can change sign across a thermodynamic phase
transition. One may also consider finite regenerator effectiveness, additional
thermodynamic cycles, and finite-time or irreversible processes. \textcolor{black}{These extensions would test the generality of the present
results beyond the equilibrium Schwarzschild cavity system.}
}\\

\noindent \textbf{Acknowledgments.} MRV is grateful to S. Borsboom, G. Ela\c{c}maz, N. Koppen, N. Lilani, S.E. \"{O}zkan, and F. Tuncer for discussions and collaborations on related topics. He also thanks the audience and organizers of the Peyresq  Spacetime Meeting~2026, where this work was presented.
This work is supported in part by the NWO  Spinoza Grant  awarded to Klaas Landsman.

\bibliography{bibliography}

\begin{figure*}[!ht]
    \centering
    \begin{minipage}{0.33\textwidth}
        \centering
        \includegraphics[width=\linewidth]{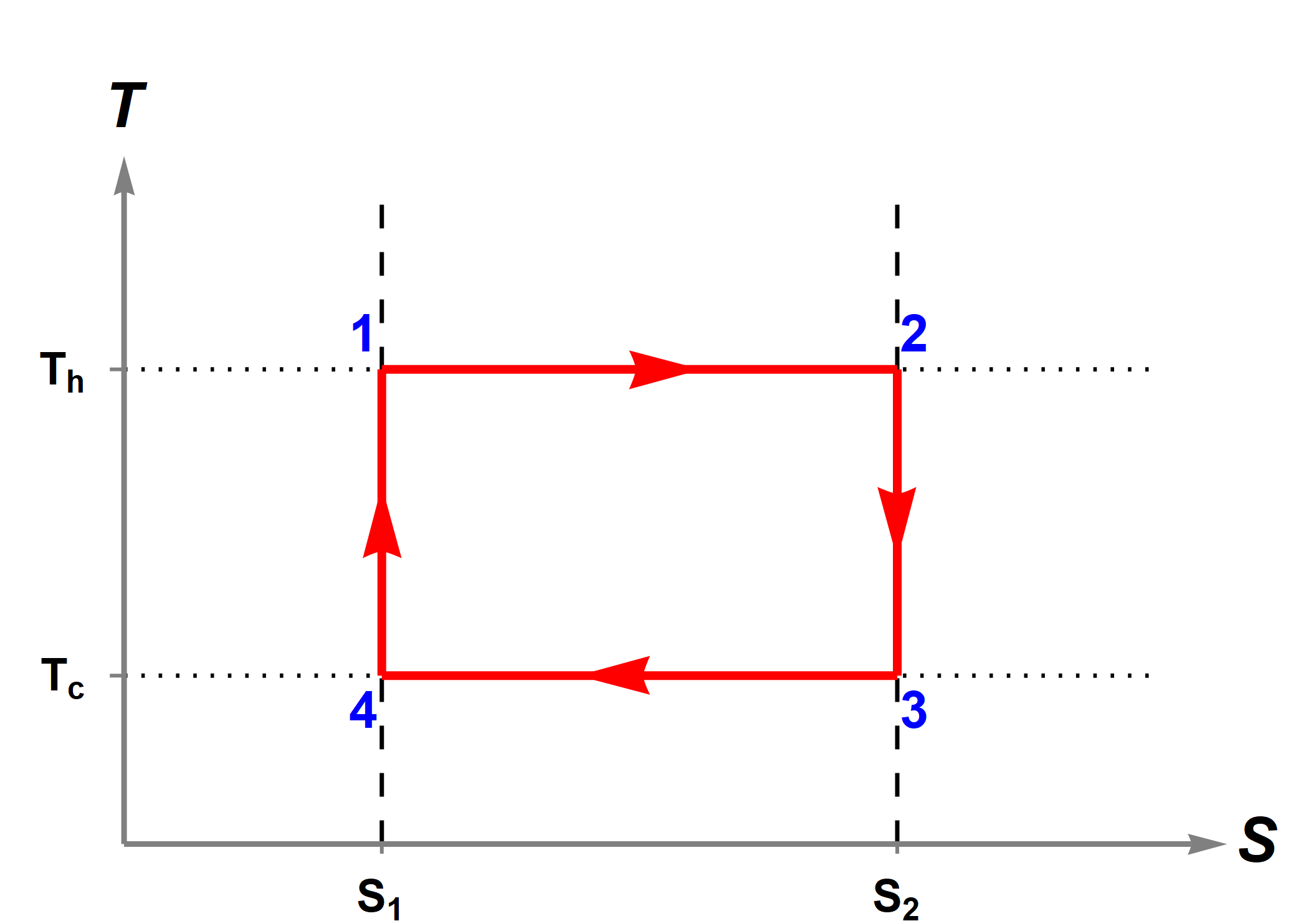}
        \par\vspace{4pt}
        (a) Carnot cycle
    \end{minipage}\hfill
    \begin{minipage}{0.33\textwidth}
        \centering
        \includegraphics[width=\linewidth]{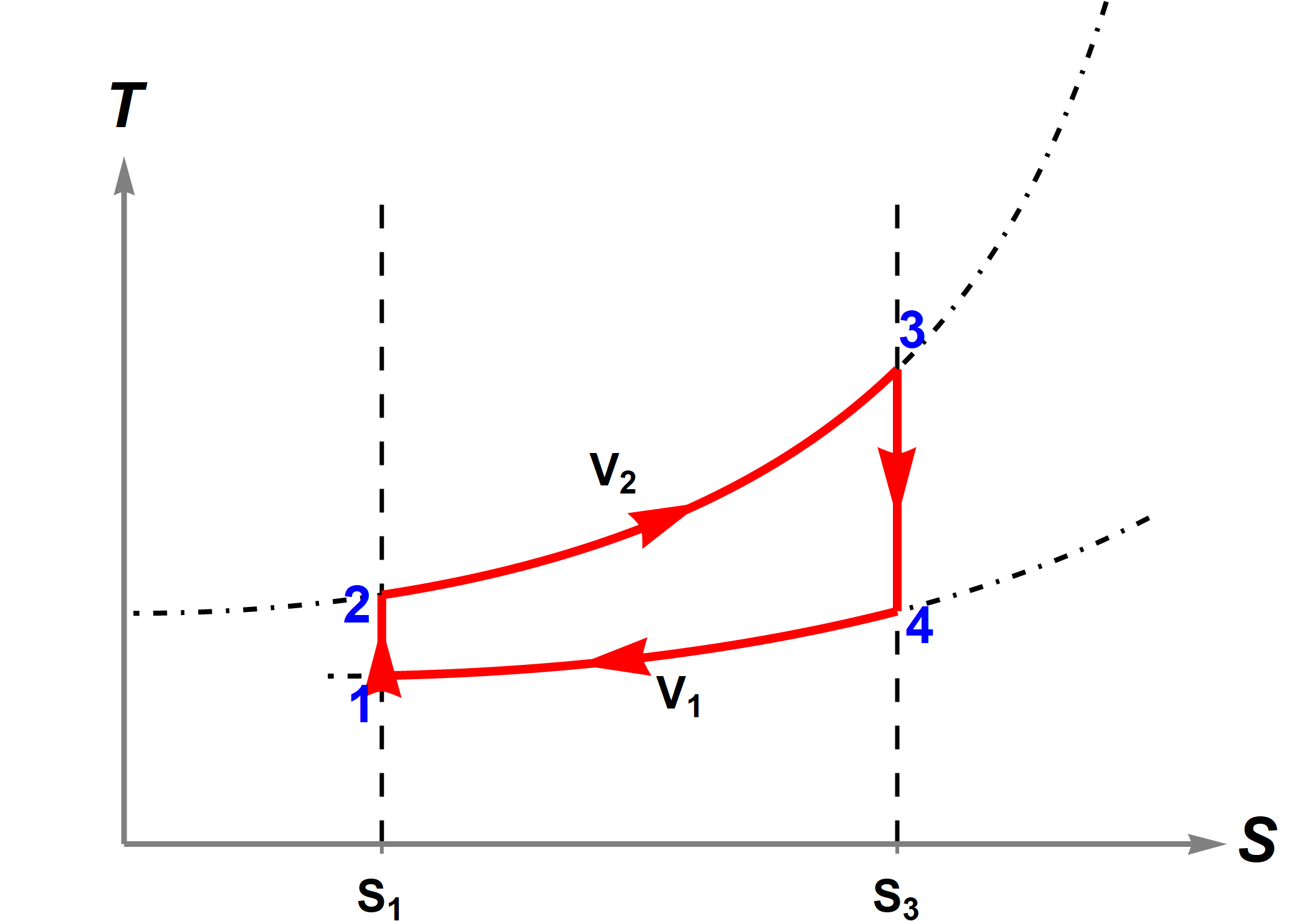}
        \par\vspace{4pt}
     (b) Otto cycle
    \end{minipage}\hfill
    \begin{minipage}{0.33\textwidth}
        \centering
        \includegraphics[width=\linewidth]{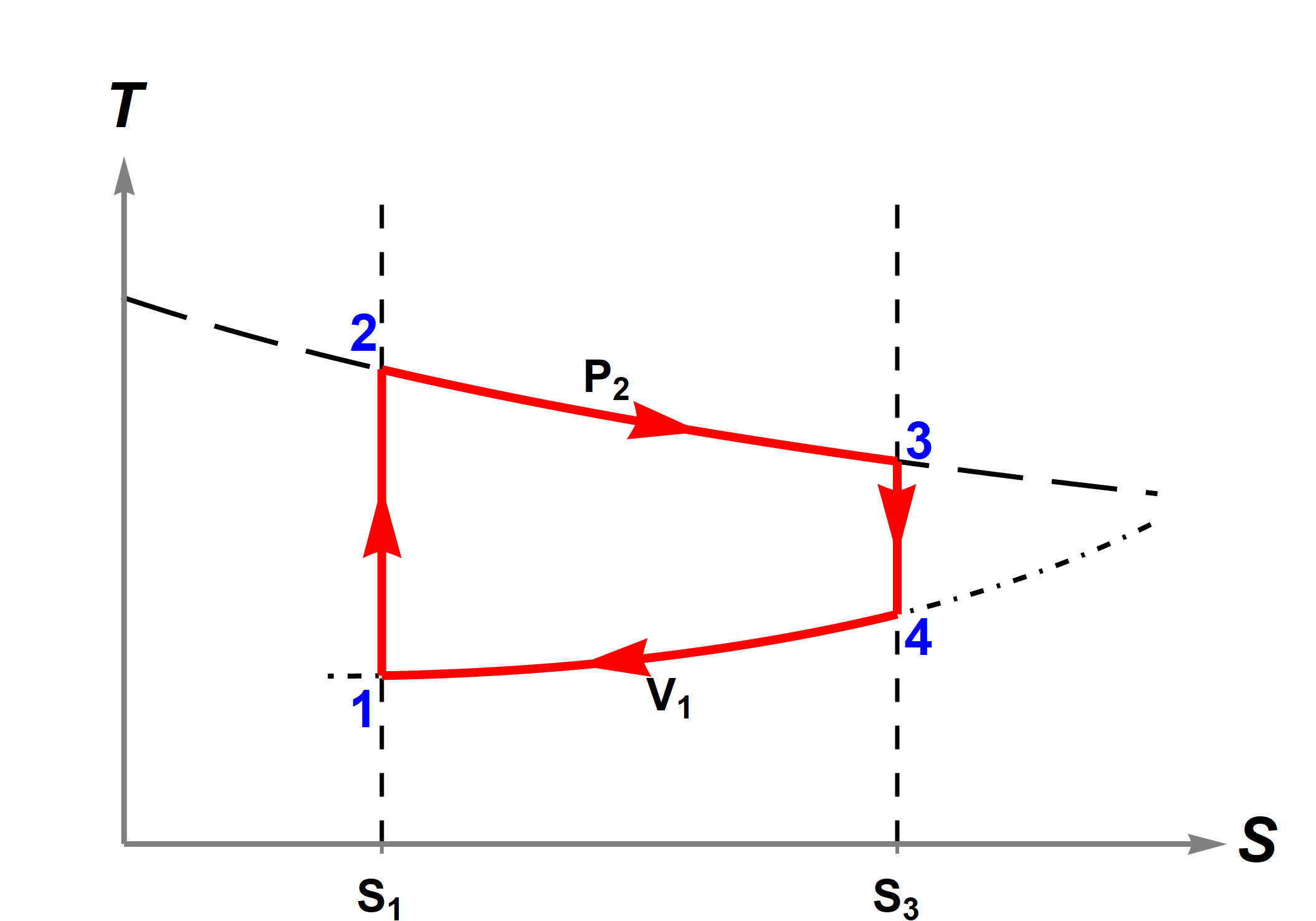}
        \par\vspace{4pt}
      (c) Diesel cycle
    \end{minipage}

    \vspace{6mm} 
    \begin{minipage}{0.33\textwidth}
        \centering
        \includegraphics[width=\linewidth]{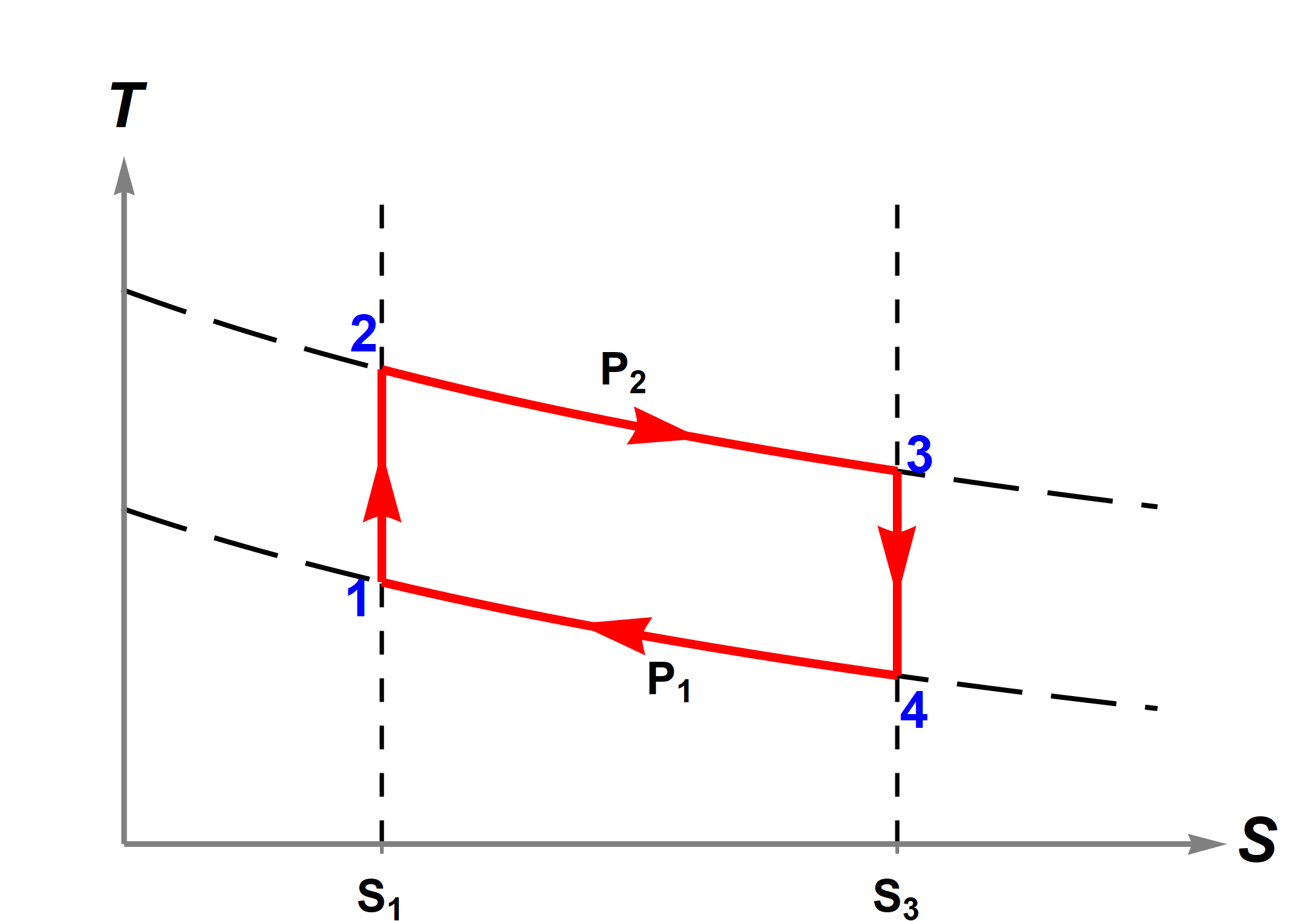}
        \par\vspace{4pt}
        (d) Brayton cycle
    \end{minipage}\hspace{0.034\textwidth}%
    \begin{minipage}{0.33\textwidth}
        \centering
        \includegraphics[width=\linewidth]{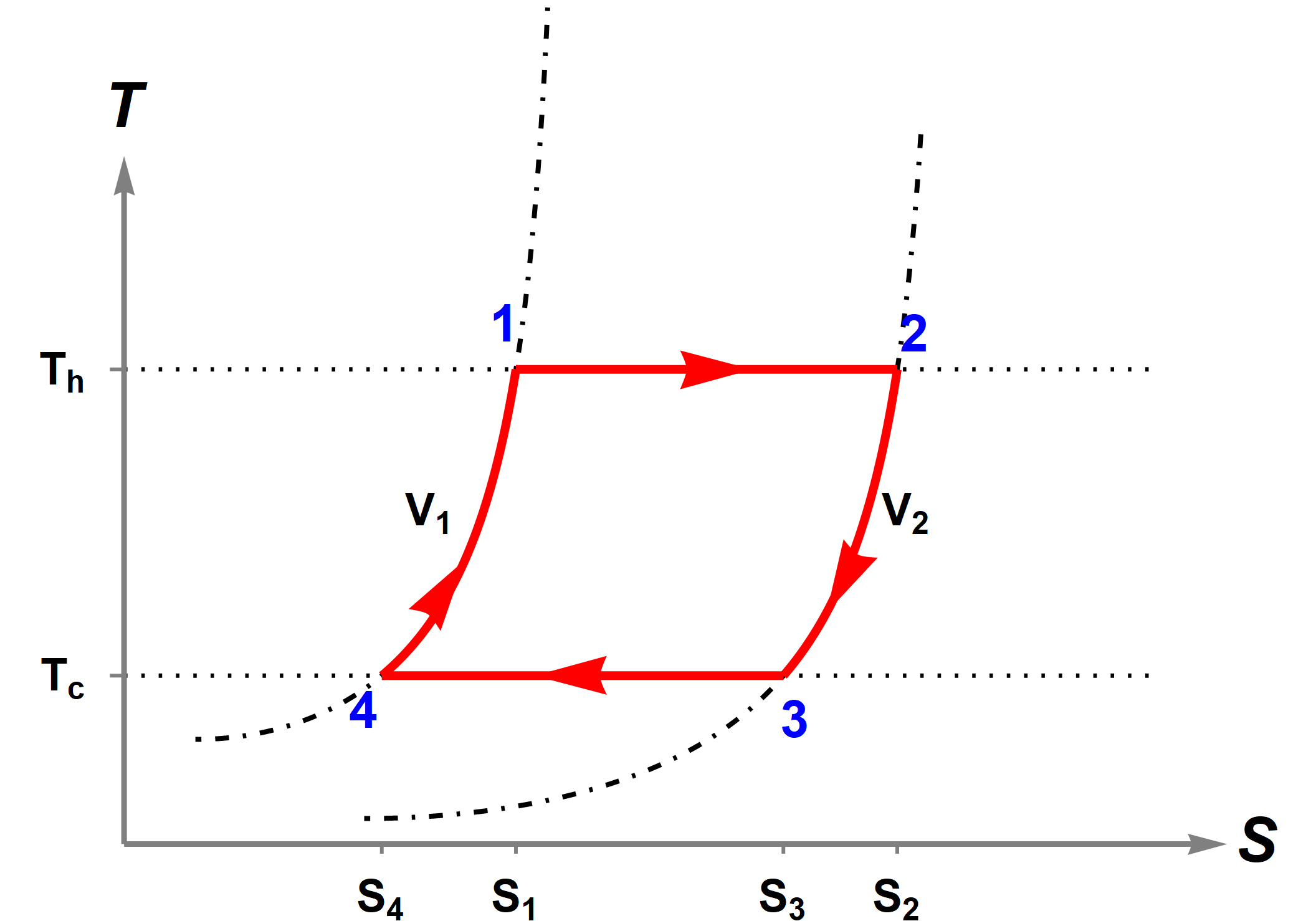}
        \par\vspace{4pt}
        (e) Stirling cycle
    \end{minipage}
 \caption{\textit{Temperature-entropy diagrams for the large black hole branch.}
Exact $T$-$S$ cycles for a thermal system on a spherical cavity that is dual to a four-dimensional Schwarzschild black hole on the large branch. The panels show, in reading order, the Carnot, Otto, Diesel, Brayton and Stirling cycles. In each panel, the solid red curve traces the working substance through the vertices $1\to2\to3\to4\to1$ clockwise. Isothermal and adiabatic strokes appear as horizontal segments of constant $T$ and vertical segments of constant $S$, respectively. The auxiliary constant-property curves are shown with different line styles: isotherms are dotted, adiabats short dashed, isobars long dashed, and isochores dot-dashed. The curved strokes in the $T$-$S$ plane follow the corresponding isochoric or isobaric curves.}
    \label{fig:ts_cycles}
\end{figure*}

\clearpage
\onecolumngrid 

\appendix
 {\begin{center}
    {\bf \Large Supplemental Material}
\end{center}
}

\section{Derivations of heat engine efficiencies}
\label{app:efficiency_derivations}

This appendix derives the efficiency formulae quoted in the main text. We use the convention that all heat inputs and outputs, $Q_{\mathrm{in}}$ and $Q_{\mathrm{out}}$, are positive quantities. Each cycle path follows the vertices $1\to2\to3\to4\to1$.

\vspace{0.3 cm}

\noindent\textbf{Carnot cycle.}
The Carnot cycle consists of two isotherms at $T_{\rm h}$ and $T_{\rm c}<T_{\rm h}$, connected by two adiabats. Since the adiabatic strokes have fixed entropy, $S_2=S_3$ and $S_4=S_1$. The heat absorbed along the hot isotherm and the heat rejected along the cold isotherm are therefore
\begin{equation}
Q_{\mathrm{in}}^{1\to2}
=
T_{\rm h}(S_2-S_1)\,,
\qquad
Q_{\mathrm{out}}^{3\to4}
=
T_{\rm c}(S_3-S_4)\,.
\end{equation}
Using $S_3-S_4=S_2-S_1$, one obtains
\begin{equation}
\eta_{\mathrm{Carnot}}
=
1-\frac{Q_{\mathrm{out}}^{3\to4}}{Q_{\mathrm{in}}^{1\to2}}
=
1-\frac{T_{\rm c}}{T_{\rm h}}\,.
\end{equation}

\vspace{0.3 cm}

\noindent\textbf{Otto cycle.}
The Otto cycle has two adiabats and two isochores. We take $S_1=S_2$, $S_3=S_4$, $V_1=V_4$, and $V_2=V_3$, with $S_3>S_1$ and $V_1>V_2$. Along an isochore $dV=0$, so the heat exchanged is equal to the change in the quasi-local energy. Hence
\begin{equation}
Q_{\mathrm{in}}^{2\to3}
=
E(S_3,V_2)-E(S_1,V_2)\,,
\qquad
Q_{\mathrm{out}}^{4\to1}
=
E(S_3,V_1)-E(S_1,V_1)\,.
\end{equation}
The Otto efficiency is therefore
\begin{equation}
\eta_{\mathrm{Otto}}
=
1-\frac{Q_{\mathrm{out}}^{4\to1}}{Q_{\mathrm{in}}^{2\to3}}
=
1-
\frac{
E(S_3,V_1)-E(S_1,V_1)
}{
E(S_3,V_2)-E(S_1,V_2)
}\,.
\end{equation}

\vspace{0.3 cm}

\noindent\textbf{Diesel cycle.}
The Diesel cycle is formed by an adiabatic compression $1\to2$, an isobaric heat input stroke $2\to3$, an adiabatic expansion $3\to4$, and an isochoric heat rejection stroke $4\to1$. Thus, the two adiabatic strokes fix $S_1=S_2$ and $S_3=S_4$, the heat input stroke fixes $P_2=P_3$, and the final cooling stroke fixes $V_4=V_1$.  The heat input along the isobar is
\begin{equation}
Q_{\mathrm{in}}^{2\to3}
=
H(S_3,P_2)-H(S_1,P_2)\,,
\end{equation}
while the heat rejected along the isochore is
\begin{equation}
Q_{\mathrm{out}}^{4\to1}
=
E(S_3,V_1)-E(S_1,V_1)\,.
\end{equation}
Therefore
\begin{equation}
\eta_{\mathrm{Diesel}}
=
1-\frac{Q_{\mathrm{out}}^{4\to1}}{Q_{\mathrm{in}}^{2\to3}}
=
1-
\frac{
E(S_3,V_1)-E(S_1,V_1)
}{
H(S_3,P_2)-H(S_1,P_2)
}\,.
\end{equation}
The remaining volumes are fixed by the equation of state as $V_2=V(S_1,P_2)$ and $V_3=V(S_3,P_2)$.

\vspace{0.3 cm}

\noindent\textbf{Brayton cycle.}
The Brayton cycle has two adiabats and two isobars. We take $S_1=S_2$, $S_3=S_4$, $P_2=P_3$, and $P_4=P_1$, with $S_3>S_1$ and $P_2>P_1$. Since the enthalpy  $H = E+PV$ satisfies
$ 
dH=T\,dS+V\,dP,
 $
the heat exchanged along an isobar is equal to the change in enthalpy. Thus
\begin{equation}
Q_{\mathrm{in}}^{2\to3}
=
H(S_3,P_2)-H(S_1,P_2)\,,
\qquad
Q_{\mathrm{out}}^{4\to1}
=
H(S_3,P_1)-H(S_1,P_1)\,.
\end{equation}
The Brayton efficiency is thus
\begin{equation}
\eta_{\mathrm{Brayton}}
=
1-\frac{Q_{\mathrm{out}}^{4\to1}}{Q_{\mathrm{in}}^{2\to3}}
=
1-
\frac{
H(S_3,P_1)-H(S_1,P_1)
}{
H(S_3,P_2)-H(S_1,P_2)
}\,.
\end{equation}

\vspace{0.3 cm}

\noindent\textbf{\textcolor{black}{Non-regenerative Stirling cycle.}}
The Stirling cycle has two isotherms ($1 \to 2$ at $T_{\rm h}$ and $3 \to 4$ at $T_{\rm c}$) and two isochores ($2 \to 3$ and $4 \to 1$). On both branches, we take
\begin{equation}
T_1=T_2=T_{\rm h}\,,
\qquad
T_3=T_4=T_{\rm c}\,,
\qquad
V_1=V_4\,,
\qquad
V_2=V_3\,,
\end{equation}
with $T_{\rm h}>T_{\rm c}$, and with all four states chosen within the corresponding branch. The ordering of $V_1$ and $V_2$ is branch-dependent. We choose it so that the hot isotherm $1\to2$ is a heat input stroke, equivalently so that the entropy increases along this stroke. On the large black hole branch, this requires $V_2>V_1$, whereas on the small black hole branch, it requires $V_2<V_1$. With these choices, the cycle is clockwise in the $P$-$V$ plane on both branches.

For a branch $b=\mathrm{l},\mathrm{s}$, where $\mathrm{l}$ denotes the large branch and $\mathrm{s}$ denotes the small branch, we write
\begin{equation}
\Delta S_b|_T
\equiv
S_b(T,V_2)-S_b(T,V_1)
=
\int_{V_1}^{V_2}
\left(\frac{\partial S_b}{\partial V}\right)_T dV\,,
\qquad
E_b(T,V)\equiv E(S_b(T,V),V)\,.
\end{equation}
With the branch-dependent volume ordering specified above, $\Delta S_b|_T>0$ on both branches. This follows from the isothermal entropy response. In four dimensions, with $x=r_h/r_B$, for the Schwarzschild-cavity system
\begin{equation}
\left(\frac{\partial S}{\partial V}\right)_T
=
\frac{x^3}{4G(3x-2)}\,.
\end{equation}
This derivative is positive on the large branch ($x>2/3$), and negative on the small branch ($0<x<2/3$). Specifically, along a reversible isotherm, $\delta Q=T\,dS$, so heat input along $1\to2$ requires $S_b(T_{\rm h},V_2)>S_b(T_{\rm h},V_1)$, or equivalently $\Delta S_b|_{T_{\rm h}}>0$. On the large branch, where $(\partial S/\partial V)_T>0$, this requires $V_2>V_1$. On the small branch, where $(\partial S/\partial V)_T<0$, it instead requires $V_2<V_1$; the integration direction is reversed, so the entropy again increases along $1\to2$. Thus, the volume ordering is chosen precisely so that the hot isotherm is a heat input stroke on both branches.

The heat flow along the isochores is instead controlled by the fixed-volume heat capacity. In four dimensions,
\begin{equation}
C_V
=
T\left(\frac{\partial S}{\partial T}\right)_V
=
\frac{4S(1-x)}{3x-2}\,.
\end{equation}
Hence $C_V>0$ on the large branch and $C_V<0$ on the small branch. Specifically, along an isochore, $\delta Q=dE=C_V\,dT$. Therefore, on the large branch, heat input increases the temperature, and heat output decreases it. On the small branch, this assignment is reversed: heat input lowers the temperature, while heat output raises it.

Consequently, on the large black hole branch, the strokes $1\to2$ and $4\to1$ are heat input strokes, while $2\to3$ and $3\to4$ are heat output strokes. The heat inputs are
\begin{equation}
Q_{\mathrm{in}}^{1\to2}
=
T_{\rm h}\Delta S_{\rm l}|_{T_{\rm h}}\,,
\qquad
Q_{\mathrm{in}}^{4\to1}
=
E_{\rm l}(T_{\rm h},V_1)-E_{\rm l}(T_{\rm c},V_1)\,,
\end{equation}
and the heat outputs are
\begin{equation}
Q_{\mathrm{out}}^{2\to3}
=
E_{\rm l}(T_{\rm h},V_2)-E_{\rm l}(T_{\rm c},V_2)\,,
\qquad
Q_{\mathrm{out}}^{3\to4}
=
T_{\rm c}\Delta S_{\rm l}|_{T_{\rm c}}\,.
\end{equation}
Without regeneration, both isochoric contributions enter the external heat balance. Therefore, the \textcolor{black}{non-regenerative Stirling efficiency} on the large branch is \cite{LilaniVisser2025,LilaniVisser2026}
\begin{equation}
{\color{black}\eta_{\mathrm{Stirling},\mathrm{l}}^{\mathrm{nonreg}}}
=
1-
\frac{
Q_{\mathrm{out}}^{2\to3}+Q_{\mathrm{out}}^{3\to4}
}{
Q_{\mathrm{in}}^{1\to2}+Q_{\mathrm{in}}^{4\to1}
}
=
1-
\frac{
T_{\rm c}\Delta S_{\rm l}|_{T_{\rm c}}+E_{\rm l}(T_{\rm h},V_2)-E_{\rm l}(T_{\rm c},V_2)
}{
T_{\rm h}\Delta S_{\rm l}|_{T_{\rm h}}
+
E_{\rm l}(T_{\rm h},V_1)-E_{\rm l}(T_{\rm c},V_1)
}\,.
\end{equation}
This is the branch used for the reservoir-driven heat engines in the main text, where the subscript $\mathrm{l}$ is suppressed.

On the small black hole branch,  with the clockwise orientation specified above, the strokes $1\to2$ and $2\to3$ are heat input strokes, while $3\to4$ and $4\to1$ are heat output strokes. The heat inputs are
\begin{equation}
Q_{\mathrm{in}}^{1\to2}
=
T_{\rm h}\Delta S_{\rm s}|_{T_{\rm h}}\,,
\qquad
Q_{\mathrm{in}}^{2\to3}
=
E_{\rm s}(T_{\rm c},V_2)-E_{\rm s}(T_{\rm h},V_2)\,,
\end{equation}
and the heat outputs are
\begin{equation}
Q_{\mathrm{out}}^{3\to4}
=
T_{\rm c}\Delta S_{\rm s}|_{T_{\rm c}}\,,
\qquad
Q_{\mathrm{out}}^{4\to1}
=
E_{\rm s}(T_{\rm c},V_1)-E_{\rm s}(T_{\rm h},V_1)\,.
\end{equation}
The corresponding \textcolor{black}{non-regenerative Stirling efficiency} is therefore
\begin{equation}
{\color{black}\eta_{\mathrm{Stirling},\mathrm{s}}^{\mathrm{nonreg}}}
=
1-
\frac{
Q_{\mathrm{out}}^{3\to4}+Q_{\mathrm{out}}^{4\to1}
}{
Q_{\mathrm{in}}^{1\to2}+Q_{\mathrm{in}}^{2\to3}
}
=
1-
\frac{
T_{\rm c}\Delta S_{\rm s}|_{T_{\rm c}}
+
E_{\rm s}(T_{\rm c},V_1)-E_{\rm s}(T_{\rm h},V_1)
}{
T_{\rm h}\Delta S_{\rm s}|_{T_{\rm h}}
+
E_{\rm s}(T_{\rm c},V_2)-E_{\rm s}(T_{\rm h},V_2)
}\,.
\end{equation}
The small branch expression is formal. Since this branch has negative fixed-volume heat capacity, it is unstable in contact with a heat bath at fixed volume: a small heat input lowers its temperature, while a small heat loss raises it, so thermal fluctuations are amplified rather than damped. Consequently, the small branch is not used as a stable reservoir-driven working branch in the main text.

\vspace{0.3cm}

\noindent\textbf{\textcolor{black}{Regenerative Stirling cycle.}}
We now include an ideal regenerator. The regenerator does not change the path in the $P$-$V$ plane and therefore does not change the net work. It only changes the external heat balance by storing heat rejected along one isochore and returning it along the other.

{\color{black}
The only branch-dependent point is which isochore rejects heat. On the large
branch, heat is rejected during $2\to3$ and required during $4\to1$. On the
small branch, the roles of the isochores are reversed: heat is rejected during
$4\to1$ and required during $2\to3$. Following the convention of
\cite{LilaniVisser2026}, we define the signed local isochoric heat mismatch as
the heat available from the isochoric heat output stroke minus the heat required
by the isochoric heat input stroke. For either branch this gives
\begin{equation}\label{eq:local-heat-mismatch-branches}
    dQ_{\mathrm{loc},b}(T)
    \equiv
    \left[
        C_{V,b}(T,V_2)-C_{V,b}(T,V_1)
    \right]dT\,,
    \qquad
    b=\mathrm{l},\mathrm{s}\,.
\end{equation}
On the large branch this is
$dQ_{\mathrm{out}}^{2\to3}-dQ_{\mathrm{in}}^{4\to1}$, whereas on the small
branch it is $dQ_{\mathrm{out}}^{4\to1}-dQ_{\mathrm{in}}^{2\to3}$. Thus, a positive value denotes a local heat surplus and a negative value
denotes a local heat deficit. As shown below, with the branch-dependent
volume orderings specified above, the local isochoric mismatch has the
deficit sign on both Schwarzschild branches.

The corresponding non-negative integrated mismatches are
\begin{equation}
\begin{aligned}
Q_{\mathrm{mis},\mathrm{l}}
&\equiv
\left|
Q_{\mathrm{out}}^{2\to3}-Q_{\mathrm{in}}^{4\to1}
\right|,
\\
Q_{\mathrm{mis},\mathrm{s}}
&\equiv
\left|
Q_{\mathrm{out}}^{4\to1}-Q_{\mathrm{in}}^{2\to3}
\right|.
\end{aligned}
\end{equation}
Equivalently,
\begin{equation}\label{eq:branch-mismatch-integral}
Q_{\mathrm{mis},b}
=
\left|
\int_{T_{\rm c}}^{T_{\rm h}}dQ_{\mathrm{loc},b}(T)
\right|
=
\left|
\int_{T_{\rm c}}^{T_{\rm h}}
\left[
C_{V,b}(T,V_2)-C_{V,b}(T,V_1)
\right]dT
\right|\,,
\qquad b=\mathrm{l},\mathrm{s}\,.
\end{equation}
}
The sign of the local mismatch follows from the volume dependence of the fixed-volume heat capacity $C_V$. In four dimensions,
\begin{equation} \label{cvgo}
C_{V,b}(T,V)
=
\frac{Vx_b^2(1-x_b)}{G(3x_b-2)}
=
\frac{1}{4\pi G T^2(3x_b-2)}\,,
\end{equation}
where $x_b=x_b(T,V)$ denotes the corresponding branch of the Schwarzschild solution. At fixed $T$, the Tolman relation can be written as
\begin{equation}
2\sqrt{\pi}\,T\sqrt{V}
=
\frac{1}{x_b\sqrt{1-x_b}}\,.
\end{equation}
On the large branch, increasing $V$ at fixed $T$ increases $x_{\mathrm{l}}$. Since $C_{V,\mathrm{l}}$ decreases with $x_{\mathrm{l}}$, it decreases with $V$. Thus, for $V_2>V_1$,
we have  $
C_{V,\mathrm{l}}(T,V_1)>C_{V,\mathrm{l}}(T,V_2).
$
On the small branch, increasing $V$ at fixed $T$ decreases $x_{\mathrm{s}}$. Since $C_{V,\mathrm{s}}$ also decreases as a function of $x_{\mathrm{s}}$, it increases with $V$ on the small branch. With the small branch ordering $V_1>V_2$, one again has
$
C_{V,\mathrm{s}}(T,V_1)>C_{V,\mathrm{s}}(T,V_2).
$
{\color{black}
Therefore, on both branches,
\begin{equation}
    dQ_{\mathrm{loc},b}(T)<0,
    \qquad
    T_{\rm c}\leq T\leq T_{\rm h},
    \qquad
    b=\mathrm{l},\mathrm{s},
\end{equation}
so the signed local mismatch has a fixed deficit sign. The absolute value in
Eq.~\eqref{eq:branch-mismatch-integral} can therefore be evaluated explicitly:
\begin{equation}\label{eq:branch-positive-deficit}
\begin{aligned}
Q_{\mathrm{mis},b}
&=
-\int_{T_{\rm c}}^{T_{\rm h}}dQ_{\mathrm{loc},b}(T)
 =
\int_{T_{\rm c}}^{T_{\rm h}}
\left[
C_{V,b}(T,V_1)-C_{V,b}(T,V_2)
\right]dT
\\
&=
\left[
E_b(T_{\rm h},V_1)-E_b(T_{\rm c},V_1)
\right]
-
\left[
E_b(T_{\rm h},V_2)-E_b(T_{\rm c},V_2)
\right]
>0\,,
\qquad b=\mathrm{l},\mathrm{s}\,.
\end{aligned}
\end{equation}
Because the sign of $dQ_{\text{loc},b} $ is fixed throughout the complete temperature interval,
$Q_{\mathrm{mis},b}$ is the total local heat deficit rather than merely a net
difference obtained after cancellations between different temperatures. A
reversible implementation therefore requires an external heat supply
$-dQ_{\mathrm{loc},b}(T)>0$ at each temperature at which the deficit occurs.
This fixed-sign property is what permits the regenerative efficiency to be
written solely in terms of the single integrated quantity
$Q_{\mathrm{mis},b}$.
}
Equivalently,
\begin{equation}
Q_{\mathrm{in}}^{4\to1}
=
Q_{\mathrm{out}}^{2\to3}
+
Q_{\mathrm{mis},\mathrm{l}}
\quad
\text{on the large branch,}
\qquad
Q_{\mathrm{in}}^{2\to3}
=
Q_{\mathrm{out}}^{4\to1}
+
Q_{\mathrm{mis},\mathrm{s}}
\quad
\text{on the small branch.}
\end{equation}
Thus, the heat stored during the isochoric heat-output stroke is not sufficient to supply the whole isochoric heat-input stroke, and the shortfall $Q_{\mathrm{mis},b}$ must be supplied externally.

The external heat input is hence the hot-isothermal heat input plus this mismatch,
\begin{equation}
Q_{\mathrm{in},b}^{\mathrm{reg}}
=
Q_{\mathrm{in}}^{1\to2}
+
Q_{\mathrm{mis},b}\,,
\end{equation}
while the externally rejected heat is only the heat rejected along the cold isotherm,
\begin{equation}
Q_{\mathrm{out},b}^{\mathrm{reg}}
=
Q_{\mathrm{out}}^{3\to4}\,.
\end{equation}
The \textcolor{black}{regenerative Stirling efficiency} on either branch is thus \cite{LilaniVisser2026}
\begin{equation}\label{regeneffst}
\eta_{\mathrm{Stirling},b}^{\mathrm{reg}}
=
1-
\frac{
Q_{\mathrm{out}}^{3\to4}
}{
Q_{\mathrm{in}}^{1\to2}+Q_{\mathrm{mis},b}
}
=
1-
\frac{
T_{\rm c}\Delta S_b|_{T_{\rm c}}
}{
T_{\rm h}\Delta S_b|_{T_{\rm h}}+Q_{\mathrm{mis},b}
}\,,
\qquad b=\mathrm{l},\mathrm{s}\,.
\end{equation}
For $b=\mathrm{l}$, this is the \textcolor{black}{regenerative Stirling efficiency} used in the main text, where the branch label is suppressed.  

{\color{black}
A sufficient condition for pointwise matching of the two isochoric heat
exchanges is \cite{LilaniVisser2026}
\begin{equation}\label{eq:pointwise-cv-matching}
    C_V(T,V_1)=C_V(T,V_2)
    \qquad
    \text{for every }T\in[T_{\rm c},T_{\rm h}]\,.
\end{equation}
A fixed-volume heat capacity that is independent of $V$ guarantees this
condition. It then follows from
Eq.~\eqref{eq:local-heat-mismatch-branches} that
$dQ_{\mathrm{loc},b}(T)=0$ at every temperature, and hence
$Q_{\mathrm{mis},b}=0$. The regenerator therefore returns to its initial
energy and entropy without requiring external heat exchange along either
isochore.

Moreover, since
$ 
    C_V
    =
    T\left(\frac{\partial S}{\partial T}\right)_V,
$
condition~\eqref{eq:pointwise-cv-matching} implies
  $\Delta S_b|_{T_{\rm h}}=\Delta S_b|_{T_{\rm c}}$, and
the efficiency~\eqref{regeneffst} reduces to the Carnot value \cite{LilaniVisser2026}
\begin{equation}
    \eta_{\mathrm{Stirling},b}^{\mathrm{reg}}
    =
    1-\frac{T_{\rm c}}{T_{\rm h}}
    =
    \eta_{\rm Carnot}\,.
\end{equation}
For the Schwarzschild cavity working substance $C_V$ depends on $V$, as can
be seen from \eqref{cvgo}, so the regenerative Stirling cycle remains
sub-Carnot at finite temperature. The high-temperature behavior of the regenerative and non-regenerative
cycles on the large branch, together with the   high-temperature
limit of the regenerative Stirling efficiency on the small 
branch, is analyzed in Appendix~\ref{app:stirling_limit}.
}

\section{\textcolor{black}{Efficiency comparisons on the small black hole branch}}
\label{app:small_branch_efficiencies}

\begin{figure*}[!ht]
    \centering
    \begin{minipage}[t]{0.48\textwidth}
        \vspace{0pt}\centering
        \includegraphics[height=0.22\textheight]{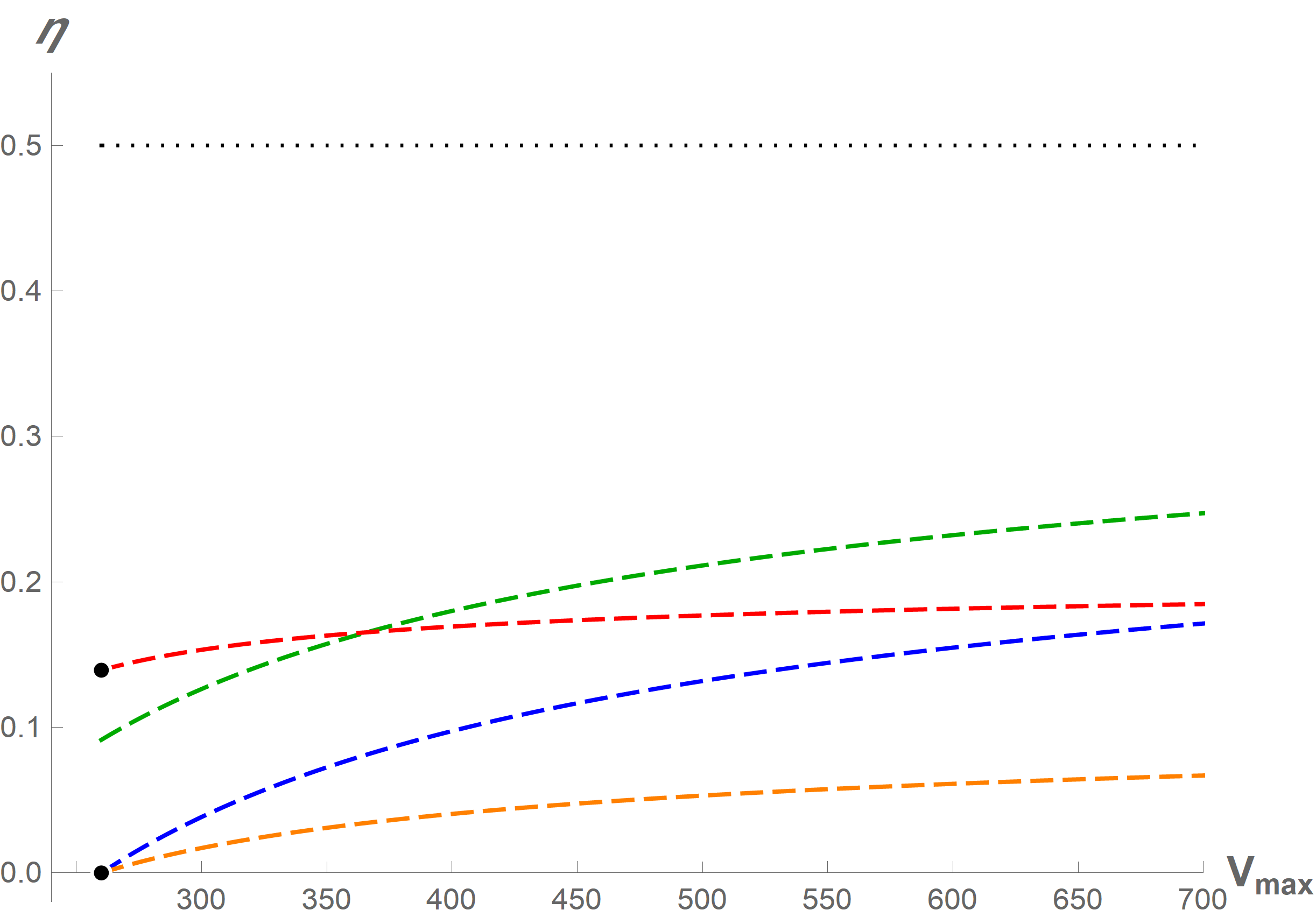}
        \caption{\textcolor{black}{\textit{Efficiency vs.\ maximum boundary volume for the small black hole branch.}
The Otto (blue), Diesel (green), and regenerative (red) and non-regenerative (orange) Stirling efficiencies are shown as functions of $V_{\max}$ using the small branch parameters specified above. The black dotted line is the Carnot efficiency associated with the Stirling reservoirs; it is not a bound for the Otto or Diesel cycles. All displayed cycle vertices remain on the small branch. As $V_{\max}\to V_{\min}^{+}$, the Otto and non-regenerative Stirling efficiencies vanish as their cycles become degenerate, whereas the regenerative Stirling efficiency approaches a finite limiting ratio. The lower boundary of the Diesel domain,
$V_{\max}=V(S_{\max},P_2)=252$, lies below the displayed range.}}
        \label{fig:eff_vmax_small}
    \end{minipage}\hfill
    \begin{minipage}[t]{0.48\textwidth}
        \vspace{0pt}\centering
        \includegraphics[height=0.22\textheight]{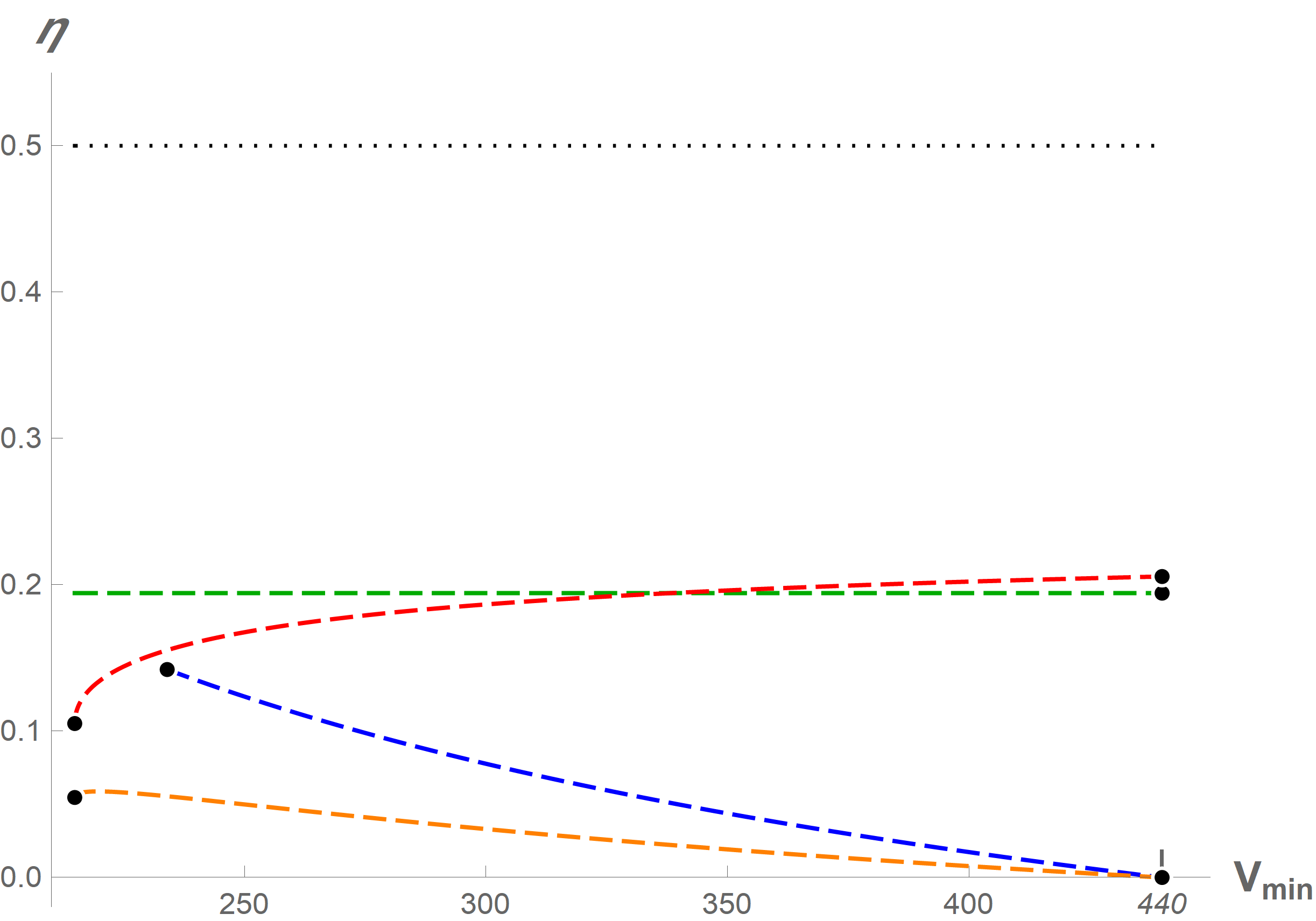}
        \caption{\textcolor{black}{\textit{Efficiency vs.\ minimum boundary volume for the small black hole branch.}
The Otto (blue) and regenerative (red) and non-regenerative (orange) Stirling efficiencies are shown as functions of $V_{\min}$, with the Diesel efficiency (green) included as a horizontal comparison curve because it is independent of $V_{\min}$. The Stirling curves begin when the state $(T_{\rm c},V_{\min})$
approaches the branch-merger point $x=2/3$, while the Otto curve
begins when the state $(S_{\max},V_{\min})$ approaches the same 
branch-merger point. As $V_{\min}\to V_{\max}^{-}$, the Otto and non-regenerative Stirling efficiencies vanish as their cycles become degenerate, whereas the regenerative Stirling efficiency approaches a finite limiting ratio; the exactly equal volume cycle is degenerate. The black dotted line is the  Carnot bound for the Stirling cycles.}}
        \label{fig:eff_vmin_small}
    \end{minipage}
\end{figure*}

\begin{figure*}[!ht]
    \centering
    \begin{minipage}[t]{0.48\textwidth}
        \vspace{0pt}\centering
        \includegraphics[height=0.22\textheight]{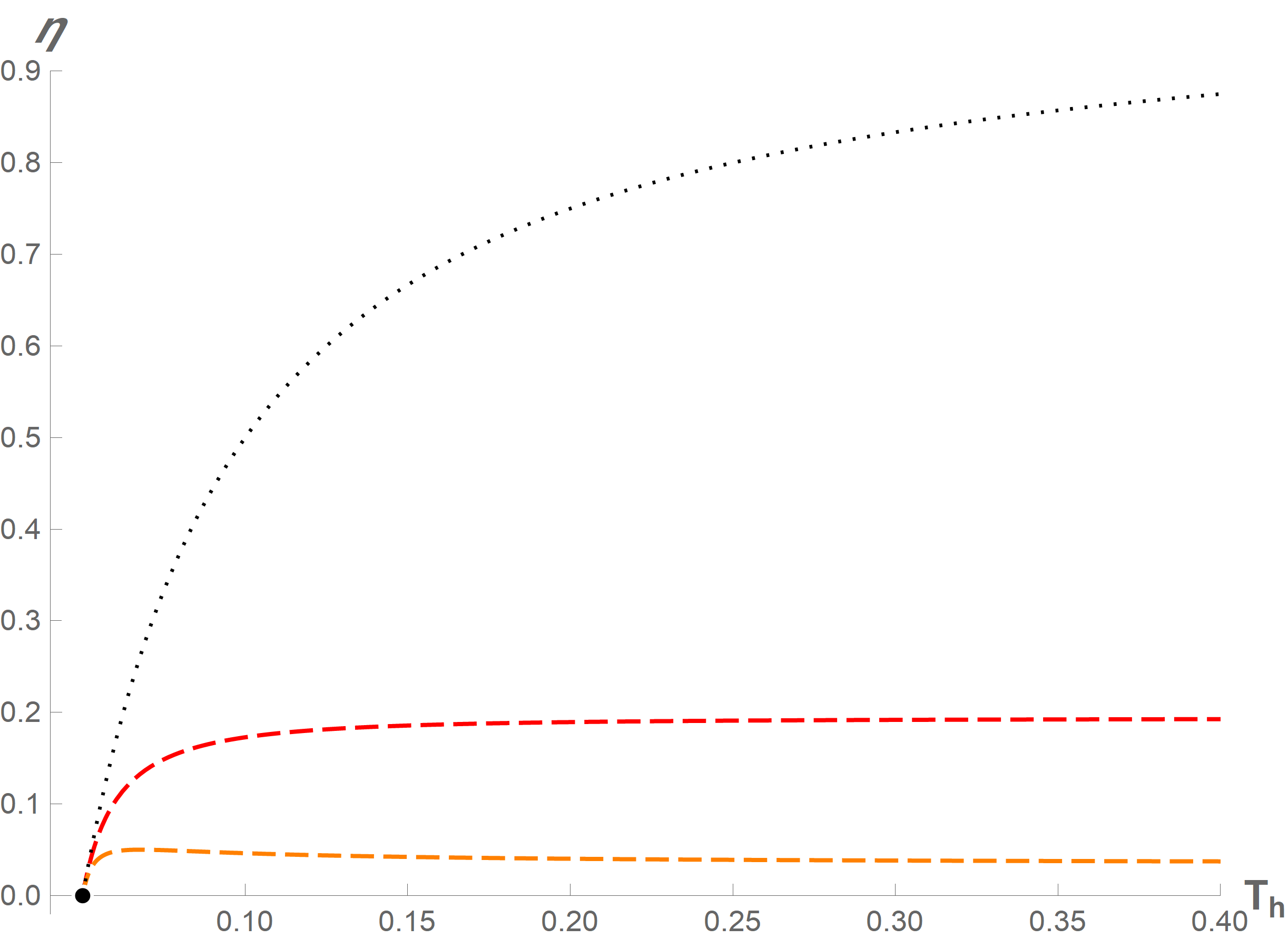}
        \caption{\textcolor{black}{\textit{Efficiency vs.\ hot reservoir temperature for the small black hole branch.}
The regenerative (red) and non-regenerative (orange) Stirling efficiencies are shown as functions of $T_{\rm h}$ at fixed $T_{\rm c}$ and fixed boundary volumes. The states are evaluated on the small branch root of the horizon cubic, with the branch appropriate ordering $V_1=V_{\max}$ and $V_2=V_{\min}$ so that the hot isotherm is a heat-input stroke and the $P$-$V$ loop is clockwise. The black dotted curve is the Carnot bound. Both efficiencies vanish at $T_{\rm h}=T_{\rm c}$. As $T_{\rm h}$ increases, the regenerative efficiency approaches a finite value below the Carnot bound, while the non-regenerative efficiency is non-monotonic.}}
        \label{fig:eff_th_small}
    \end{minipage}\hfill
    \begin{minipage}[t]{0.48\textwidth}
        \vspace{0pt}\centering
        \includegraphics[height=0.22\textheight]{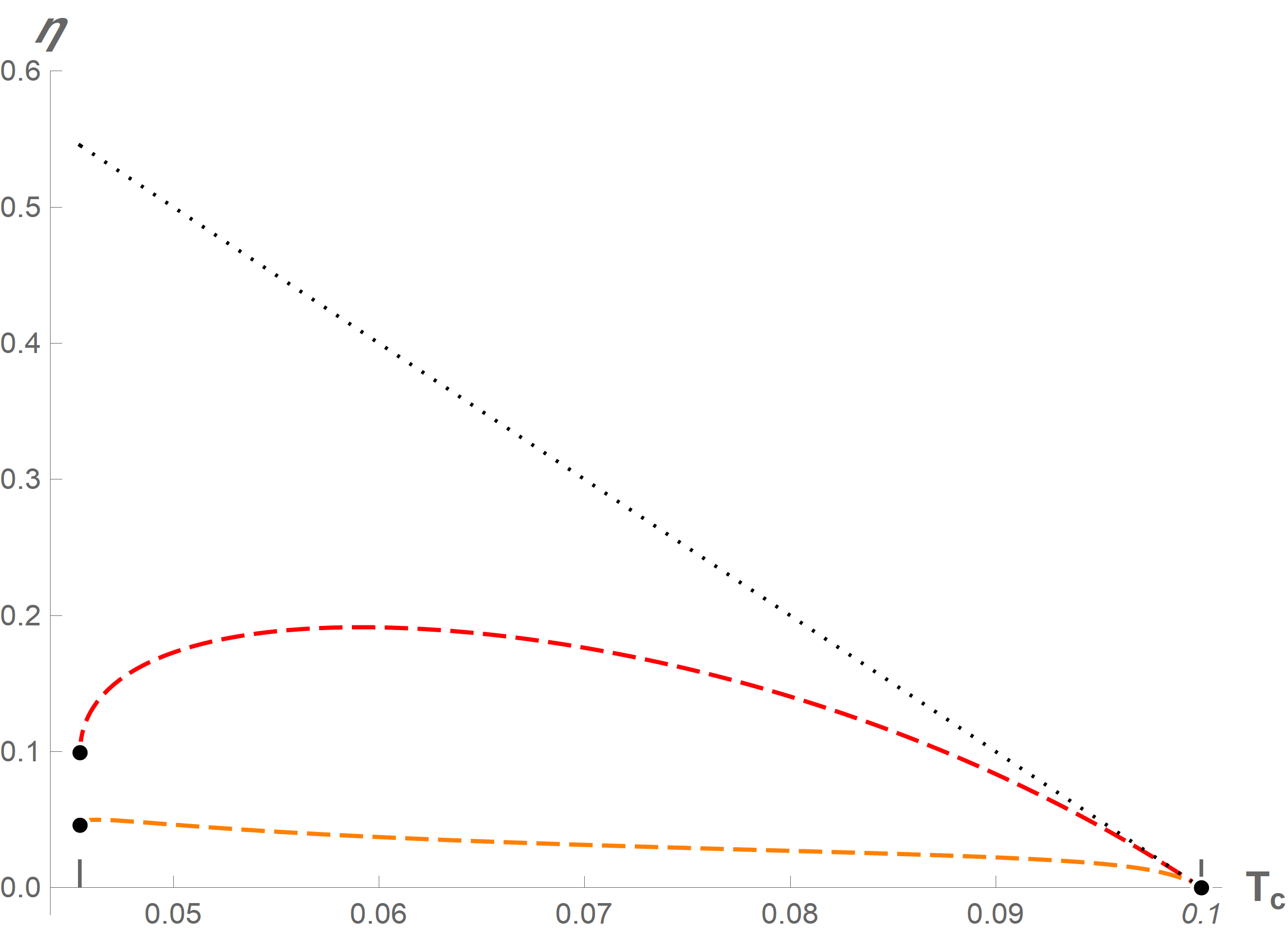}
        \caption{\textcolor{black}{\textit{Efficiency vs.\ cold reservoir temperature for the small black hole branch.}
The regenerative (red) and non-regenerative (orange) Stirling efficiencies are shown as functions of $T_{\rm c}$ at fixed $T_{\rm h}$ and fixed boundary volumes, using the   volume ordering described in Figure~\ref{fig:eff_th_small}. The black dotted curve is the Carnot bound. The curves begin when the cold isothermal state at the smaller volume approaches the branch-merger point $x=2/3$, and both efficiencies vanish at $T_{\rm c}=T_{\rm h}$. The regenerative efficiency is non-monotonic and remains strictly below the Carnot bound throughout the displayed domain.}}
        \label{fig:eff_tc_small}
    \end{minipage}
\end{figure*}

\begin{figure*}[!ht]
    \centering
    \begin{minipage}[t]{0.48\textwidth}
        \vspace{0pt}\centering
        \includegraphics[height=0.22\textheight]{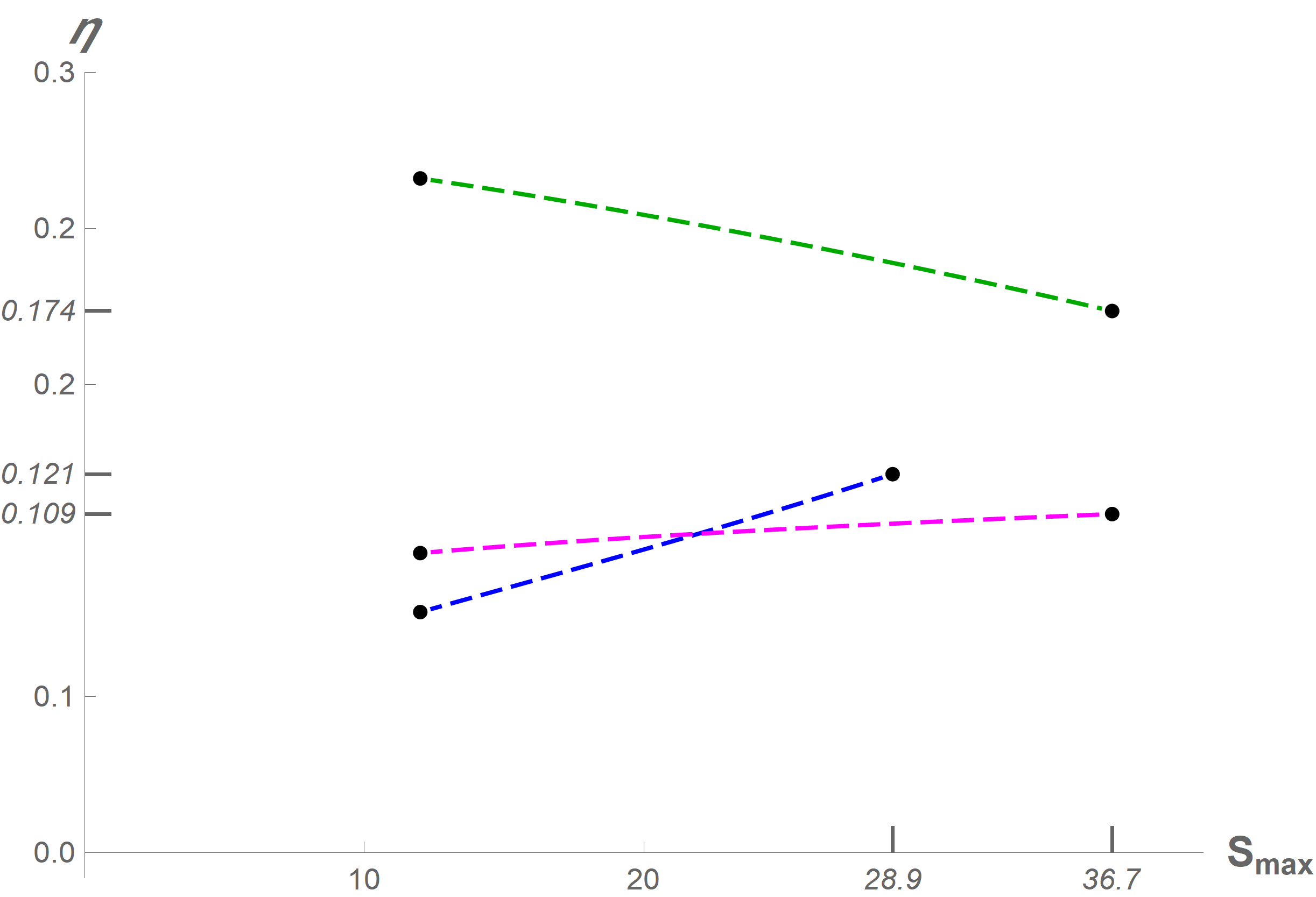}
        \caption{\textcolor{black}{\textit{Efficiency vs.\ maximum entropy for the small black hole branch.}
The Otto (blue), Diesel (green), and Brayton (magenta) efficiencies are shown as functions of $S_{\max}$ using the small branch parameters specified above. All three curves have finite lower and upper endpoints in $S_{\max}$. As $S_{\max}\to S_{\min}^{+}$, the work and heat input vanish together while the efficiency ratios approach finite limiting values. At the upper endpoints, the Otto curve terminates when the state
$(S_{\max},V_{\min})$ approaches the branch-merger point $x=2/3$,
whereas the Diesel and Brayton curves terminate when the corresponding
high-pressure state $(S_{\max},P_2)$ approaches that same branch-merger
point.}}
        \label{fig:eff_smax_small}
    \end{minipage}\hfill
    \begin{minipage}[t]{0.48\textwidth}
        \vspace{0pt}\centering
        \includegraphics[height=0.22\textheight]{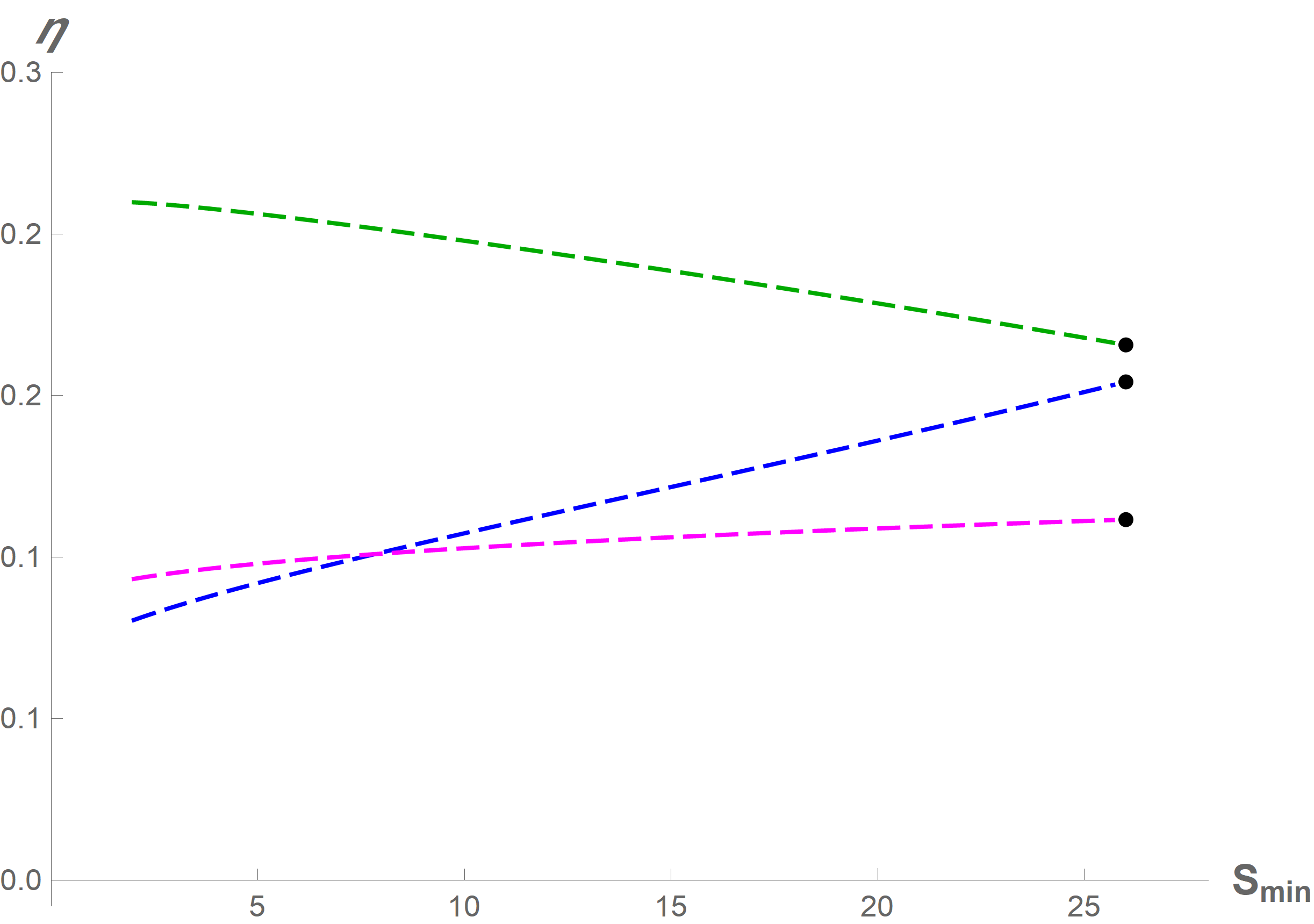}
        \caption{\textcolor{black}{\textit{Efficiency vs.\ minimum entropy for the small black hole branch.}
The Otto (blue), Diesel (green), and Brayton (magenta) efficiencies are shown as functions of $S_{\min}$ at fixed $S_{\max}$, with the other parameters specified above. Every cycle vertex remains on the small branch throughout
$0<S_{\min}<S_{\max}$, so the curves extend to the state space boundary
$S_{\min}\to0^{+}$ without encountering the branch-merger point
$x=2/3$. In this limit, the Otto, Diesel, and Brayton efficiencies
approach approximately $0.064$, $0.203$, and $0.083$, respectively.
At $S_{\min}=0$ the horizon disappears, so these are limiting values
rather than efficiencies of cycles whose vertices all represent
Schwarzschild black holes. As $S_{\min}\to S_{\max}^{-}$, the work and
heat input vanish for all three cycles while the efficiency ratios
approach finite limiting values; the exactly equal entropy cycles are
degenerate.}}
        \label{fig:eff_smin_small}
    \end{minipage}
\end{figure*}

{\color{black}
For completeness, this appendix presents numerical efficiency comparisons
obtained by evaluating the formulas derived in
Appendix~\ref{app:efficiency_derivations} on the small black hole branch. This branch has negative fixed-volume heat capacity and is
therefore not used as a passively stable reservoir-driven working substance
in the main text. Nevertheless, the corresponding quasi-static cycle paths
and efficiency ratios can be evaluated using the same thermodynamic
equations of state.

For the Stirling plots, the large  and small branch results are obtained
from the two roots of the horizon cubic at the same temperature and volume
values. The vertices are ordered separately on each branch so that the hot
isotherm is a heat-input stroke and the $P$-$V$ loop is clockwise. Thus, the Stirling plots use the same displayed volume ranges,
reservoir temperatures, and Carnot reference as their large-branch
counterparts, although the volume ordering and the direction of
traversal along the isotherms are branch dependent.

For the Otto, Diesel, and Brayton plots, by contrast, every small branch
vertex must satisfy $V>9GS$. The entropy and pressure parameters used for
the large-branch plots therefore cannot in general be reused. We instead
choose the representative small branch values
$S_{\min}=12$, $S_{\max}=26$, and $P_2=1.2\times10^{-3}$ for the
volume-dependent comparisons, and
$P_1=6\times10^{-4}$ and $P_2=1.2\times10^{-3}$ for the
entropy-dependent comparisons, with $S_{\min}=12$ when varying
$S_{\max}$ and $S_{\max}=26$ when varying $S_{\min}$. The displayed
volume ranges are kept the same as in the corresponding large-branch
figures. In all numerical plots, we set $G=1$. The curves in
Figures~\ref{fig:eff_vmax_small}--\ref{fig:eff_smin_small} are dashed to
emphasize the negative fixed-volume heat capacity of the small branch.

}

   \clearpage

\newpage

\section{Equations for thermodynamic processes in the $P$-$V$ and $T$-$S$ diagrams}
\label{app:cycle_paths}

To construct the exact $P$-$V$ and $T$-$S$ diagrams for the holographic Schwarzschild heat engines, we need explicit equations for the constant-property curves corresponding to adiabats, isotherms, isochores, and isobars. In four spacetime dimensions, these curves can be written in closed form.  The only non-trivial steps are the inversion of the Tolman temperature equation $T(S,V)=T_0$, needed for isotherms in the $P$-$V$ plane, and the inversion of the pressure equation $P(S,V)=P_0$, needed for isobars in the $T$-$S$ plane. We use the closed-form inversions summarized in Section~\ref{sec:thermo} in the main text and derived in \cite{York1990, Borsboom:2026sex}. 

\vspace{0.3 cm}

\noindent\textbf{Paths in the $P$-$V$ plane.}
In the $P$-$V$ plane, isochores and isobars are vertical and horizontal lines, respectively. The nontrivial curves are the adiabats and isotherms. Let an adiabat have fixed entropy $S_0$, and define
\begin{equation}
x_{S_0}(V)\equiv \frac{r_{h0}}{r_B(V)}
=2\sqrt{\frac{GS_0}{V}}\,,
\end{equation}
then the pressure as a function of the volume, at fixed \textcolor{black}{$S_0$,} is given by 
\begin{equation} \label{eq:adiabat_PV}
    P(V; S_0) =
    \frac{1}{4 G \sqrt{\pi V}}
    \left[
    \frac{1-\tfrac12 x_{S_0}(V)}
    {\sqrt{1-x_{S_0}(V)}}-1
    \right]\,,
\end{equation}
\textcolor{black}{defined on the full physical domain $V>4GS_0$. Its large-black-hole portion has $4GS_0<V<9GS_0$, while its small-black-hole portion has $V>9GS_0$; the two portions meet at $V=9GS_0$, where $x=2/3$. Since fixing $S_0$ fixes the horizon radius, both portions belong to the same constant entropy curve, with the branch selected by the range of $V$.}

\begin{figure*}[!ht]
 \vspace*{2cm} 
    \centering

    \begin{minipage}[t]{0.49\textwidth}
        \centering
        \vspace{0pt}
        \includegraphics[height=6cm,keepaspectratio]{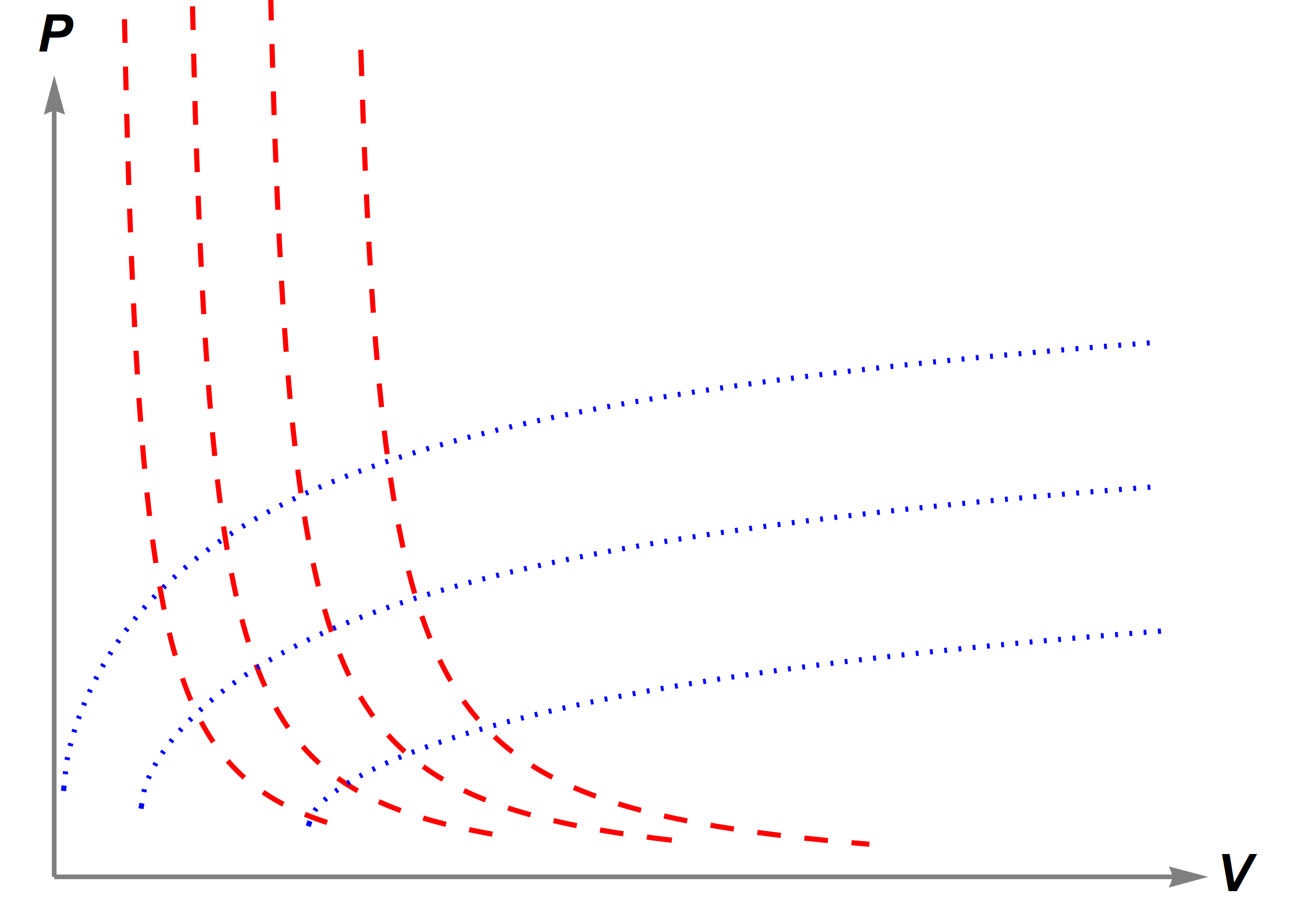}
        \par\vspace{1mm}
         (a) Adiabats and isotherms in the $P$-$V$ plane.
    \end{minipage}
    \hfill
    \begin{minipage}[t]{0.49\textwidth}
        \centering
        \vspace{0pt}
        \includegraphics[height=6cm,keepaspectratio]{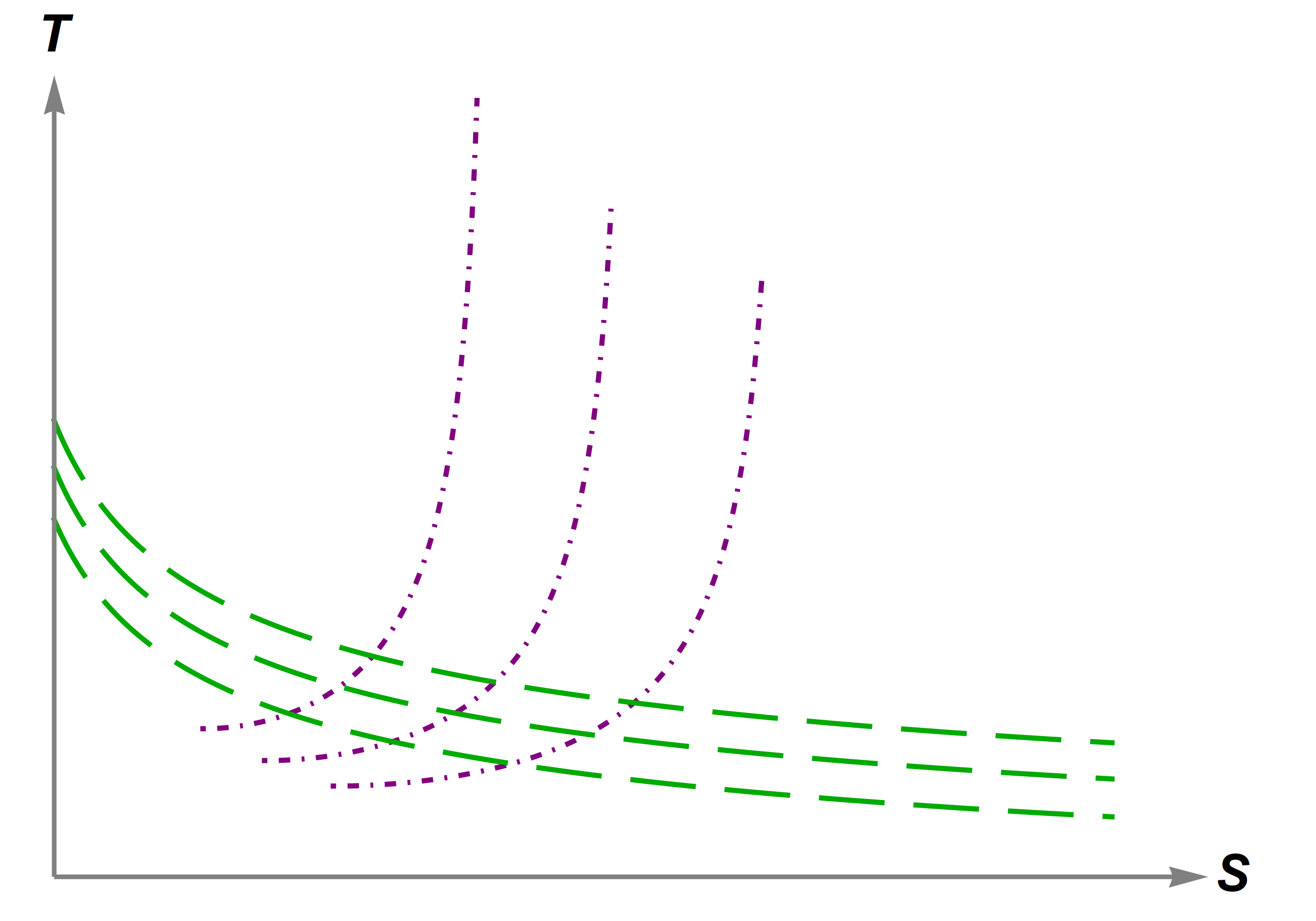}
        \par\vspace{1mm}
         (b) Isochores and isobars in the $T$-$S$ plane.
    \end{minipage}
   \caption{\textit{Constant-property paths for the large black hole
branch.}
(a) $P$-$V$ plane with representative large branch portions of the
adiabats (red dashed) and isotherms (blue dotted) for a
four-dimensional Schwarzschild black hole in a cavity.
For an adiabat of fixed entropy $S_0$, the large branch segment lies
in $4GS_0<V<9GS_0$. The pressure diverges as $V\to(4GS_0)^+$, where the horizon approaches
the cavity wall, and the large branch segment terminates at the
branch-merger point $V=9GS_0$, where $x=2/3$.
Each large branch isotherm begins at
$V=V_{\min}(T_0)$, where the two isothermal branches meet.
(b) $T$-$S$ plane with representative large branch portions of the
isochores (purple dot-dashed) and isobars (green dashed) for the same
thermal system. The isochores diverge as
$S\to V_0/(4G)$, where the horizon approaches the cavity wall.
These curves provide the curved strokes from which the $P$-$V$ and
$T$-$S$ cycles in Figures~\ref{fig:pv_cycles} and~\ref{fig:ts_cycles} are assembled.}
    \label{fig:constant_property_paths_large}
\end{figure*}

\begin{figure*}[!ht]
    \centering
    \begin{minipage}[t]{0.49\textwidth}
        \centering
        \vspace{0pt}
        \includegraphics[height=6cm,keepaspectratio]{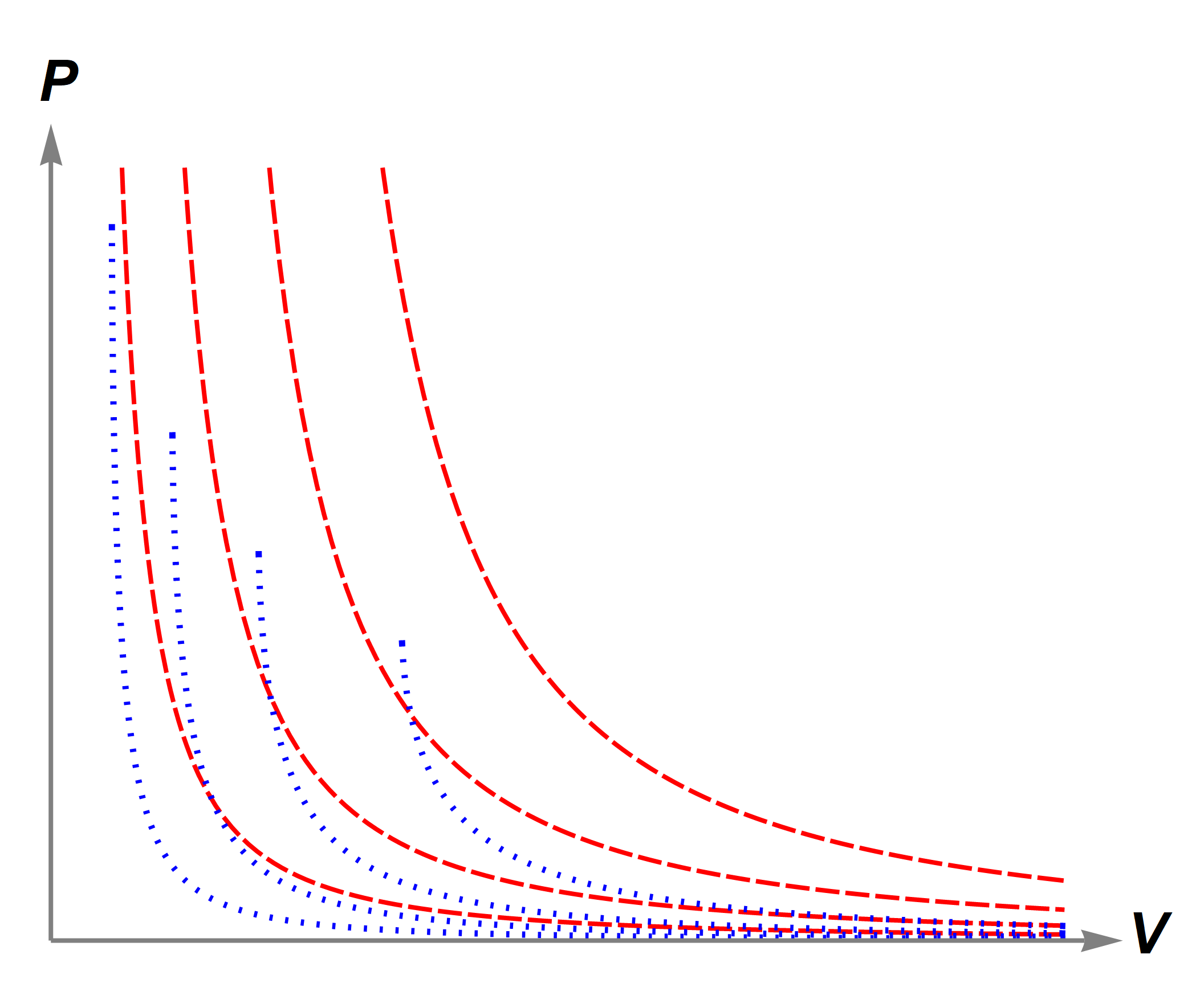}
         \par\vspace{1mm}
         (a) Adiabats and isotherms in the $P$-$V$ plane.
    \end{minipage}\hfill
    \begin{minipage}[t]{0.51\textwidth}
        \centering
        \vspace{0pt}
        \includegraphics[height=6cm,keepaspectratio]{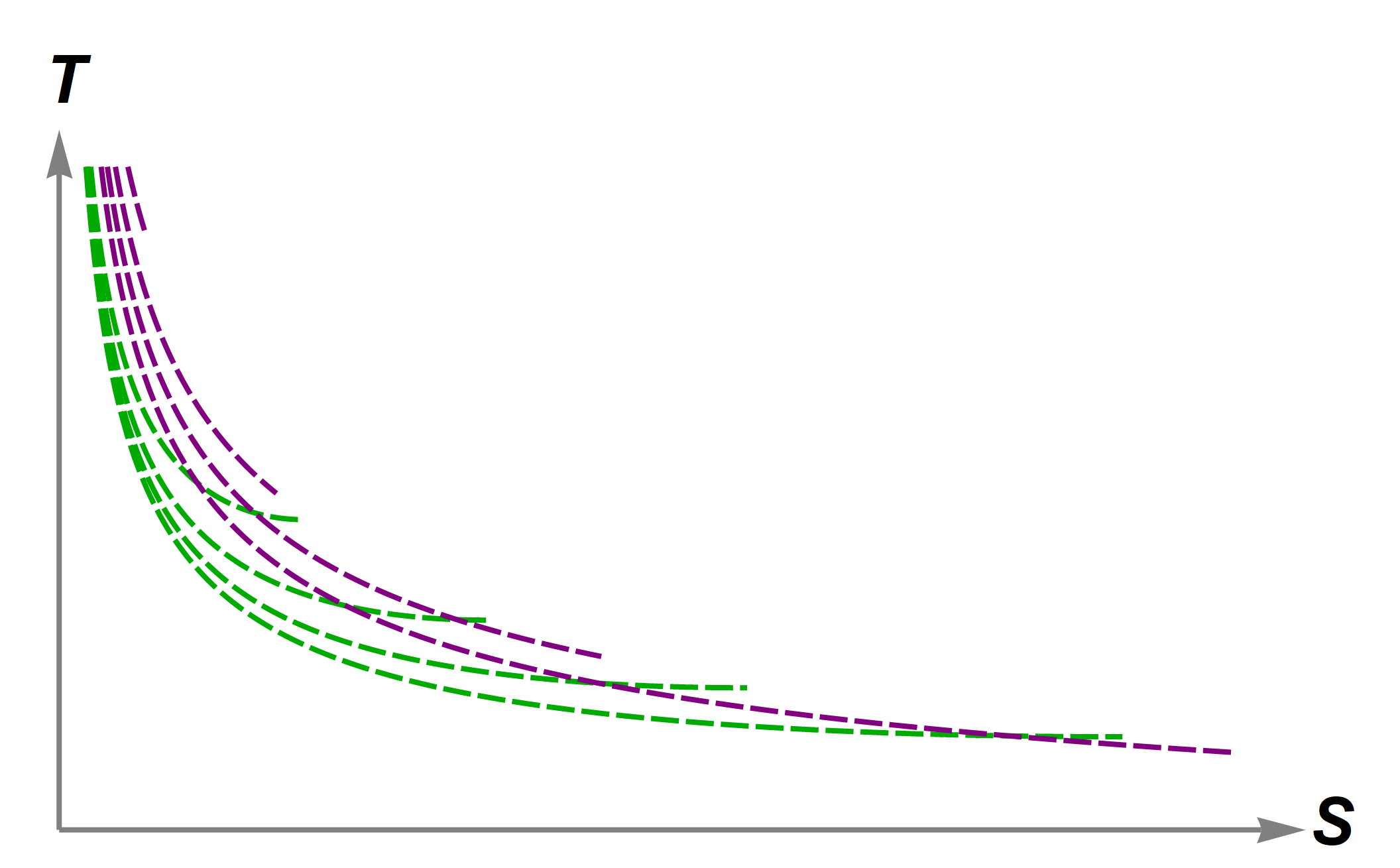}
         \par\vspace{1mm}
         (b) Isochores and isobars in the $T$-$S$ plane.
    \end{minipage}
  \caption{\textit{Constant-property paths for the small black hole branch.}
(a) $P$-$V$ plane with representative small-branch portions of the adiabats (red dashed) and isotherms (blue dotted) for a four-dimensional Schwarzschild black hole in a cavity. The adiabatic curves are determined by $S_0$ alone, and their small branch segments lie in $V>9GS_0$. For each temperature $T_0$, the small branch isotherm approaches its branch-merger point as $V\to V_{\min}(T_0)^+$, where $V_{\min}(T_0)=27/(16\pi T_0^2)$, and approaches $P\to0$ as the cavity grows  and the horizon radius approaches
$r_h=1/(4\pi T_0)$.
(b) $T$-$S$ plane with representative small branch portions of the
isochores (purple dot-dashed) and isobars (green dashed).
Both families diverge as $S\to0$ and terminate at their respective
branch-merger points, where $x=2/3$.}
    \label{fig:paths_small}
\end{figure*}

For an isotherm, the boundary temperature is fixed to $T_0$. The horizon radius is obtained from the corresponding large or small branch root $r_h(T_0,V)$ of the Tolman temperature equation. Defining
\begin{equation}
x_{T_0}(V)\equiv \frac{r_h(T_0,V)}{\sqrt{V/(4\pi)}}\,,
\end{equation}
the isotherm on either branch  is
\begin{equation} \label{eq:isotherm_PV}
    P(V; T_0) =
    \frac{1}{4 G \sqrt{\pi V}}
    \left[
    \frac{1-\tfrac12 x_{T_0}(V)}
    {\sqrt{1-x_{T_0}(V)}}-1
    \right]\,.
\end{equation}
The two isothermal branches meet at
\begin{equation}\label{eq:Vmin_iso}
    V_{\min}(T_0)=\frac{27}{16\pi T_0^2},
\end{equation}
where $x_{T_0}=2/3$. For $V>V_{\min}(T_0)$, equation \eqref{eq:isotherm_PV} gives either the large branch or small branch isotherm, depending on the root chosen for $r_h(T_0,V)$. Representative large branch curves are shown in Figure~\ref{fig:constant_property_paths_large}, while the corresponding small branch curves are shown in Figure~\ref{fig:paths_small}.\\

\noindent\textbf{Paths in the $T$-$S$ plane.}
In the $T$-$S$ plane, isotherms and adiabats are horizontal and vertical lines, respectively. The non-trivial curves are the isochores and isobars. Since both can be parametrized directly by the entropy,   no separate branch inversion is required. The small and large black hole branches are the two portions of the same curve.

For an isochore the volume is fixed, $V=V_0$. Defining
\begin{equation}
x_{V_0}(S)\equiv \frac{r_h(S)}{r_{B0}}
=
2\sqrt{\frac{GS}{V_0}}\,,
\end{equation}
the Tolman temperature gives
\begin{equation}\label{eq:isochore_TS}
    T(S;V_0)
    =
    \frac{1}{4\sqrt{\pi G S}\sqrt{1-x_{V_0}(S)}}\,,
\end{equation}
with domain $0<S<V_0/(4G)$. The branch-separation point is $x_{V_0}=2/3$, or $S=V_0/(9G)$.

\begin{figure}[p]
 \vspace*{4cm} 
    \centering
    \begin{minipage}{0.33\textwidth}
        \centering
        \includegraphics[width=\linewidth]{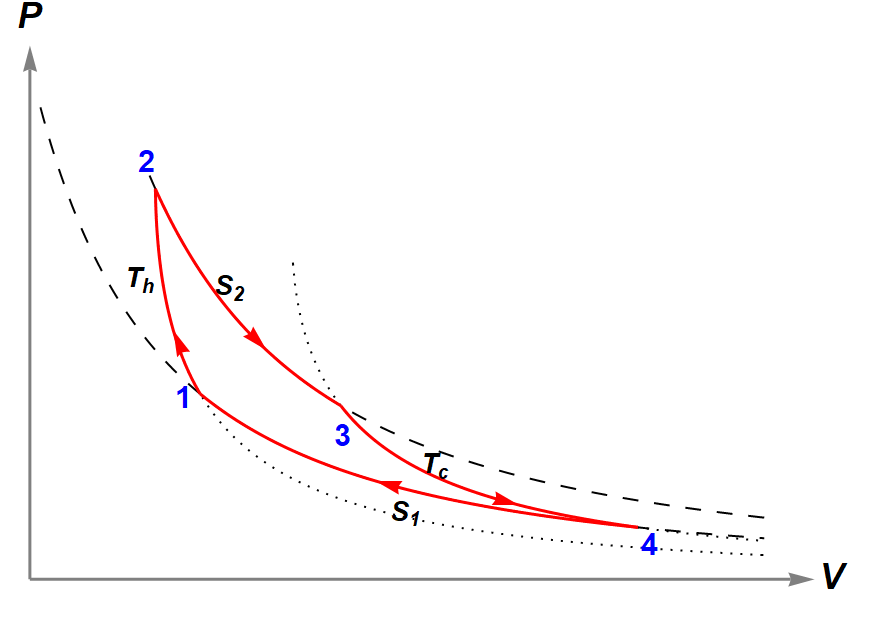}
        \par\vspace{4pt}
        (a) Carnot cycle ($P$-$V$)
    \end{minipage}\hfill
    \begin{minipage}{0.33\textwidth}
        \centering
        \includegraphics[width=\linewidth]
        {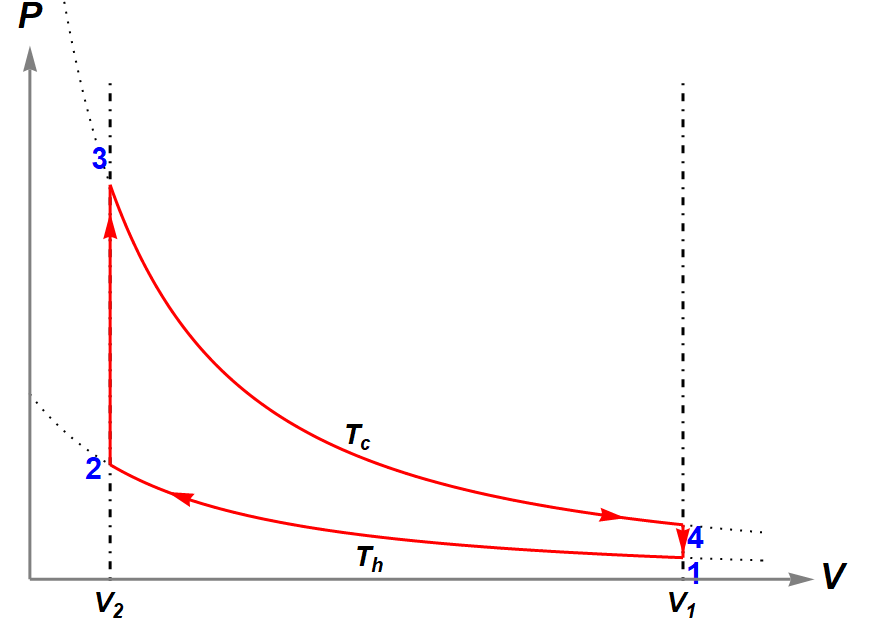}
      \par\vspace{4pt}
        (b)  Stirling cycle ($P$-$V$)      \end{minipage}\hfill
    \begin{minipage}{0.33\textwidth}
        \centering
        \includegraphics[width=\linewidth]
          {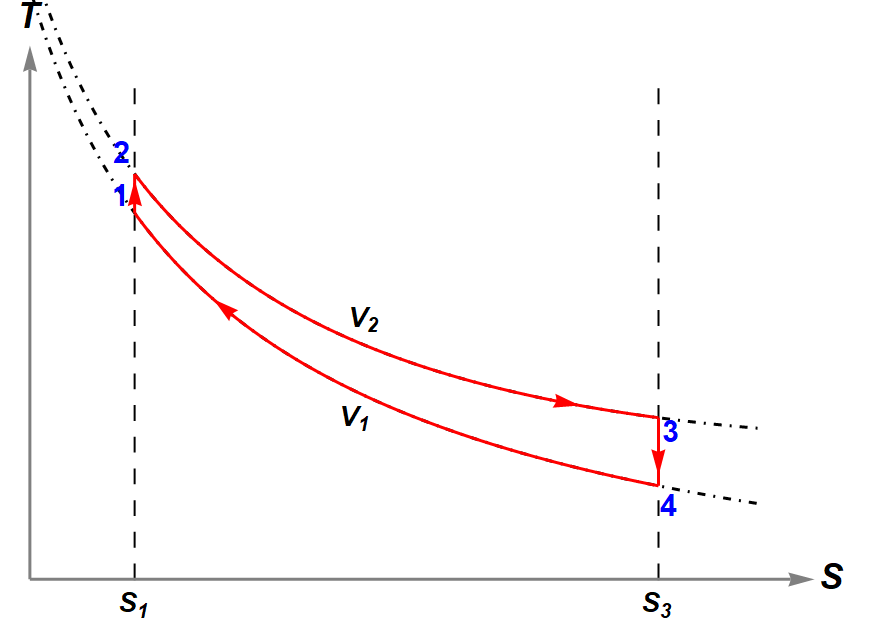}
        \par\vspace{4pt}
        (c) Otto cycle ($T$-$S$)
    \end{minipage}

        \vspace{6mm}
    \begin{minipage}{0.33\textwidth}
        \centering
        \includegraphics[width=\linewidth]{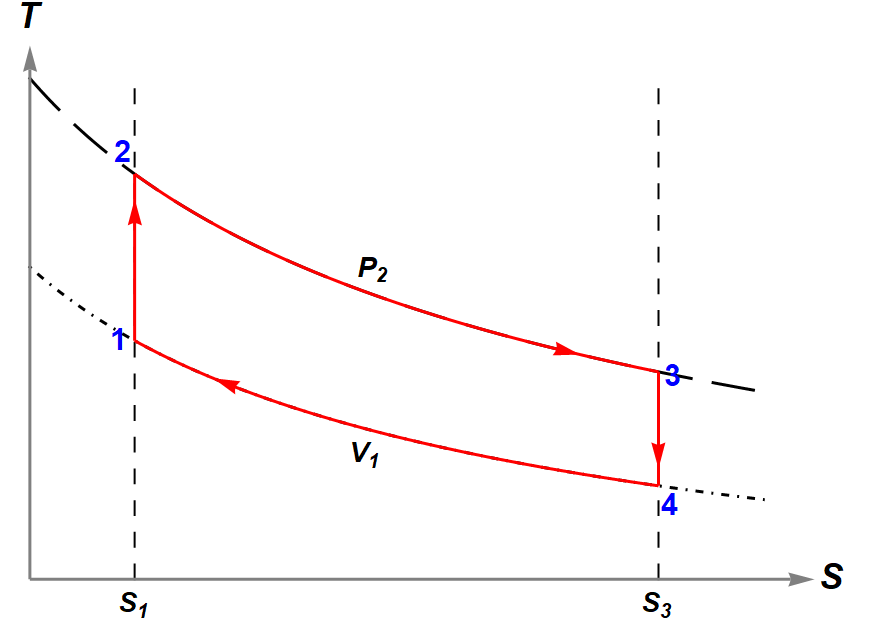}
        \par\vspace{4pt}
        (d) Diesel cycle ($T$-$S$)
    \end{minipage}\hspace{0.034\textwidth}%
    \begin{minipage}{0.33\textwidth}
        \centering
        \includegraphics[width=\linewidth]{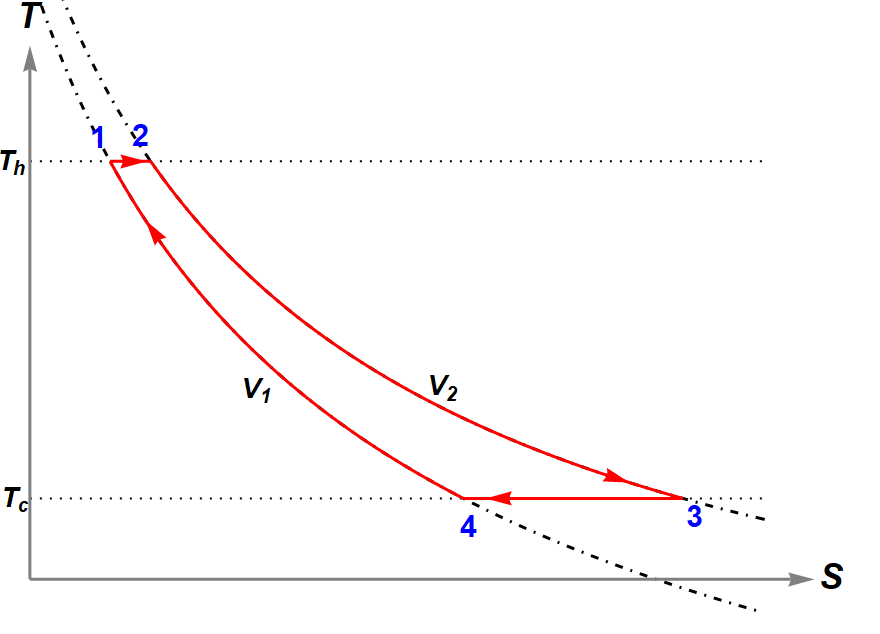}
        \par\vspace{4pt}
        (e) Stirling cycle ($T$-$S$)  
    \end{minipage}
\caption{\textit{Pressure-volume and temperature-entropy diagrams for the small black hole branch.}
Representative cycle diagrams for a thermal system on a spherical cavity dual to a four-dimensional Schwarzschild black hole on the small branch. Only those cycle diagrams that differ qualitatively from their
large branch counterparts are shown. The omitted diagrams (Otto, Diesel,
and Brayton in the $P$-$V$ plane, and Carnot and Brayton in the $T$-$S$
plane) have the same qualitative form as the corresponding large branch
diagrams in Figures~\ref{fig:pv_cycles} and~\ref{fig:ts_cycles}. In each panel, the solid red curves trace the working substance through the vertices $1\to2\to3\to4\to1$ in the heat engine orientation, clockwise in the $P$-$V$ plane. The auxiliary constant-property curves use the same line styles as before: isotherms are dotted, adiabats short dashed, isobars long dashed, and isochores dot-dashed. The qualitative differences from the large branch arise from the reversed small branch response to heat exchange. Isothermal heat input still increases the entropy, but it is accompanied by a decrease of the \textcolor{black}{boundary volume}. Isothermal heat rejection decreases the entropy, but is accompanied by an increase of the \textcolor{black}{boundary volume}. Thus, in cycles with isothermal heat input, the hot isotherm is traversed toward smaller $V$. In the $P$-$V$ plane, the small branch  also reverses the ordering of the isotherms: at fixed $V$, the hot isotherm lies below the cold isotherm over their common range. For the Stirling cycle this implies the opposite volume ordering from the large branch, $V_1>V_2$, so that the hot isotherm $1\to2$ is a heat input stroke and the $P$-$V$ loop is clockwise. In the $T$-$S$ plane the temperature ordering remains $T_{\rm h}>T_{\rm c}$, but the isochores slope downward with increasing entropy, because the fixed-volume heat capacity is negative.}
    \label{fig:small_branch_cycles}
\end{figure}

For an isobar the pressure is fixed, $P=P_0$. The constant-pressure curve is
\begin{equation}\label{eq:isobar_TS}
    T(S;P_0)
    =
    \frac{1}{4\sqrt{\pi G S}\;y(S,P_0)}\,.
\end{equation}
For $S>0$ and $P_0>0$, $y(S,P_0)$ denotes the unique admissible redshift parameter in the interval $0<y<1$, given in closed form by \eqref{eq:ySP} in the main text. The branch is identified by $x_P=1-y^2$: the small branch has $x_P<2/3$, while the large branch has $x_P>2/3$. Representative isochores and isobars are shown in Figure~\ref{fig:constant_property_paths_large}, with the small branch portions displayed separately in Figure~\ref{fig:paths_small}.
Figure~\ref{fig:small_branch_cycles} shows the   $P$-$V$ and
$T$-$S$ cycle diagrams for the standard engines on the small
Schwarzschild black hole branch. These diagrams complement the
large branch cycles shown in Figures~\ref{fig:pv_cycles} and
\ref{fig:ts_cycles}, but should be interpreted with the caveat that
the small branch has negative \textcolor{black}{$C_V$.}

\newpage

\clearpage

 \section{\textcolor{black}{High-temperature behavior of the Stirling cycle}}
\label{app:stirling_limit}

{\color{black}
This appendix first derives the high-temperature behavior of the Stirling cycle
on the large black hole branch. We keep $T_{\rm c}$, $V_1$, and $V_2$ fixed,
take $V_2>V_1$, and send $T_{\rm h}\to\infty$. We then determine the
corresponding limit for the   small branch regenerative cycle and show
explicitly that its efficiency remains separated from the Carnot bound.
}

\par\medskip
\noindent\textbf{\textcolor{black}{Regenerative Stirling cycle for the large branch.}}
We use the notation introduced before for the entropy difference at fixed temperature,
\begin{equation}
\Delta S|_T
\equiv
S(T,V_2)-S(T,V_1)\,,
\end{equation}
with all quantities evaluated on the large black hole branch. Starting from the \textcolor{black}{regenerative Stirling efficiency} derived in \eqref{regeneffst}, one finds
\begin{equation}
\eta_{\rm Carnot}
-
\eta_{\mathrm{Stirling}}^{\mathrm{reg}}
=
\frac{
T_{\rm c}\left[
T_{\rm h}(\Delta S|_{T_{\rm c}}-\Delta S|_{T_{\rm h}})
-
Q_{\mathrm{mis}}
\right]
}{
T_{\rm h}
\left(
T_{\rm h}\Delta S|_{T_{\rm h}}
+
Q_{\mathrm{mis}}
\right)
}\,.
\end{equation}
{\color{black}
To determine the sign and asymptotic size of this expression, we use the
fixed-deficit form of the mismatch. With the signed convention
\eqref{eq:local-isochoric-mismatch}, this is
\begin{equation}
Q_{\mathrm{mis}}
=
-\int_{T_{\rm c}}^{T_{\rm h}}dQ_{\mathrm{loc}}(T)
=
\int_{T_{\rm c}}^{T_{\rm h}}
\left[
C_V(T,V_1)-C_V(T,V_2)
\right]dT\,.
\end{equation}
}
Since $C_V=T(\partial S/\partial T)_V$, this can be written as
\begin{equation}
Q_{\mathrm{mis}}
=
-\int_{T_{\rm c}}^{T_{\rm h}}
T\,\frac{d}{dT}\Delta S|_T dT\,.
\end{equation}
Integrating by parts gives
\begin{equation}
Q_{\mathrm{mis}}
=
T_{\rm c}\Delta S|_{T_{\rm c}}
-
T_{\rm h}\Delta S|_{T_{\rm h}}
+
\int_{T_{\rm c}}^{T_{\rm h}}\Delta S|_T dT\,.
\end{equation}
Therefore
\begin{equation}\label{effgapapp}
\eta_{\rm Carnot}
-
\eta_{\mathrm{Stirling}}^{\mathrm{reg}}
=
\frac{
T_{\rm c}
}{
T_{\rm h}
\left(
T_{\rm h}\Delta S|_{T_{\rm h}}
+
Q_{\mathrm{mis}}
\right)
}
\int_{T_{\rm c}}^{T_{\rm h}}
\left[
\Delta S|_{T_{\rm c}}-\Delta S|_T
\right]dT\,.
\end{equation}
This form makes the sign transparent. On the large branch, $C_V(T,V)$ decreases with $V$ at fixed $T$. Hence, for $V_2>V_1$,
\begin{equation}
\frac{d}{dT}\Delta S|_T
=
\frac{1}{T}
\left[
C_V(T,V_2)-C_V(T,V_1)
\right]
<0\,.
\end{equation}
Thus $\Delta S|_T<\Delta S|_{T_{\rm c}}$ for all $T>T_{\rm c}$, and the integral is strictly positive. Hence
\begin{equation}
\eta_{\mathrm{Stirling}}^{\mathrm{reg}}
<
\eta_{\rm Carnot}
\end{equation}
\textcolor{black}{for every finite large branch cycle with finite $T_{\rm h}>T_{\rm c}$ and $V_2>V_1$.}

We now extract the leading high-temperature behavior. Let
\begin{equation}
r_{B i}
\equiv
\sqrt{\frac{V_i}{4\pi}}\,,
\qquad
i=1,2.
\end{equation}
On the large black hole branch, the solution of the Tolman equation satisfies
\begin{equation}
r_h(T,V_i)\to r_{B i}\,,
\qquad
T\to\infty.
\end{equation}
The   energy therefore approaches its geometric upper bound,
\begin{equation}
E(T,V_i)\to \frac{r_{B i}}{G}\,,
\qquad
T\to\infty.
\end{equation}
Consequently, the regenerator mismatch approaches the finite limit
\begin{equation}
Q_{\mathrm{mis}}^\infty
\equiv
\lim_{T_{\rm h}\to\infty}Q_{\mathrm{mis}}
=
\frac{r_{B1}-r_{B2}}{G}
+
E(T_{\rm c},V_2)-E(T_{\rm c},V_1)\,.
\end{equation}
{\color{black}
This limit is strictly positive because $Q_{\mathrm{mis}}$ is the integral of
a positive integrand over $[T_{\rm c},T_{\rm h}]$ and increases monotonically
with $T_{\rm h}$.}
Similarly,
{\color{black}
\begin{equation}
\Delta S|_{T_{\rm h}}
\to
\Delta S_{\infty}
\equiv
\frac{\pi}{G}
\left(
r_{B2}^2-r_{B1}^2
\right)
=
\frac{V_2-V_1}{4G}\,.
\end{equation}
}
The heat absorbed along the hot isotherm therefore grows linearly,
{\color{black}
\begin{equation}
Q_{\mathrm{in}}^{1\to2}
=
T_{\rm h}\Delta S|_{T_{\rm h}}
\sim
T_{\rm h}\Delta S_{\infty}\,,
\qquad
T_{\rm h}\to\infty\,.
\end{equation}
}
{\color{black}
Since $\Delta S|_T$ decreases monotonically with $T$ and approaches its
limiting value $\Delta S_{\infty}$, one has
$\Delta S|_{T_{\rm c}}>\Delta S_{\infty}$. Substituting the limiting behavior into the expression \eqref{effgapapp} for the
efficiency gap gives
\begin{equation}
\eta_{\rm Carnot}
-
\eta_{\mathrm{Stirling}}^{\mathrm{reg}}
=
\frac{
T_{\rm c}
\left(
\Delta S|_{T_{\rm c}}-\Delta S_{\infty}
\right)
}{
\Delta S_{\infty}
}
\frac{1}{T_{\rm h}}
+
\mathcal{O}\!\left(\frac{1}{T_{\rm h}^2}\right)\,.
\end{equation}
}
Thus
\begin{equation}
\eta_{\rm Carnot}
-
\eta_{\mathrm{Stirling}}^{\mathrm{reg}}
=
\mathcal{O}\!\left(\frac{1}{T_{\rm h}}\right)\,,
\qquad
T_{\rm h}\to\infty\,.
\end{equation}
{\color{black}
The physical origin of this limit is that, at fixed boundary volume, the large
black hole horizon approaches the cavity wall:
\begin{equation}
    r_h(T_{\rm h},V_i)\longrightarrow r_{B i}\,,
    \qquad
    T_{\rm h}\to\infty\,.
\end{equation}
Thus, the high-temperature limit is simultaneously a  
near-horizon  limit of the Schwarzschild cavity system. The result
should be understood within the leading semiclassical approximation used
throughout this paper. Consequently, the large  branch regenerative Stirling
efficiency approaches the Carnot bound from below.
}

{\color{black}
\par\medskip
\noindent\textbf{Regenerative Stirling cycle for the small branch.}
For completeness, consider the   small branch cycle with $V_1>V_2$,
which is the volume ordering required for heat input along the hot isotherm.
The same integration-by-parts argument used above shows that this cycle is
\textcolor{black}{strictly sub-Carnot for  every finite small branch cycle with finite $T_{\rm h}>T_{\rm c}$ and $V_1>V_2$.} Indeed, on the small branch
$C_{V,\mathrm{s}}(T,V)$ increases with $V$ at fixed $T$, and therefore
\begin{equation}
\frac{d}{dT}\Delta S_{\mathrm{s}}|_T
=
\frac{1}{T}
\left[
C_{V,\mathrm{s}}(T,V_2)-C_{V,\mathrm{s}}(T,V_1)
\right]
<0
\end{equation}
for $V_2<V_1$. It follows that
\begin{equation}
\eta_{\mathrm{Stirling},\mathrm{s}}^{\mathrm{reg}}
<
\eta_{\rm Carnot}
\end{equation}
\textcolor{black}{for every  finite $T_{\rm h}>T_{\rm c}$ and $V_1>V_2$.}

At fixed boundary volume, the small root of the Tolman equation has the
large-temperature expansion
\begin{equation}
r_{h,\mathrm{s}}(T,V_i)
=
\frac{1}{4\pi T}
+
\frac{1}{32\pi^2 r_{B i}T^2}
+
\mathcal{O}\!\left(\frac{1}{T^3}\right).
\end{equation}
Consequently,
\begin{equation}
S_{\mathrm{s}}(T,V_i)
=
\frac{1}{16\pi G T^2}
+
\frac{1}{64\pi^2G r_{B i}T^3}
+
\mathcal{O}\!\left(\frac{1}{T^4}\right),
\end{equation}
and hence
\begin{equation}
\Delta S_{\mathrm{s}}|_T
=
\frac{1}{64\pi^2G}
\left(
\frac{1}{r_{B2}}-\frac{1}{r_{B1}}
\right)
\frac{1}{T^3}
+
\mathcal{O}\!\left(\frac{1}{T^4}\right).
\end{equation}
Since $r_{B1}>r_{B2}$, this entropy difference is positive, while
\begin{equation}
T_{\rm h}\Delta S_{\mathrm{s}}|_{T_{\rm h}}
\longrightarrow 0\,, \qquad T_{\rm h} \to \infty.
\end{equation}
The quasi-local energy also vanishes on the small branch,
$E_{\mathrm{s}}(T,V_i)\to0$, as $T\to\infty$. Therefore, the isochoric
mismatch approaches
\begin{equation}
Q_{\mathrm{mis},\mathrm{s}}^\infty
\equiv
\lim_{T_{\rm h}\to\infty}Q_{\mathrm{mis},\mathrm{s}}
=
E_{\mathrm{s}}(T_{\rm c},V_2)
-
E_{\mathrm{s}}(T_{\rm c},V_1)
>0.
\end{equation}
The strict positivity also follows directly from
$(\partial E_{\mathrm{s}}/\partial V)_T
=
T(\partial S_{\mathrm{s}}/\partial V)_T-P<0$
on the small branch. Substituting these limits into
equation~\eqref{regeneffst} gives
\begin{equation}
\lim_{T_{\rm h}\to\infty}
\eta_{\mathrm{Stirling},\mathrm{s}}^{\mathrm{reg}}
=
1-
\frac{
T_{\rm c}\Delta S_{\mathrm{s}}|_{T_{\rm c}}
}{
E_{\mathrm{s}}(T_{\rm c},V_2)
-
E_{\mathrm{s}}(T_{\rm c},V_1)
}.
\end{equation}
Moreover, along the cold isotherm,
\begin{equation}
E_{\mathrm{s}}(T_{\rm c},V_2)
-
E_{\mathrm{s}}(T_{\rm c},V_1)
=
T_{\rm c}\Delta S_{\mathrm{s}}|_{T_{\rm c}}
-
\int_{V_1}^{V_2}P(T_{\rm c},V)\,dV
>
T_{\rm c}\Delta S_{\mathrm{s}}|_{T_{\rm c}},
\end{equation}
because $V_2<V_1$ and $P>0$. The limiting small branch efficiency is therefore
strictly between zero and one. Since
$\eta_{\rm Carnot}\to1$, its asymptotic separation from the Carnot bound is
\begin{equation}
\lim_{T_{\rm h}\to\infty}
\left(
\eta_{\rm Carnot}
-
\eta_{\mathrm{Stirling},\mathrm{s}}^{\mathrm{reg}}
\right)
=
\frac{
T_{\rm c}\Delta S_{\mathrm{s}}|_{T_{\rm c}}
}{
E_{\mathrm{s}}(T_{\rm c},V_2)
-
E_{\mathrm{s}}(T_{\rm c},V_1)
}
>0.
\end{equation}
Thus, unlike the large branch regenerative efficiency, the   small branch
regenerative efficiency remains a finite distance below the Carnot bound in
the high-temperature limit.
}

\par\medskip
\noindent\textbf{\textcolor{black}{Large-branch comparison with the non-regenerative cycle.}}
The \textcolor{black}{non-regenerative Stirling cycle} follows the same path in the $P$-$V$ plane and therefore has the same net work. The difference is that, without a regenerator, the full isochoric heat input must be supplied externally. On the large branch,
\begin{equation}
Q_{\mathrm{mis}}
=
Q_{\mathrm{in}}^{4\to1}
-
Q_{\mathrm{out}}^{2\to3}\,,
\qquad
Q_{\mathrm{out}}^{2\to3}>0\,,
\end{equation}
so $Q_{\mathrm{mis}}<Q_{\mathrm{in}}^{4\to1}$. Hence
\begin{equation}
{\color{black}\eta_{\mathrm{Stirling}}^{\mathrm{nonreg}}}
=
\eta_{\mathrm{Stirling}}^{\mathrm{reg}}\,
\frac{
Q_{\mathrm{in}}^{1\to2}+Q_{\mathrm{mis}}
}{
Q_{\mathrm{in}}^{1\to2}+Q_{\mathrm{in}}^{4\to1}
}
<
\eta_{\mathrm{Stirling}}^{\mathrm{reg}}\,.
\end{equation}
Together with the sub-Carnot result above, this gives the hierarchy
\begin{equation}
{\color{black}\eta_{\mathrm{Stirling}}^{\mathrm{nonreg}}}
<
\eta_{\mathrm{Stirling}}^{\mathrm{reg}}
<
\eta_{\rm Carnot}
\end{equation}
\textcolor{black}{for every finite large branch cycle with finite $T_{\rm h}>T_{\rm c}$ and $V_2>V_1$.}

The two Stirling efficiencies nevertheless have the same high-temperature limit. Indeed,
\begin{equation}
\lim_{T_{\rm h}\to\infty}
Q_{\mathrm{in}}^{4\to1}
=
\frac{r_{B1}}{G}
-
E(T_{\rm c},V_1)\,,
\end{equation}
is finite, whereas \textcolor{black}{$Q_{\mathrm{in}}^{1\to2}\sim T_{\rm h}\Delta S_{\infty}$}.

{\color{black}
The leading approach of the non-regenerative efficiency can also be obtained
explicitly. Since
\begin{equation}
    Q_{\mathrm{out}}^{2\to3}
    \longrightarrow
    \frac{r_{B2}}{G}-E(T_{\rm c},V_2),
    \qquad
    T_{\rm h}\to\infty,
\end{equation}
one finds
{\color{black}
\begin{align}\label{eq:nonreg-high-temperature-deficit}
    \eta_{\rm Carnot}
    -
    \eta_{\mathrm{Stirling}}^{\mathrm{nonreg}}
    &=
    \left[
        \frac{
        T_{\rm c}\left(\Delta S|_{T_{\rm c}}-\Delta S_{\infty}\right)
        +
        \dfrac{r_{B2}}{G}-E(T_{\rm c},V_2)
        }{
        \Delta S_{\infty}
        }
    \right]\frac{1}{T_{\rm h}}
    +
    \mathcal{O}\!\left(\frac{1}{T_{\rm h}^2}\right).
\end{align}
}
Comparing with the regenerative result gives
{\color{black}
\begin{align}\label{eq:regeneration-leading-improvement}
    \eta_{\mathrm{Stirling}}^{\mathrm{reg}}
    -
    \eta_{\mathrm{Stirling}}^{\mathrm{nonreg}}
    &=
    \frac{
        \dfrac{r_{B2}}{G}-E(T_{\rm c},V_2)
    }{
        \Delta S_{\infty}
    }
    \frac{1}{T_{\rm h}}
    +
    \mathcal{O}\!\left(\frac{1}{T_{\rm h}^2}\right).
\end{align}
}
Since $E(T_{\rm c},V_2)<r_{B2}/G$ on the large branch, the coefficient is
positive. Regeneration therefore removes a strictly positive contribution to
the leading deviation from Carnot, even though both efficiencies have
corrections of order $1/T_{\rm h}$.
}

Therefore
\begin{equation}
\frac{Q_{\mathrm{in}}^{4\to1}}{Q_{\mathrm{in}}^{1\to2}}
\to0\,,
\qquad
\frac{Q_{\mathrm{mis}}}{Q_{\mathrm{in}}^{1\to2}}
\to0\,,
\qquad
T_{\rm h}\to\infty\,.
\end{equation}
It follows that
\begin{equation}
\lim_{T_{\rm h}\to\infty}
{\color{black}\eta_{\mathrm{Stirling}}^{\mathrm{nonreg}}}
=
\lim_{T_{\rm h}\to\infty}
\eta_{\mathrm{Stirling}}^{\mathrm{reg}}
=
\lim_{T_{\rm h}\to\infty}
\eta_{\rm Carnot}
=
1\,.
\end{equation}

\end{document}